\documentclass[useAMS,usenatbib]{mn2e}
\usepackage{myaasmacros}
\usepackage{graphicx}
\usepackage{amssymb}
\usepackage{color}

\def\ltsima{$\; \buildrel < \over \sim \;$}
\def\simlt{\lower.5ex\hbox{\ltsima}}   
\def\gtsima{$\; \buildrel > \over \sim \;$}
\def\simgt{\lower.5ex\hbox{\gtsima}}

\DeclareMathAlphabet{\mathitbf}{OML}{cmm}{b}{it}

\title[Constraining the Milky Way Halo Shape Using Thin Streams]{Constraining the Milky Way Halo Shape Using Thin Streams}

\author[H. Lux et al.]{H. Lux$^{1,2}$\thanks{Lux.Hanni@gmail.com}, J. I. Read$^{3}$, G. Lake$^4$, K. V. Johnston$^5$\\  
$^1$School of Physics and Astronomy, University of Nottingham, University Park, Nottingham, NG7 2RD, UK\\ 
$^2$Department of Physics, University of Oxford, Denys Wilkinson Building, Keble Road, Oxford, OX1 3RH, UK\\
$^3$Department of Physics, University of Surrey, Guildford, GU2 7XH, Surrey, United Kingdom\\
$^4$Department of Theoretical Physics, University of Z\"urich, Winterthurerstr. 190, CH-8057 Z\"urich, Switzerland\\ 
$^5$Department of Astronomy, Columbia University, Pupin Physics Laboratories, 550 West 120th Street, New York, New York 10027, USA\\\\}
 
\begin{document}

\date{Accepted XXX. Received XXX; in original form XXX}

\maketitle

\begin{abstract}
Tidal streams are a powerful probe of the Milky Way (MW) potential shape. In this paper, we introduce a simple test particle method to fit stream data, using a Markov Chain Monte Carlo technique to marginalise over uncertainties in the progenitor's orbit and the Milky Way halo shape parameters. Applying it to mock data of thin streams in the MW halo, we show that, even for very cold streams, stream-orbit offsets -- not modelled in our simple method -- introduce systematic biases in the recovered shape parameters. For the streams that we consider, and our particular choice of potential parameterisation, these errors are of order $\sim 20$\% on the halo flattening parameters. However, larger systematic errors can arise for more general streams and potentials; such offsets need to be correctly modelled in order to obtain an unbiased recovery of the underlying potential. 

Assessing which of the known Milky Way streams are most constraining, we find NGC 5466 and Pal 5 are the most promising candidates. These form an interesting pair as their orbital planes are both approximately perpendicular to each other and to the disc, giving optimal constraints on the MW halo shape. We show that - while with current data their constraints on potential parameters are poor - good radial velocity data along the Pal 5 stream will provide constraints on $q_z$ -- the flattening perpendicular to the disc. Furthermore, as discussed in a companion paper, NGC 5466 can provide rather strong constraints on the MW halo shape parameters, if the tentative evidence for a departure from the smooth orbit  towards its western edge is confirmed. 
\end{abstract}


\section{Introduction}\label{sec:intro}
The shape of galaxy dark matter halos encodes information both about our cosmological model and galaxy formation. In the absence of baryons, numerical simulations of structure formation in $\Lambda$CDM predict triaxial dark matter halos with the shape parameters that vary with radius \cite[e.g.][]{1991ApJ...378..496D,2002ApJ...574..538J}. However, in disc galaxies gas cooling leads to an axisymmetric dark matter potential that is aligned with the disc -- at least out to $\sim 10$ disc scale lengths \citep[e.g.][]{1994ApJ...431..617D,2008ApJ...681.1076D,2010ApJ...720L..62K}. The shape of galaxy halos can also be used to constrain alternative gravity models \citep{2004ApJ...610L..97H,2005MNRAS.361..971R}.

Our own Galaxy provides a unique opportunity to measure the shape of a dark matter halo due to the wealth of phase space data from halo stars \citep[e.g.][]{2012MNRAS.424L..44D}, and kinematic streams \citep[e.g.][]{2001ApJ...551..294I,2006ApJ...642L.137B}. Most attempts to date have focussed on the Sagittarius stream \citep{2001ApJ...547L.133I, 2010ApJ...714..229L, 2013arXiv1301.7069B} due to its length and the quality of available data \citep{2001ApJ...547L.133I,2003ApJ...599.1082M,2004AJ....128..245M,2006ApJ...642L.137B,2009ApJ...700.1282Y, 2010ApJ...714..229L}. However, the stream has consistently defied attempts to build a unified model \citep{2001ApJ...551..294I,2005ApJ...619..800J,2004ApJ...610L..97H,2006ApJ...651..167F,2006ApJ...642L.137B}. \citet{2009ApJ...703L..67L} and \citet{2010ApJ...714..229L} were the first to present a model consistent with both position and velocity data along the stream. They required a mildly triaxial potential with potential flattening $(c/a)_\Phi \sim 0.72$ in the disc plane and $(b/a)_\Phi \sim 0.99$ aligned with the symmetry axis of the Galactic disc\footnote{Note, that $(b/a)_\Phi = 0.99$ implies an oblate potential rather than a triaxial one. However, this refers only to the dark matter halo. Since this is aligned with the symmetry axis of the Galactic disc, the full gravitational potential -- disc plus dark halo -- is triaxial.}. However, such a model is dynamically unstable \citep{2013arXiv1301.2670D}, suggesting un-modelled stream complexity or incomplete data \citep{2013arXiv1301.7069B}. One particular concern is that for streams of the width of Sagittarius, the properties of the progenitor -- whether disc-like, or in a binary pair -- matter \citep{2010MNRAS.408L..26P,2011ApJ...727L...2P}. The orbit can also significantly evolve over time if Sagittarius fell inside a larger `loose group' of galaxies \citep{2008ApJ...686L..61D,2008MNRAS.385.1365L,2008MNRAS.389.1041R,2010MNRAS.406.2312L}, or if Sagittarius were significantly more massive in the past \citep{2004MNRAS.351..891Z}. Thinner, colder streams in the Galactic halo coming from, for example, globular clusters are by contrast significantly simpler to model and understand.

\begin{figure}
\begin{center}
\includegraphics[width=0.22\textwidth]{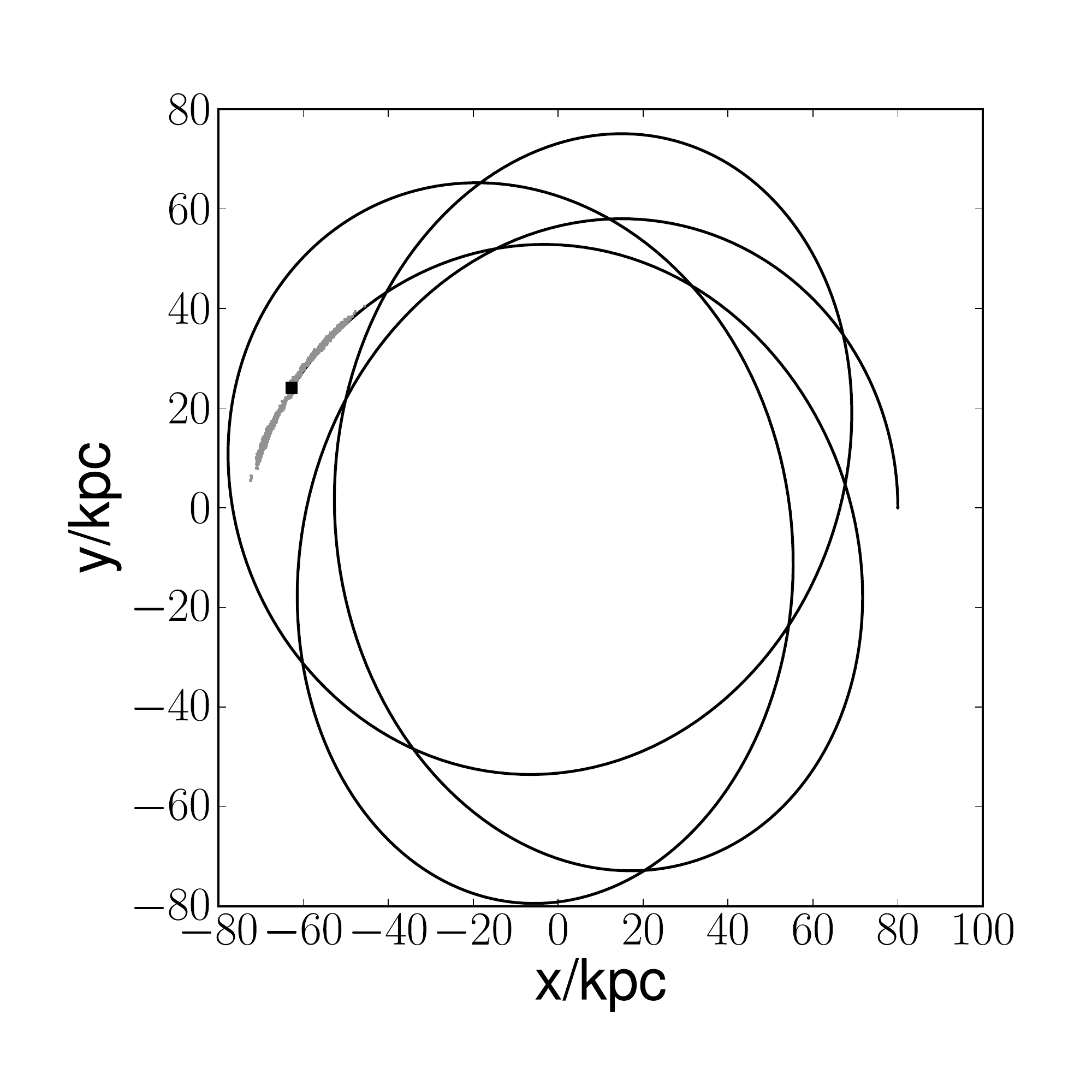}
\includegraphics[width=0.22\textwidth]{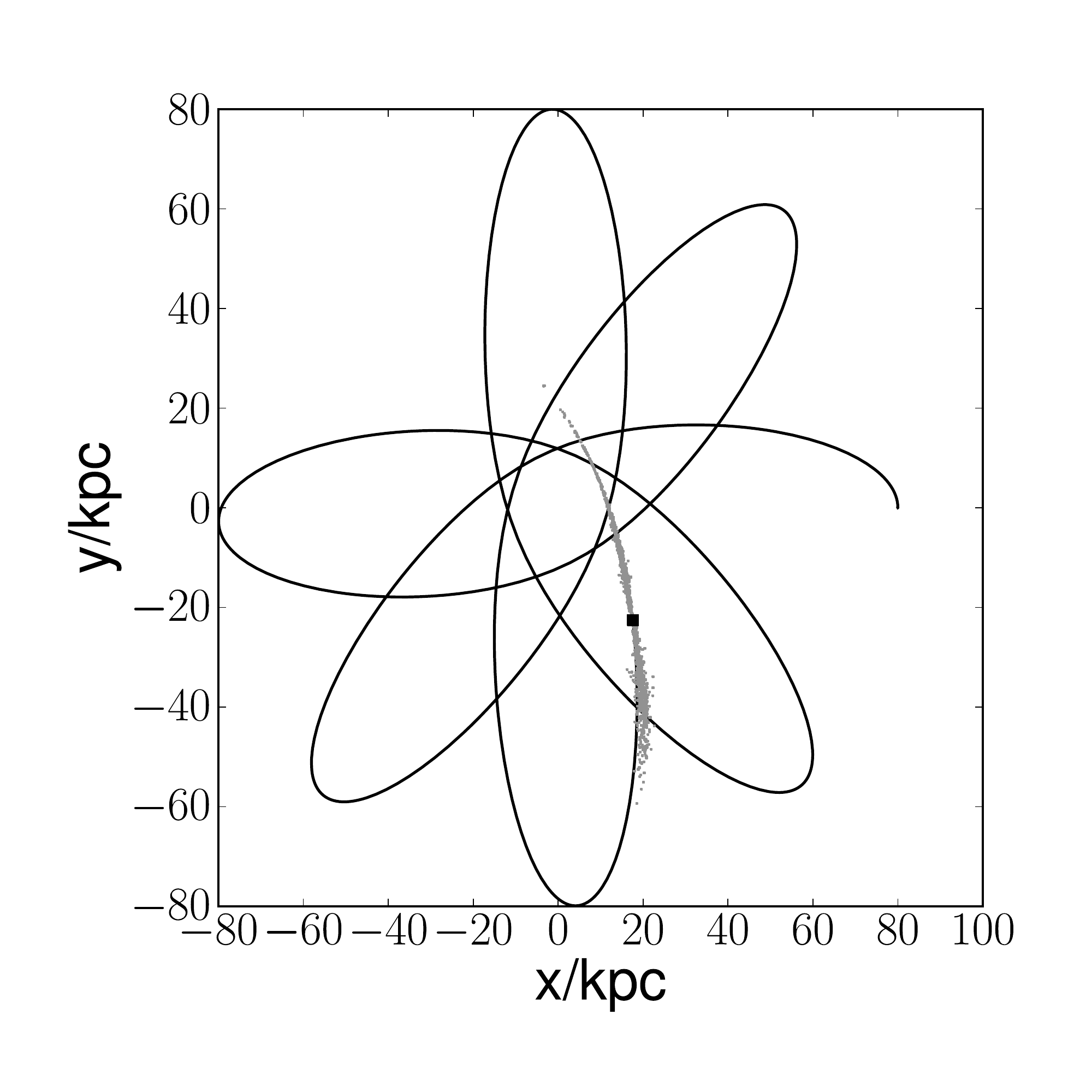}
\includegraphics[width=0.22\textwidth]{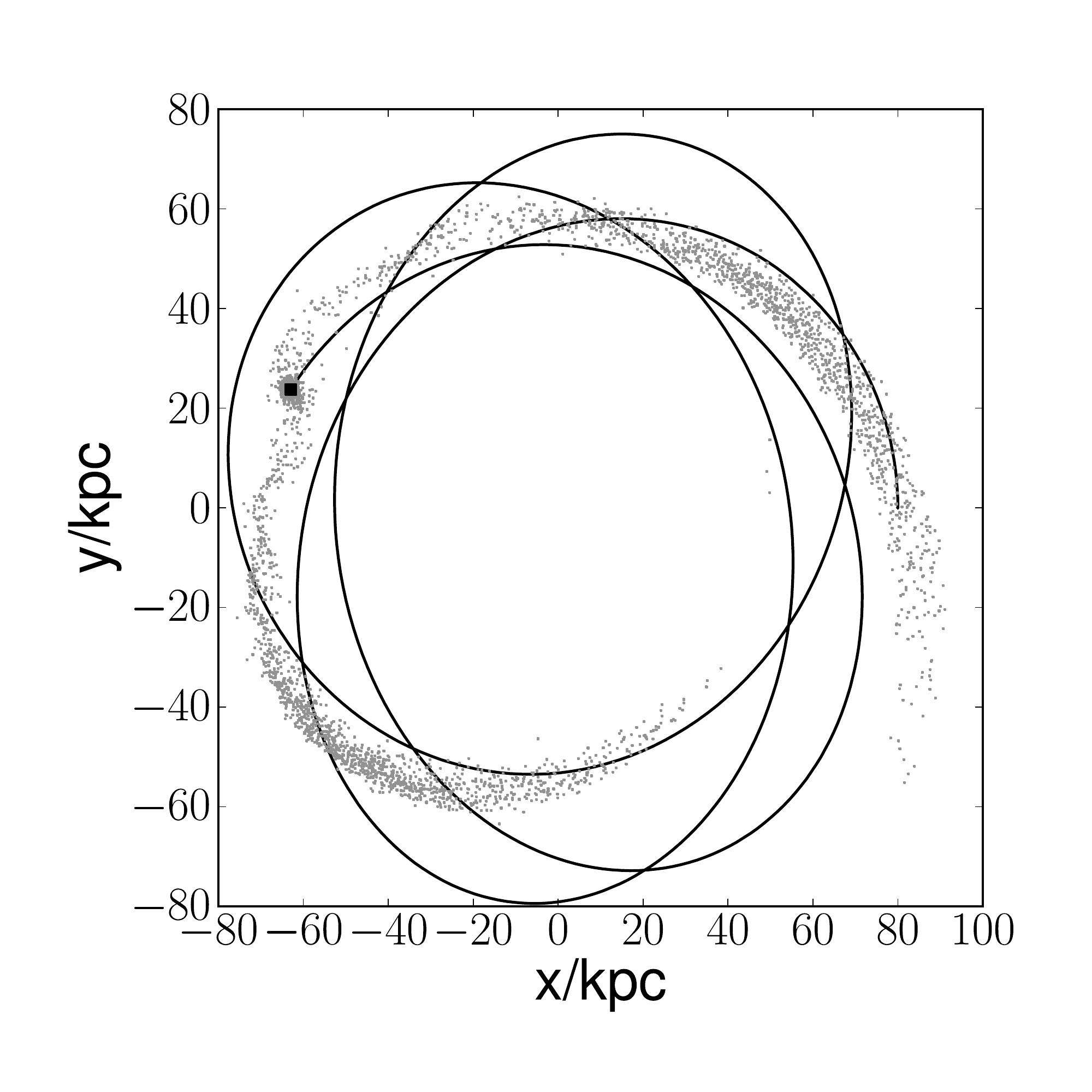}
\includegraphics[width=0.22\textwidth]{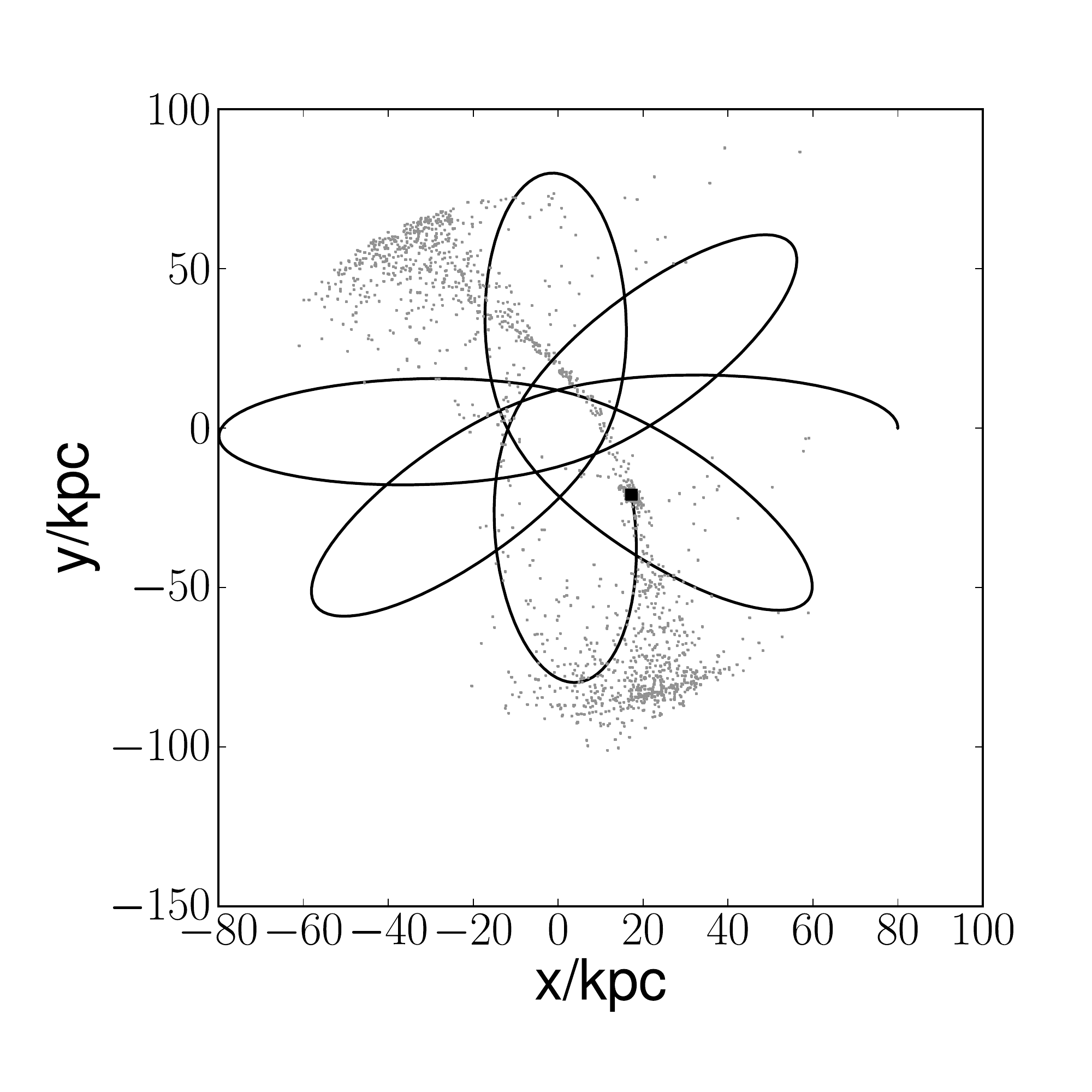}
\caption{The orbit (solid line), tidal debris (grey dots) and current position (square) of a $M_s =10^5$\,M$_\odot$ (upper panels) and a $M_s = 10^8$\,M$_\odot$ satellite (lower panels) on a mildly (left) and a highly (right) eccentric orbit, viewed in the orbital plane of the satellites.} 
\label{fig:SimsPlane}
\end{center}
\end{figure}

With the advent of several large scale stellar surveys such as the Sloan Digital Sky Survey (SDSS) many new streams of different thickness and morphology have been detected around the Milky Way (MW) \citep[e.g.][]{2003AJ....126.2385O,2006ApJ...642L.137B,2006ApJ...643L..17G,2006ApJ...639L..17G,2009ApJ...693.1118G}. This provides us with a wealth of opportunities to study the potential of the Milky Way halo at a variety of distances. Several groups have already fit simple test particle models to some of these streams to determine the Milky Way halo shape and mass (e.g., \citet{2010ApJ...711...32N} for the Orphan stream; and \cite{2009ApJ...697..207W} and \cite{2010ApJ...712..260K} for the GD1 stream). Such `test-particle' approaches are advantageous in that they are extremely fast to run, allowing a large range of potential models for the Milky Way to be rapidly explored. However, real streams do not exactly follow a true orbit \citep{1998ApJ...495..297J,1999MNRAS.307..495H,2001ApJ...557..137J,2011MNRAS.417..198V,2011MNRAS.413.1852E}. As such, fitting simple test particle trajectories will lead to biases on the recovered potential parameters \citep{2013MNRAS.433.1813S}. More sophisticated methods that model such offsets have been explored in the literature, from tracing known satellite debris stars back to a common tidal radius \citep{1999ApJ...512L.109J}; modelling stream tracks using action-angle variables \cite{2011MNRAS.413.1852E}; and modelling tidally stripped debris using test particles \citet{2011MNRAS.417..198V}. Such methods are less biased than fitting test particle orbits, but they come at the cost of increased computational complexity and expense, limiting the size of the parameter space that can be explored.

In this paper, we consider what can be learned from modelling thin cold streams in the Galactic halo. We introduce a fast test particle method that uses a Markov Chain Monte Carlo to marginalise over the measurement uncertainties and model parameters. We use this, applied to mock data generated from N-body models, to determine the magnitude of systematic errors introduced in the model fitting if we ignore stream-orbit offsets. We then use our method to determine which of the known streams in the Milky Way halo are most constraining today, and which will be most promising given improved data in the future. 

This paper is organised as follows. In \S\ref{sec:tracing}, we briefly review the mechanics of thin streams. In \S\ref{sec:data}, we determine which of the currently known streams are suitable candidates for constraining the Milky Way halo shape. In \S\ref{sec:method}, we describe the details of our Markov Chain Monte Carlo (MCMC) method. In \S\ref{sec:results}, we generate mock N-body data with similar properties to known Milky Way streams. We use these data to determine the magnitude of systematic biases arising from neglecting stream-orbit offsets, and to determine which of the known thin streams in the MW halo can provide useful constraints on the halo shape. Finally, in \S\ref{sec:conclusion}, we present our conclusions. 

\begin{figure*}
\begin{center}
$M_s=10^5$\,M$_\odot$, e = 0.21 \hspace{0.12\textwidth} $M_s=10^8$\,M$_\odot$, e = 0.21\hspace{0.12\textwidth} $M_s=10^8$\,M$_\odot$, e = 0.75   \\
\includegraphics[width=0.3\textwidth]{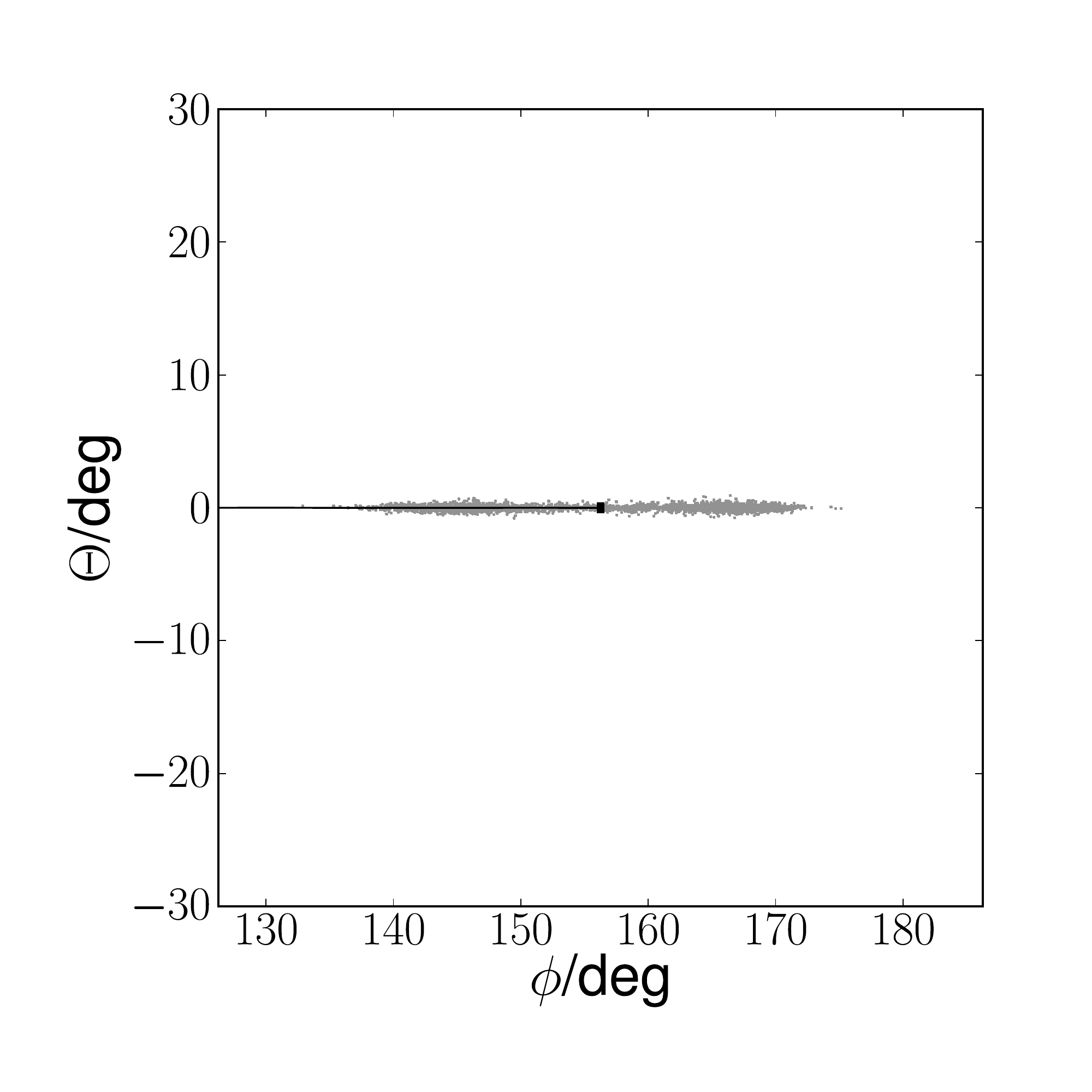}
\includegraphics[width=0.3\textwidth]{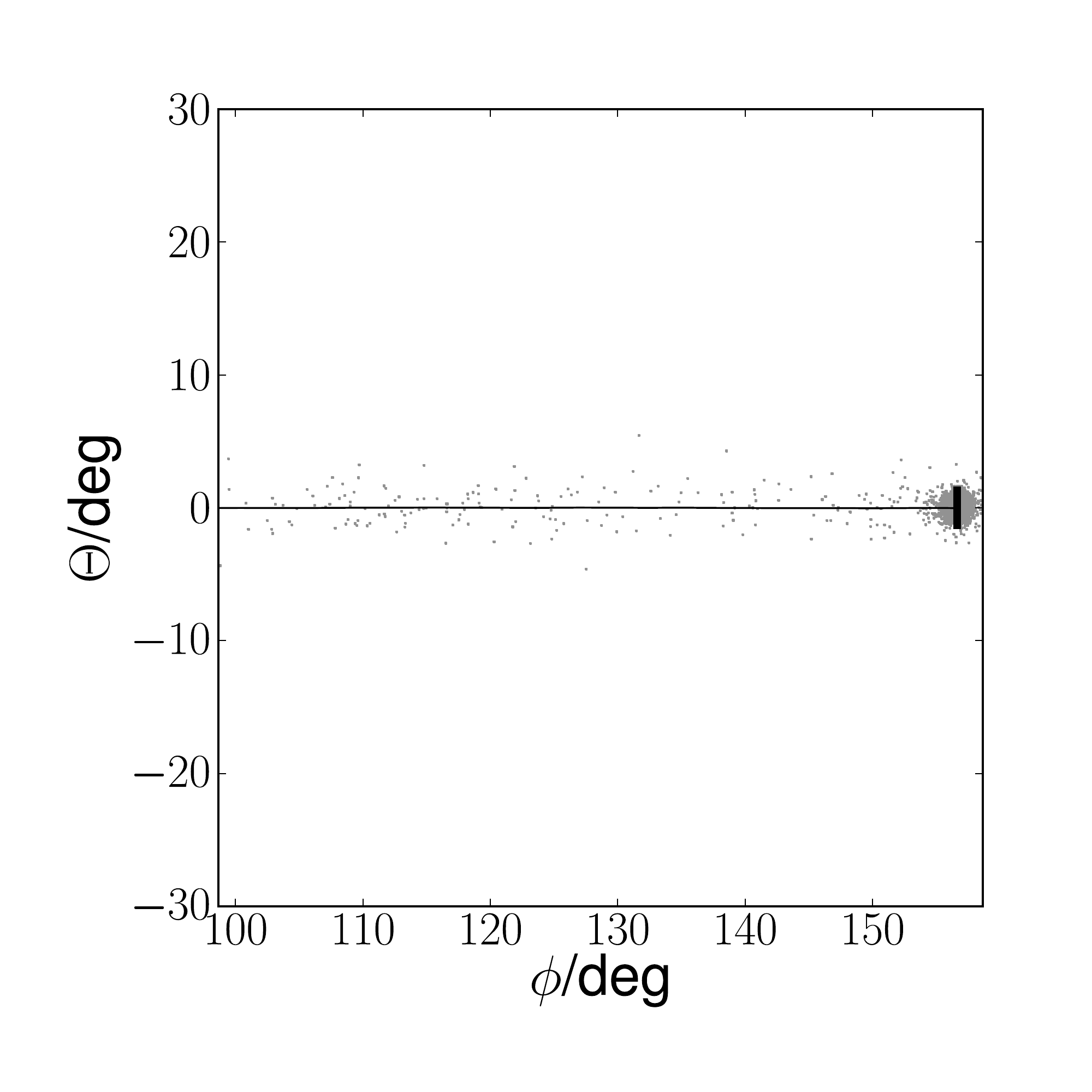}
\includegraphics[width=0.3\textwidth]{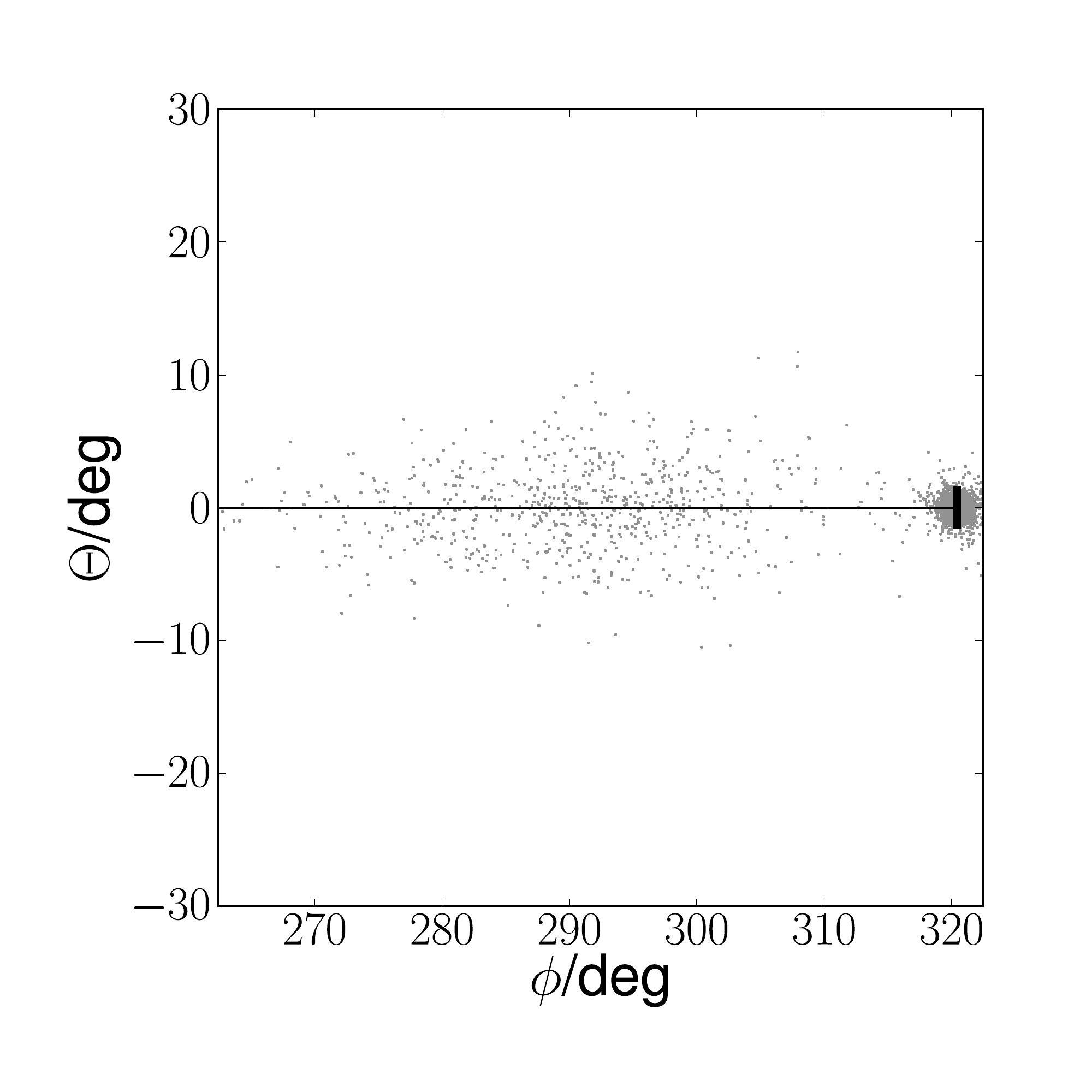}
\includegraphics[width=0.3\textwidth]{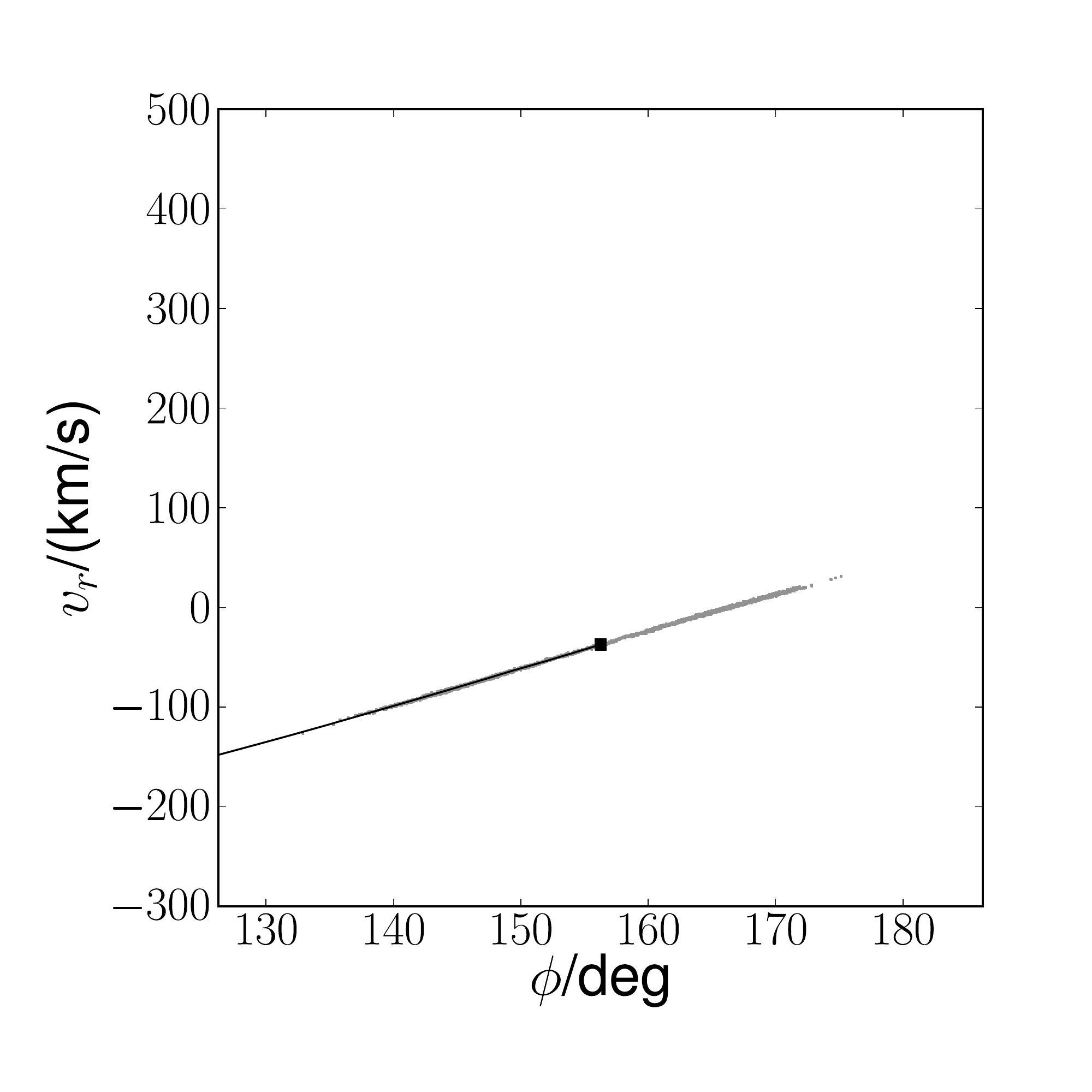}
\includegraphics[width=0.3\textwidth]{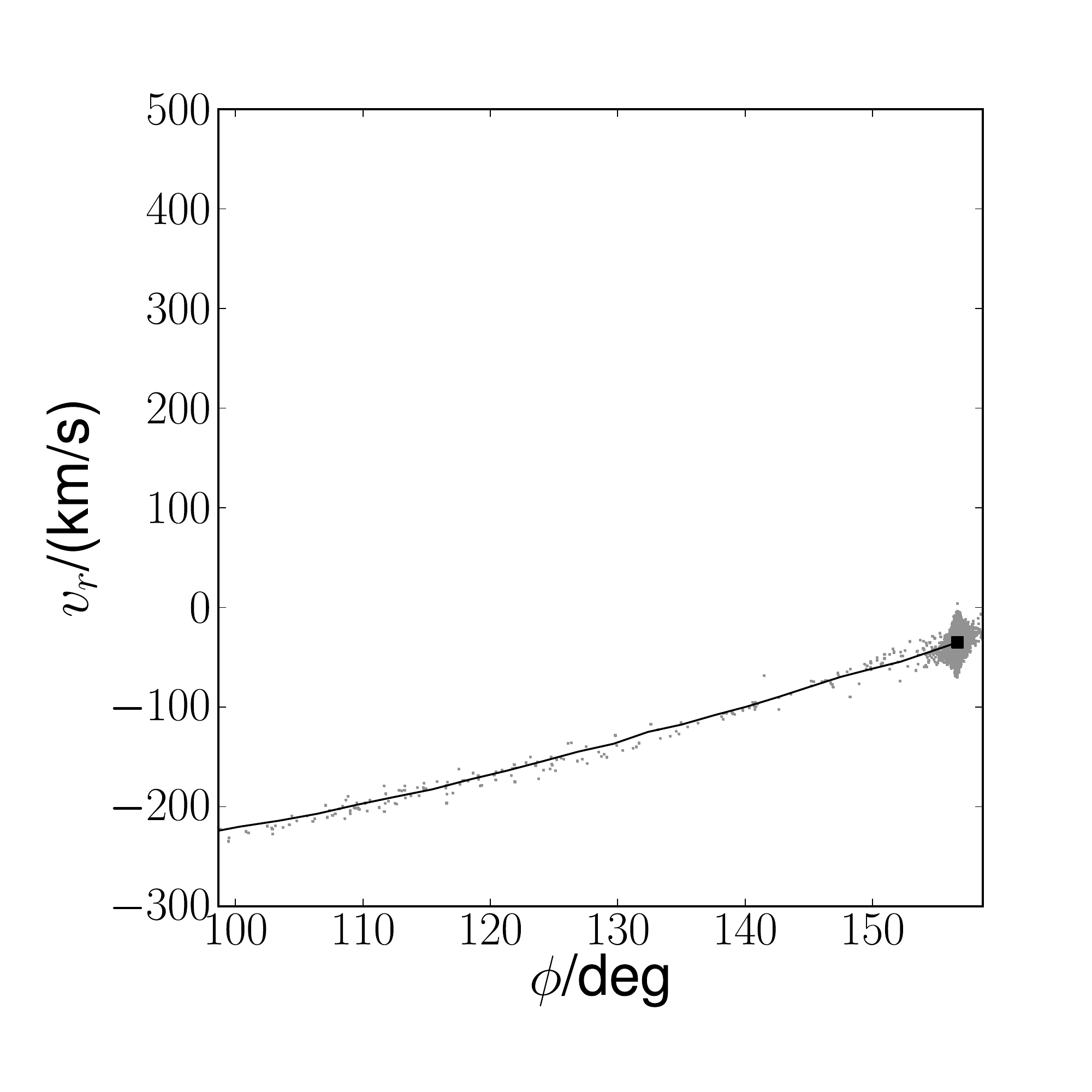}
\includegraphics[width=0.3\textwidth]{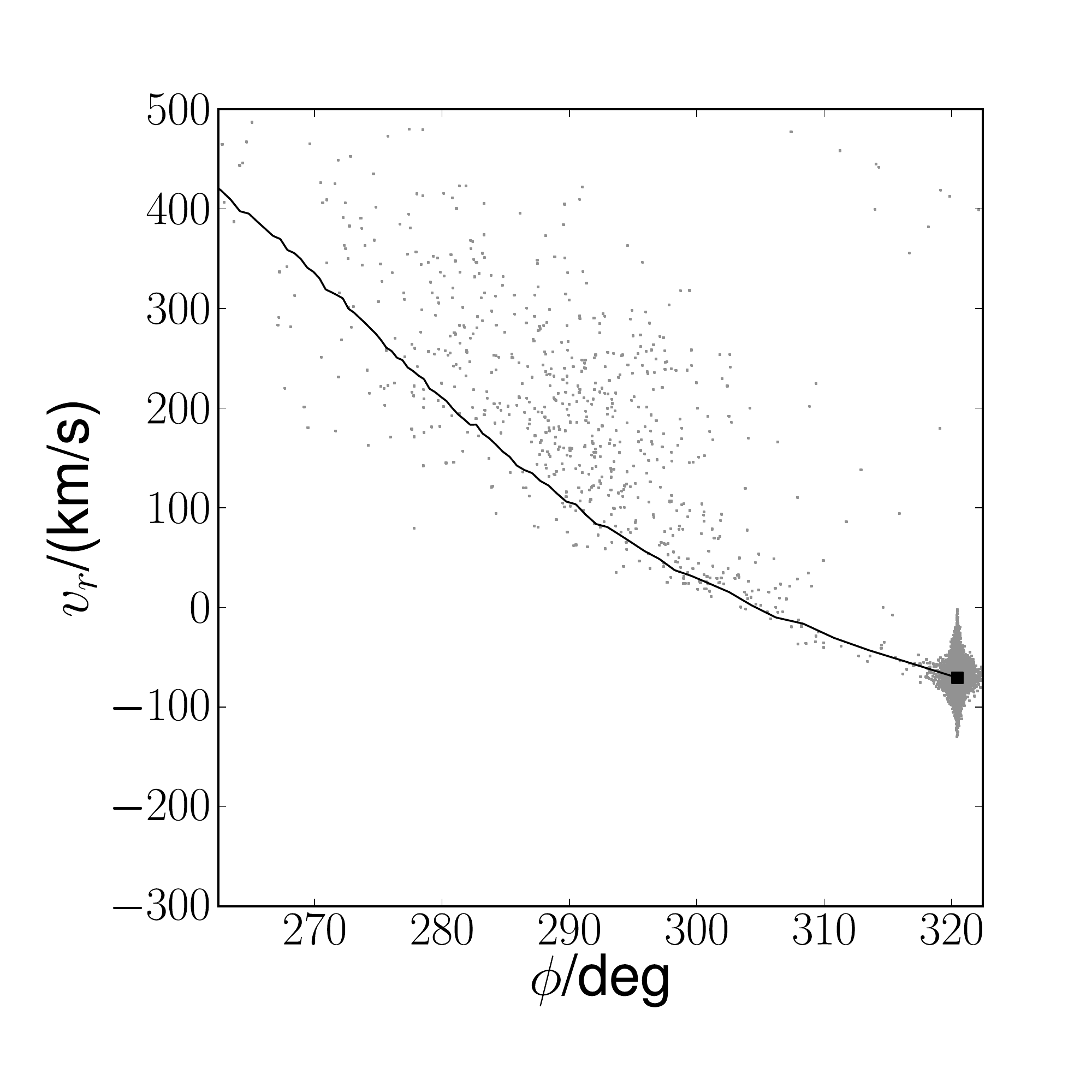}
\includegraphics[width=0.3\textwidth]{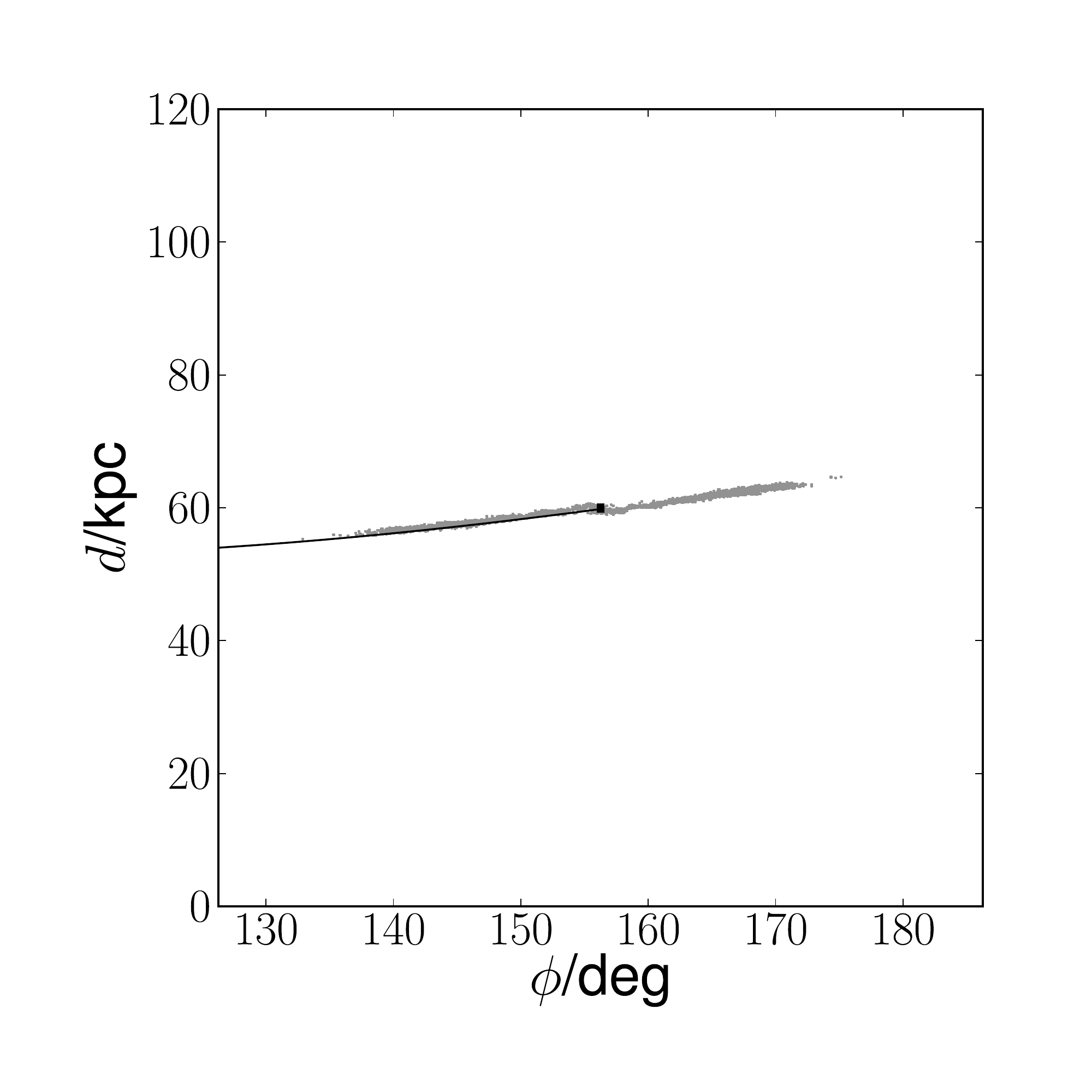}
\includegraphics[width=0.3\textwidth]{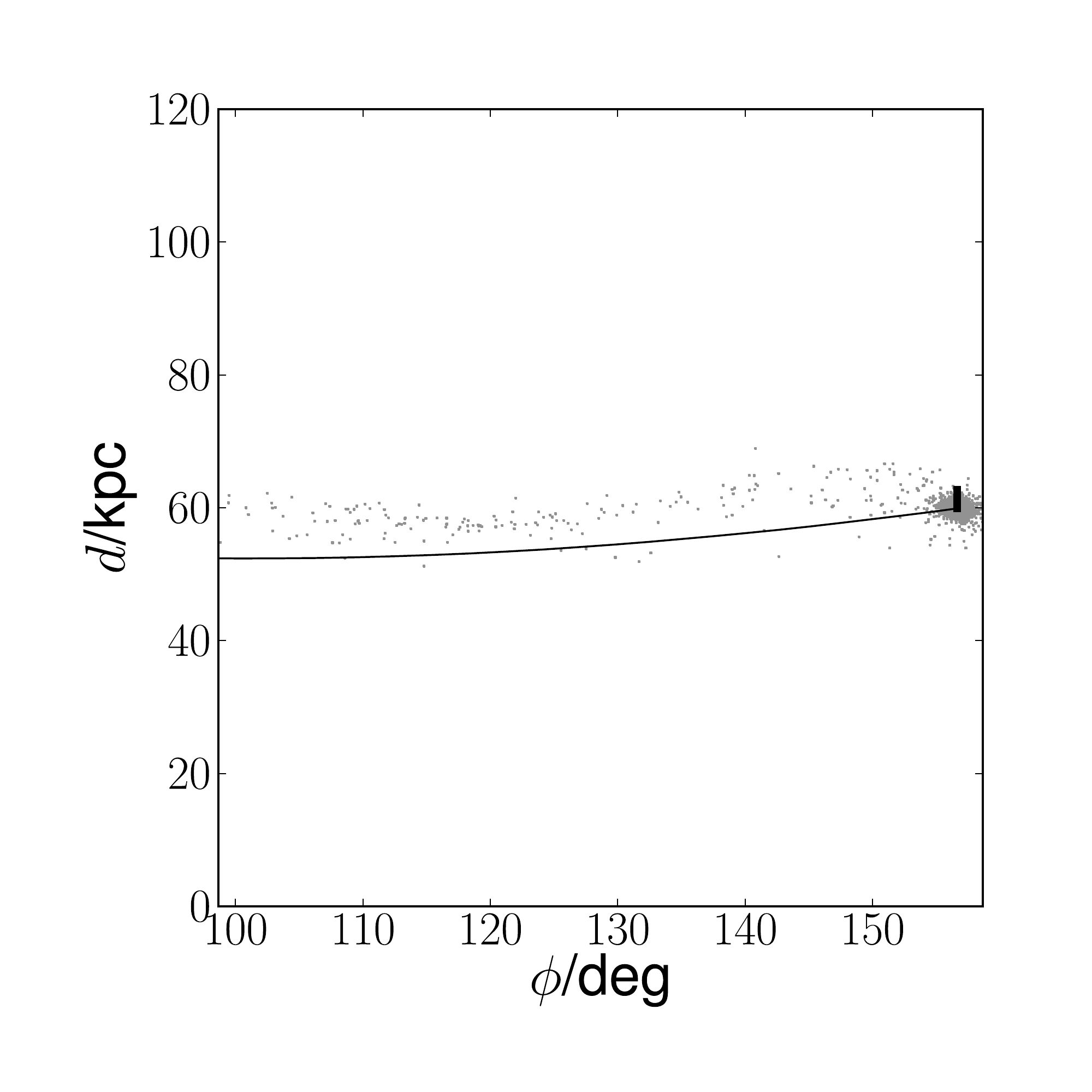}
\includegraphics[width=0.3\textwidth]{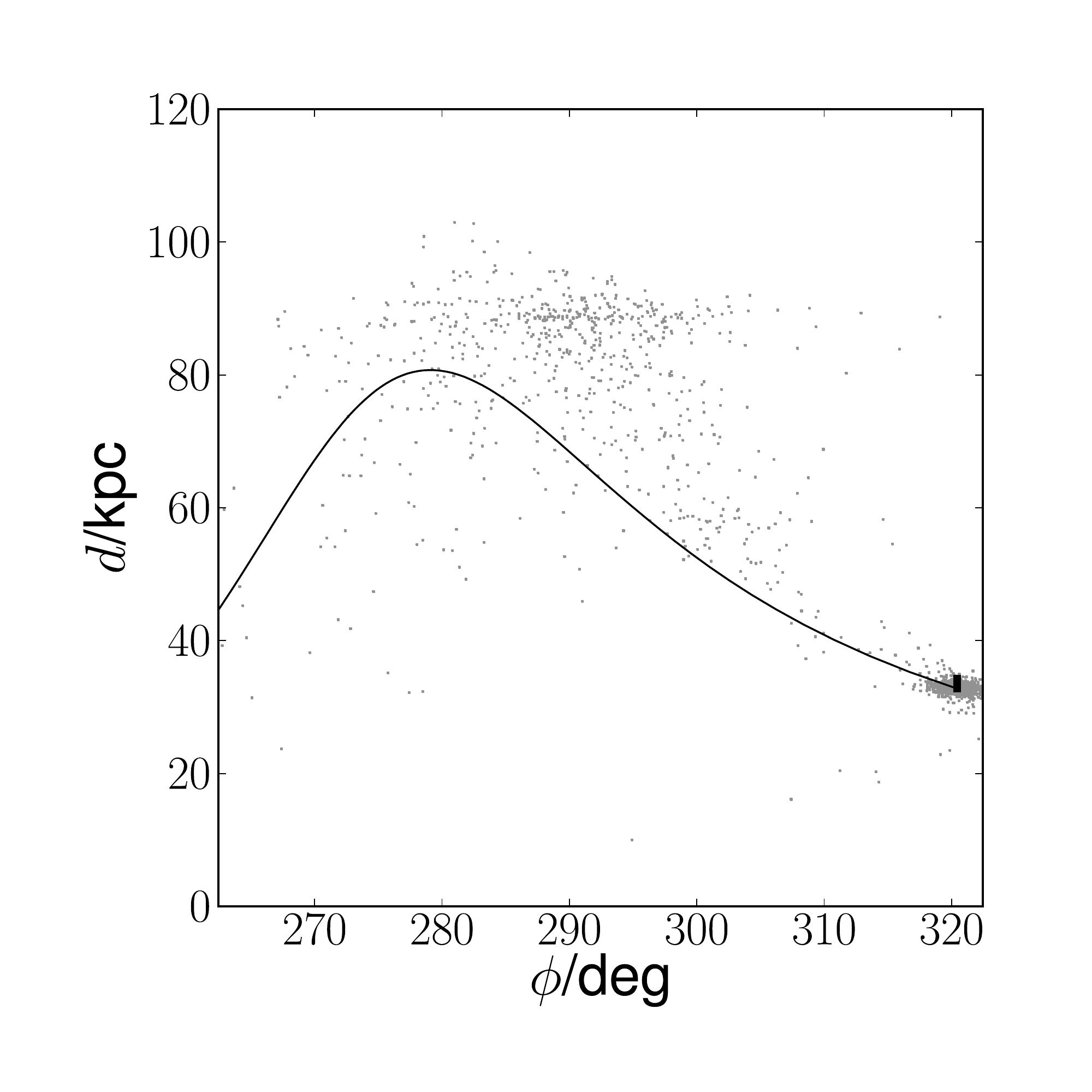}
\caption{Galactic latitude vs. longitude (upper row), line-of-sight velocity $v_r$ (middle row) and distance $d$ with respect to a virtual observer at ${\mathitbf r}=(8,0,0)$\,kpc moving like the Sun for a $10^5$\,M$_\odot$ satellite on a mildly eccentric orbit (left column), as well as a $10^8$\,M$_\odot$ satellite on a mildly (middle column) and on a highly (right column) eccentric orbit. Note that the interval in $\phi$ has been chosen to show the most interesting part of each individual stream, while keeping the interval length fixed to $60^\circ$ in all three cases. The thin solid line describes the orbit of the progenitor, while the tidal debris is traced by the N-body particles (grey dots). The thick vertical line denotes the position of the progenitor. For the angular positions and distances, the length of this line denotes the stream width/stream-orbit-offset as predicted by equation \ref{eqn:streamwidth}. Note that the width/offset is always (slightly) underestimated.}
\label{fig:obsSims}
\end{center}
\end{figure*}

\begin{table}
\begin{center}
\caption{Scale length of the Hernquist profiles in the different simulations of streams with different eccentricity $e$ and satellite mass $M_s$.}
\begin{tabular}{ rl l l}
\hline
			&$M_s = 10^5\,$M$_\odot$	& $M_s = 10^8$\,M$_\odot$\\ 
\hline
$e = 0.21$	&$r_c$ = 0.03 kpc		&$r_c =$ 0.3 kpc\\
$e = 0.75$	&$r_c$ = 0.01 kpc		&$r_c =$ 0.1 kpc\\
\hline
\end{tabular}
\label{tab:Sims1}
\end{center}
\end{table}

\section{Theory}\label{sec:tracing}
For debris generated from objects disrupting along mildly eccentric orbits, both the width of a stream and the fractional offset of a stream from its orbit -- $\Delta R/R$ -- is of order the tidal radius \citep{1998ApJ...495..297J,2001ApJ...557..137J}:
\begin{equation}
	\frac{\Delta R}{R}=\left({m_{\rm sat} \over M_{\rm Gal}}\right)^{1/3},
\label{eqn:streamwidth}		
\end{equation}
where $\Delta R$ is the distance from the centre to the edge of the stream; $R$ is the Galactocentric distance to the disrupting satellite; and the above has a mild dependence on time that becomes important after many orbits \citep{1999MNRAS.307..495H}.

\newlength{\myspace}
\setlength{\myspace}{0.013\textwidth}

\begin{table*}
\setlength{\arrayrulewidth}{0.3mm}
\caption[]{Table of selected streams. Columns show from left to right: name; angular extent on the sky $\phi$ in degrees; angular width on the sky $\Delta \phi$ in degrees; type of progenitor with mass estimate (GC - globular cluster, DW - dwarf galaxy); stream morphology in categories as defined in \cite{2008ApJ...689..936J} (str.-like - a thin continuous stream rather than a cloudy/plumy structure; b. - bound progenitor is observed); inclination to the disc plane; distance to the Sun; viewing angle of the orbital plane ($\alpha=0^\circ$ $=$ edge on; $\alpha=90^\circ$ $=$ face on); the dimensionality of the data; the `candidate score' (see text for details); and the data references.}
\label{tab:streams}
\begin{tabular*}{\textwidth}
{
@{\hspace{\myspace}}l
@{\hspace{\myspace}}r
@{\hspace{\myspace}}r
@{\hspace{\myspace}}l
@{\hspace{\myspace}}l
@{\hspace{\myspace}}r
@{\hspace{\myspace}}r
@{\hspace{\myspace}}r
@{\hspace{\myspace}}c
@{\hspace{\myspace}}c
@{\hspace{\myspace}}c
}
\hline
Stream 			&$ \phi(^\circ)$	& $\Delta \phi(^\circ)$	&progenitor 				& morphology		&incl.$(^\circ)$			& $d$\,(kpc) 		& $\alpha(^\circ)$ & data dim.& candidate score  &  Refs.\\ 
\hline
NGC 5466 Stream	& $\sim 45$	&$1.4$				& GC $(\sim5\times 10^4$\,M$_\odot)$	& str.-like/b.	&$\sim 90$	& $15.9\pm1.6^\star$&$\sim10$& 2	         	& Thin stream	& 1 \\
Pal 5	 Stream		& $\sim22$	&$0.7$				& GC $(\sim 1.3\times10^4$\,M$_\odot)$	& str.-like/b.	&$\sim 90$	& 23.2-23.9  		&$\sim60$& 3         	& Thin stream & 2 \\
GD1 				& $>100$		&$\lesssim0.25$		& GC $(\sim2\times 10^4$\,M$_\odot)\dagger$& str.-like		&$\sim35$	& 7-11		    	&$\sim20$ & 6         	& Too close		& 3\\
Triagulum Stream     & $\sim 12$       &$0.2$                                 & GC $\dagger$                                              & str.-like                    & -                       & $26 \pm 4$             &-                & 2            & Poor data                 & 4 \\
Acheron			& 37			& $\sim 0.9$			& GC$\dagger$					& str.-like 			&$\sim90$	& 3.5-3.8		   	& -		& 2		& Poor data		& 5\\ 
Cocytos			& 80			& $\sim 0.7$			& GC$\dagger$					& str.-like 			&$\sim60$	& $11\pm2$	   	& - 		& 2		& Poor data		& 5\\ 
Lethe	 		& $\sim 80$	& $\sim 0.4$			& GC$\dagger$					& str.-like			&$\sim60$	&12.2-13.4	   	& -		& 2		& Poor data 		& 5\\ 
Styx 				& $\sim 65$	& $\sim 3.3$			& DW$\dagger$					& str.-like			&$\sim90$	&38-50	   		& -		& 2		& N-body		& 5\\ 
Orphan Stream		& $>100$ 		& $\sim 2$			& DW$(\sim 10^7$\,M$_\odot)\dagger$& str.-like		&$\sim 34$	& 20-50    	      	     		& $\sim40$& 4 	 	& N-body 		& 6 \\
Monoceros Stream	& $> 100$		& -					& DW $(\gtrsim 10^8$\,M$_\odot)\dagger$ & str.-like	&$\sim25$	& $\sim 15$       		& -		& 6		& N-body 		& 7 \\
Sagittarius Stream  	& $>360$		&$\sim10$			& DW $(\sim 10^8$\,M$_\odot)$		& str.-like/b.       	&$\sim90$	& 20-60	   		& $\sim 5$& 4		& N-body  	& 8 \\
\hline

\end{tabular*}
$^1$ \cite{2006ApJ...639L..17G,2006ApJ...637L..29B,2007MNRAS.380..749F};  $^2$ \cite{1977AJ.....82..459S,2001ApJ...548L.165O,2002AJ....124.1497O,2003AJ....126.2385O,2009AJ....137.3378O,2006ApJ...641L..37G}; $^3$ \cite{2006ApJ...643L..17G,2009ApJ...697..207W,2010ApJ...712..260K}; $^4$ \cite{2012ApJ...760L...6B}; $^5$ \cite{2009ApJ...693.1118G};  $^6$  \cite{2007ApJ...658..337B,2008MNRAS.389.1391S,2010ApJ...711...32N};   $^{7}$ \cite{2002ApJ...569..245N,2003MNRAS.340L..21I,2003ApJ...588..824Y,2004ApJ...605..575Y,2006ApJ...651L..29G,2008ApJ...689L.117G,2010AJ....139.1889C}; $^{8}$ \citet{2010ApJ...714..229L} and references therein; 
$^\star$ distance to progenitor; $\dagger$ tentative.
\end{table*} 

Figures \ref{fig:SimsPlane} and \ref{fig:obsSims} test equation \ref{eqn:streamwidth} using a sequence of N-body models (see also Figure 5 of \citealt{2001ApJ...557..137J}). We model three streams in a static, spherical `NFW' potential \citep{1996ApJ...462..563N} with virial mass $M_{\rm vir}=1.77\times 10^{12}$\,M$_\odot$; virial radius $389$\,kpc; and scale length $r_s=24.6$\,kpc. Each satellite was modelled as a \cite{1990ApJ...356..359H} profile represented by 10,000 particles. (Note that these low resolution simulations are sufficient as we are not investigating the nature of the disruption but merely the locus of the debris in phase space.) We compare high mass ($M_s = 10^8$\,M$_\odot$) versus low mass ($M_s =10^5$\,M$_\odot$) satellites on mildly eccentric ($e = 0.21$) and highly eccentric ($e = 0.75$) orbits. The respective scale lengths $r_c$ of the satellite Hernquist profiles are summarised in Table \ref{tab:Sims1}. These were chosen to ensure a significant amount of mass loss, while keeping the fractional mass loss in each simulation the same. All simulations were run with an SCF code \citep{1992ApJ...386..375H} using methods described in \cite{1995ApJ...451..598J} and conserved energy to better than 10\% of the initial internal satellite energy. In each case, the satellite orbit and tidal debris are marked by a thin solid line and grey dots, respectively. The position of the progenitor is denoted by a black square. Figure \ref{fig:SimsPlane} shows all four simulations in Galactocentric coordinates. Figure \ref{fig:obsSims} shows the galactic latitude vs. longitude (upper row), the line-of-sight velocity $v_r$ (middle row) and the distance $d$ (bottom row) of the stream as seen by a fictitious observer at ${\mathitbf r}=(8,0,0)$\,kpc moving like the Sun\footnote{This corrects for the circular velocity in the potential at ${\mathitbf r}=(8,0,0)$\,kpc and the Sun's peculiar motion as measured by \cite{2009MNRAS.397.1286A}.}. For the angular positions and distances, the analytical prediction of the stream width and stream-orbit-offset is indicated by the length of the thick vertical line at the progenitor's position. To emphasise the differences in morphology, the plots have been focused on the relevant part of each stream while keeping the scales for all satellites the same. The orbits have been chosen such that stream-orbit offset appears almost exclusively in the distances (bottom row). Notice that equation \ref{eqn:streamwidth} performs remarkably well, slightly under-estimating the stream-orbit offsets.

\begin{figure*}
\centering
\includegraphics[width=0.33\textwidth]{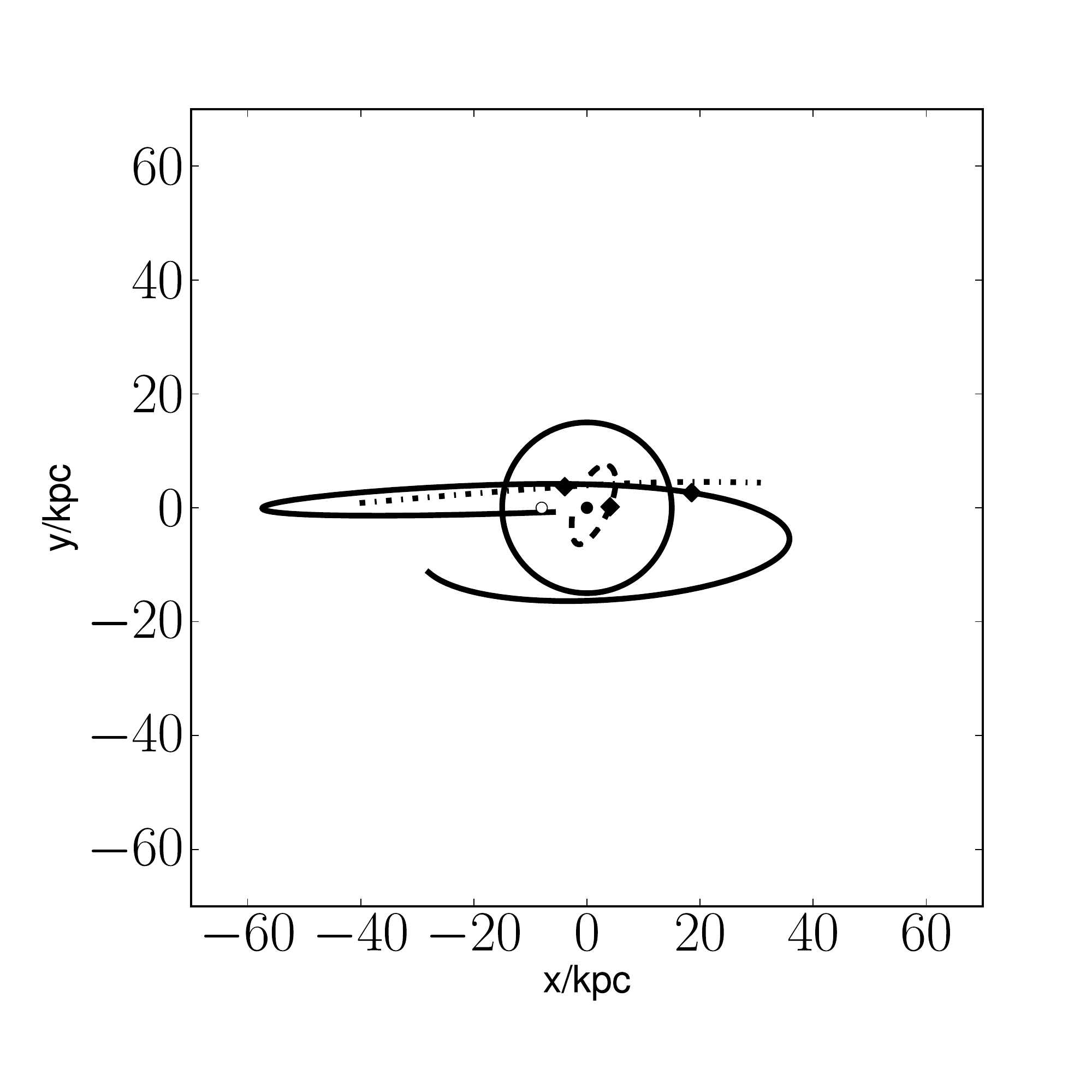}
\includegraphics[width=0.33\textwidth]{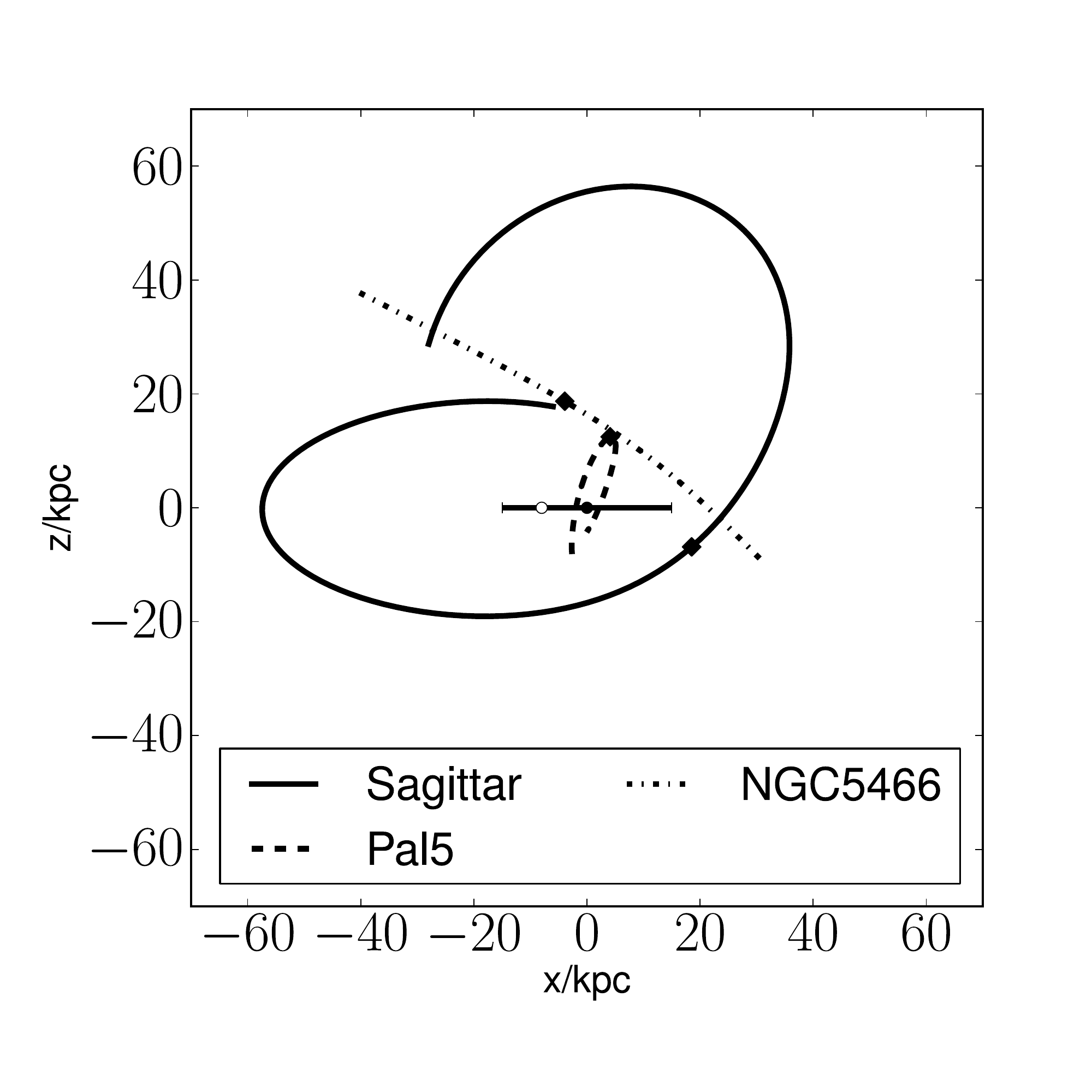}
\includegraphics[width=0.33\textwidth]{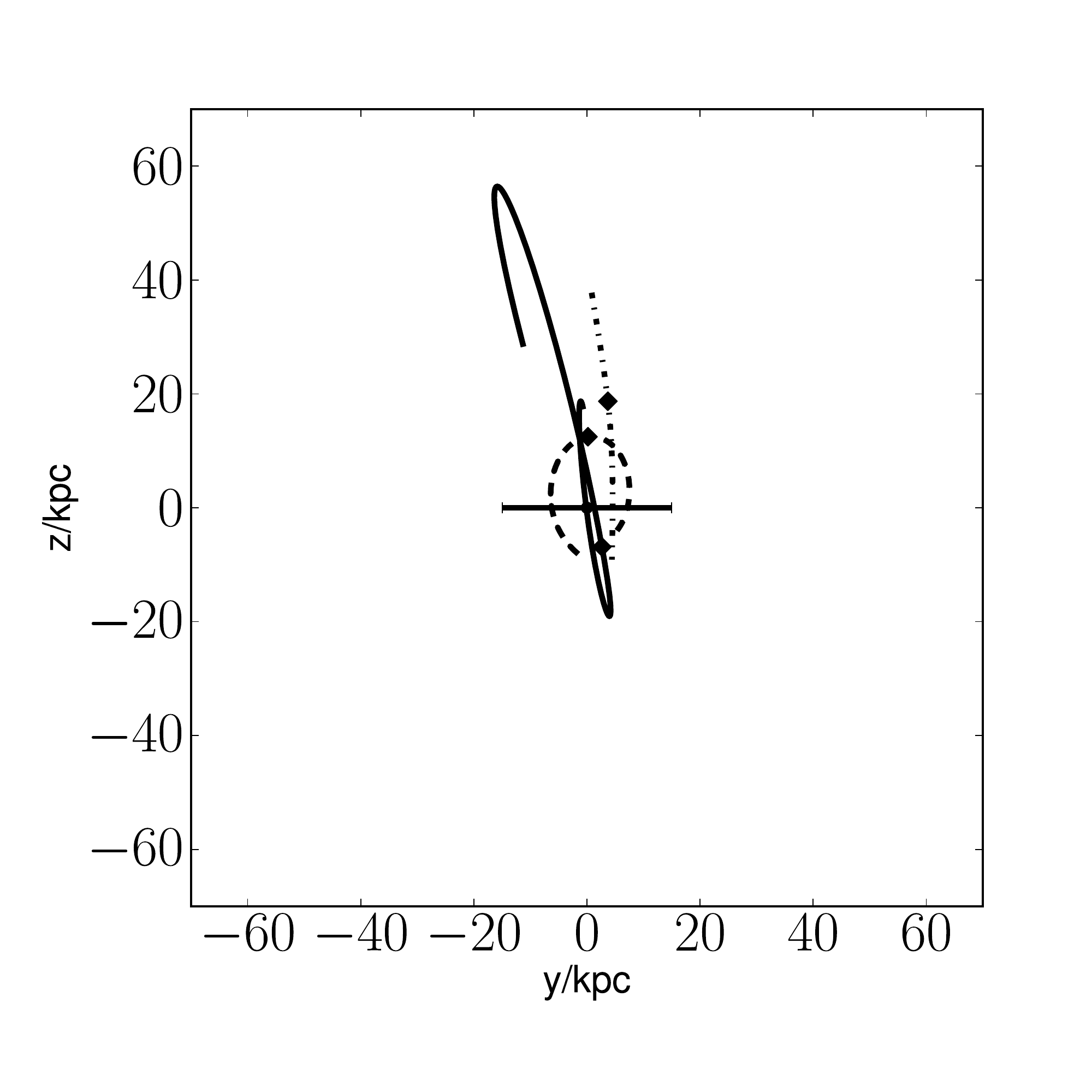}
\caption{Orbits of selected Milky Way streams: Sagittarius (solid line); Pal 5 (dash line); and NGC 5466 (dash-dotted line) in the triaxial potential of \citet{2010ApJ...714..229L}. The starting point of the integrations are marked by black diamonds. The plots show different projections in Galactic coordinates $x$, $y$, $z$. The Galactic centre is marked with a black filled circle; the Solar position with a small open circle. The Galactic disc is denoted by a circle in the $x-y$-plane and a line in the other two.}
\label{fig:XYZ}
\end{figure*}

The N-body simulations confirm that less massive objects lead to thinner, shorter streams that are less offset from their progenitor's orbit. Additionally, the eccentricity of the orbit influences the stream width/offset such that more eccentric orbits lead to thicker and more offset streams. Finally, note that the morphology of the stream becomes less stream-like and more cloudy for more eccentric orbits, particularly for the more massive satellite. It is clear that for such `cloudy' streams, a test-particle modelling approach will be inadequate. For the thinner colder streams, however, a test particle approach may be adequate. We examine this in more detail in \S\ref{sec:results}.

\section{Data}\label{sec:data}
In Table \ref{tab:streams}, we collate data for known streams in the Milky Way: their angular extent $\phi$ and width $\Delta\phi$ in degrees; the (probable) type and mass of their progenitor; the morphology of their stream based on the classification of \cite{2008ApJ...689..936J}; their inclination to the disc; distance to the Sun $d$\,(kpc); viewing angle of the orbital plane ($\alpha=0^\circ$ $=$ edge on; $\alpha=90^\circ$ $=$ face on); and the dimensionality of the data set. Viewing angles above the plane are often quite uncertain, especially where the orbit of the stream is not very well constrained. Based on these data, we evaluate how suitable a stream is for probing the Milky Way potential. We hone in on streams that are thin, with non-cloudy morphology, that lie sufficiently far from the Milky Way disc that we can hope to probe the dark matter halo potential \citep[see e.g.][]{2009ApJ...697..207W,2010ApJ...711...32N,2010ApJ...712..260K}. We apply a `candidate score' in the second to last column in the table, defined as:

\begin{itemize} 
\item {\bf Thin stream}: An archetypal thin stream;
\item {\bf Too close}: Close proximity to the Milky Way disc -- i.e. not amenable to probing the potential of the Milky Way halo;
\item {\bf Poor data}: Data quality too poor to determine usefulness;
\item {\bf N-body}: A stream of sufficient width or complex morphology that N-body models are likely required. 
\end{itemize} 

Based on the above, we immediately exclude the following from our analysis:
\begin{itemize}
\item Moving groups \citep[e.g. Hercules Corona Borealis;][]{2010MNRAS.405.1796H};
\item Streams without a stream-like morphology, or with insufficient data to determine the morphology: the Hercules-Aquila cloud \citep{2007ApJ...657L..89B}; the Cetus Polar Stream \citep{2009ApJ...700L..61N}; the Virgo Stellar Stream \citep{2002ApJ...569..245N,2007ApJ...660.1264M,2008AJ....136.1645V,2009ApJ...701L..29C,2009ApJ...691..306P}; the Virgo Overdensity \citep{2007ApJ...660.1264M,2008ApJ...673..864J}; and the Pisces Overdensity \citep{2007AJ....134.2236S,2009MNRAS.398.1757W,2009ApJ...705L.158K,2010ApJ...717..133S};  the Triangulum-Andromeda Cloud \citep{2004ApJ...615..732R,2004ApJ...615..738M,2007ApJ...668L.123M}
\end{itemize}
all other streams appear in Table \ref{tab:streams}.

With the above cuts, we are left with the globular cluster streams NGC 5466 and Pal 5. This constellation of two thin streams with perpendicular orbital planes that lie at similar Galactocentric distance makes an intriguing couple for constraining the MW halo shape. Furthermore, they have comparable distances and therefore probe the same halo shape, even if the shape has some radial dependence. We use our test particle method to determine the constraining power of these streams in \ref{sec:results}. 

\section{Method} \label{sec:method}

In this section, we describe our implementation of the test particle fitting method.

\subsection{Test particle orbit integration}
We integrate the test particle orbits using the {\tt Orbit\_Int} code described in \cite{2010MNRAS.406.2312L}. The initial starting point of the integration was chosen to be the progenitor of the stream. We transform the resulting orbits into observable coordinates using M. Metz's tool {\tt bap.coords}\footnote{{\tt http://www.astro.uni-bonn.de/$\sim$mmetz/py/docs/mkj\_libs/ public/bap.coords-module.html}} \citep{2007MNRAS.374.1125M}. We adopt $8.0$\,kpc as the distance of the Sun to the Galactic center. The velocities of the local standard of rest (LSR) are adjusted to the circular velocity at that distance in the respective potential model and the peculiar motion of the Sun has been assumed to be $(9.96,5.25,7.07)\,$km/s \citep{2009MNRAS.397.1286A}. As our fiducial model, we adopt the potential used in \citet{2010ApJ...714..229L} consisting of a \cite{1975PASJ...27..533M} disc:
\begin{equation}
        \Phi_{\rm disc}=-  {GM_{\rm disc} \over
                 \sqrt{R^{2}+(a+\sqrt{z^{2}+b^{2}})^{2}}},
\end{equation}
a \cite{1990ApJ...356..359H} bulge
\begin{equation}
        \Phi_{\rm bulge}=-{GM_{\rm bulge} \over r+c},
\end{equation}
and a cored triaxial logarithmic potential
\begin{equation}
        \Phi_{\rm halo}=v_{\rm halo}^2 \ln (C_1 x^2 + C_2 y^2 +C_3 x y + (z/q_z)^2 + r_{\rm halo}^2)
\end{equation}
where the constants are defined as
\begin{equation}
C_1 = \left(\frac{\textrm{cos}^2 \phi}{q_{1}^2} + \frac{\textrm{sin}^2 \phi}{q_{2}^2}\right),
\end{equation}
\begin{equation}
C_2 = \left(\frac{\textrm{cos}^2 \phi}{q_{2}^2} + \frac{\textrm{sin}^2 \phi}{q_{1}^2}\right),
\end{equation}
\begin{equation}
C_3 = 2 \, \textrm{sin}\phi \, \textrm{cos} \phi \left( \frac{1}{q_{1}^2} - \frac{1}{q_{2}^2}\right).
\end{equation}
The disc mass is set to $M_{\rm disc}=1.0 \times 10^{11}$\,M$_{\odot}$ with a scale length $a=6.5$\,kpc and scale height $b=0.26$\,kpc. The mass of the bulge is fixed at $M_{\rm bulge}=3.4 \times 10^{10}$\,M$_{\odot}$ and its scale length $c=0.7$ kpc. The logarithmic potential is chosen in such a way, that the circular velocity of the total potential at the position of the Sun ($R_\odot = 8$\,kpc) is equal to 220\,km/s. This means that $v_{\rm halo} = 121.86$\,km/s, for $r_{\rm halo} = 12$\,kpc, $q_{1} = 1.38$, $q_2 = 1$, $q_z = 1.36$ and $\phi = 97^\circ$.

In the above potential parametrisation, $q_1$ and $q_2$ are the Galactic halo shape parameters, fixed to lie in the disc plane. The angle between them is $\phi$. For $\phi = 0^\circ$, $q_1$ describes the shape along the $x$-axis of the halo potential. Without loss of generality we keep $q_2=1$ fixed throughout as any shape can be described by varying $q_1$ and $\phi$. Because of this special configuration, in our evaluation $\phi$ is only meaningfully recovered as long as $q_1$ is well constrained. $q_z$ denotes the shape parameter along the $z$-axis, i.e. the axis perpendicular to the disc plane.

As streams are not yet able to constrain the Milky Way mass better than other tracers \citep[e.g.]{2008ApJ...684.1143X,2009PASJ...61..227S,2010ApJ...714..229L}, we hold the mass fixed and focus on constraining the shape of the Milky Way halo potential. The mass enclosed in $r=60$\,kpc in our fiducial  model $M(<60\textnormal{kpc}) = 5.2\times 10^{11}M_\odot$ is consistent with the high mass end of recent measurements using blue horizontal branch stars selected from SDSS DR6 $M(<60\textnormal{kpc}) = (4.0\pm0.7)\cdot10^{11}M_\odot$ \citep{2008ApJ...684.1143X}. This approach is likely to underestimate the uncertainties in the derived potential shape, especially as we are using a fixed parametrisation instead of a non-parametric model \citep{2013ApJ...765L..15I}. This should be kept in mind, when interpreting our results. When modelling real stream data, all kinematic tracers should be fitted simultaneously while varying the potential mass and shape ideally in a non-parametric way. However, one should note that, when determining the mass of the individual MW components, fitting to the rotation curve in the inner parts of the Milky Way typically overestimates the enclosed mass \citep{Gomez:2006p1638}. 

Finally, note that our global potential parameterisation means that any given stream will constrain the global shape parameters $q_1$ and $q_z$. In practice, if using a basis function expansion for the potential, for example, streams will only constrain the {\it local} potential shape along their orbits. For this reason, finding many thin streams in the halo with excellent data would be advantageous. For the moment, given the paucity of data, a global model is sufficient. 

\subsection{Markov Chain Monte Carlo Method}
The Markov Chain Monte Carlo method (MCMC) is a statistical method to determine the probability distribution of a multidimensional parameter space in a very efficient way using a random walk. From the random walk, a new parameter set is always accepted if it is better than the previous set. If not, it is accepted/rejected with a probability given by the likelihood of these parameters with respect to the previous set. Assuming that the errors have a Gaussian distribution with respect to the orbit, the logarithm of the likelihood that the model orbit agrees with the data within the errors is proportional to the $\chi^2$ measure:
\begin{equation}
\chi_j^2 = \sum_{i=1}^{N_j} \frac{(y_m-y_d)^2}{\sigma_d^2} ,
\end{equation}
where $y_m$ is the $y$ value of the model, $y_d$ is the $y$ value of the data and $\sigma_d$ the error of the data. Depending on the available data, these can range from angular positions only to full 6D phase space data\footnote{Note, that a more correct approach where parallax distances are not available is fitting to the observed magnitudes as done in e.g. \cite{2010ApJ...712..260K}.}. The different data sets for each stream are then combined to give a total $\chi^2$
\begin{equation}
\chi^2 = \sum_j \chi^2_j.
\end{equation}
For better constraints, several streams at similar distances can be combined by adding their individual $\chi^2$.

Our implementation follows the algorithm detailed in \cite{Simard:2002p2258}. To efficiently search our parameter spaces, we have adopted a variable step-size MCMC. After an initial phase of constant steps that lasts for the first quarter of our iterations, the step-size (Metropolis temperature) is either shortened or broadened dependent on whether more models in the last 50 iterations have been accepted or reject, respectively. To account for the necessity of different step-size adjustments in different parameters, we keep track of the lowest $\chi^2$ value found, never allowing the step-size become smaller than twice the distance between the current position and the lowest $\chi^2$ point. This also avoids the chain becoming stuck in a local minima after the global minimum has already been found. Note that this implementation of the MCMC method only correctly samples the lowest minimum of the $\chi^2$ distribution. To correctly map more complex probability spaces, a different method should be adopted.

We fit for the distance $d$ and proper motions $\mu_l$, $\mu_b$ of the initial starting point of the integration along the stream, as well as the parameters constraining the halo shape $q_1$, $q_z$ and $\phi$. We assume throughout that the other halo potential parameters can be constrained by independent probes like the Galactic rotational curve and halo star kinematics. We consider the errors on the angular position and radial velocities at the initial conditions of the orbit integration to be negligible in comparison to the other 3 coordinates. As it is not important to constrain one specific orbit, we can assume these parameters are fixed without loss of generality. 

\begin{figure*}
\centering 
\includegraphics[width=0.44\textwidth]{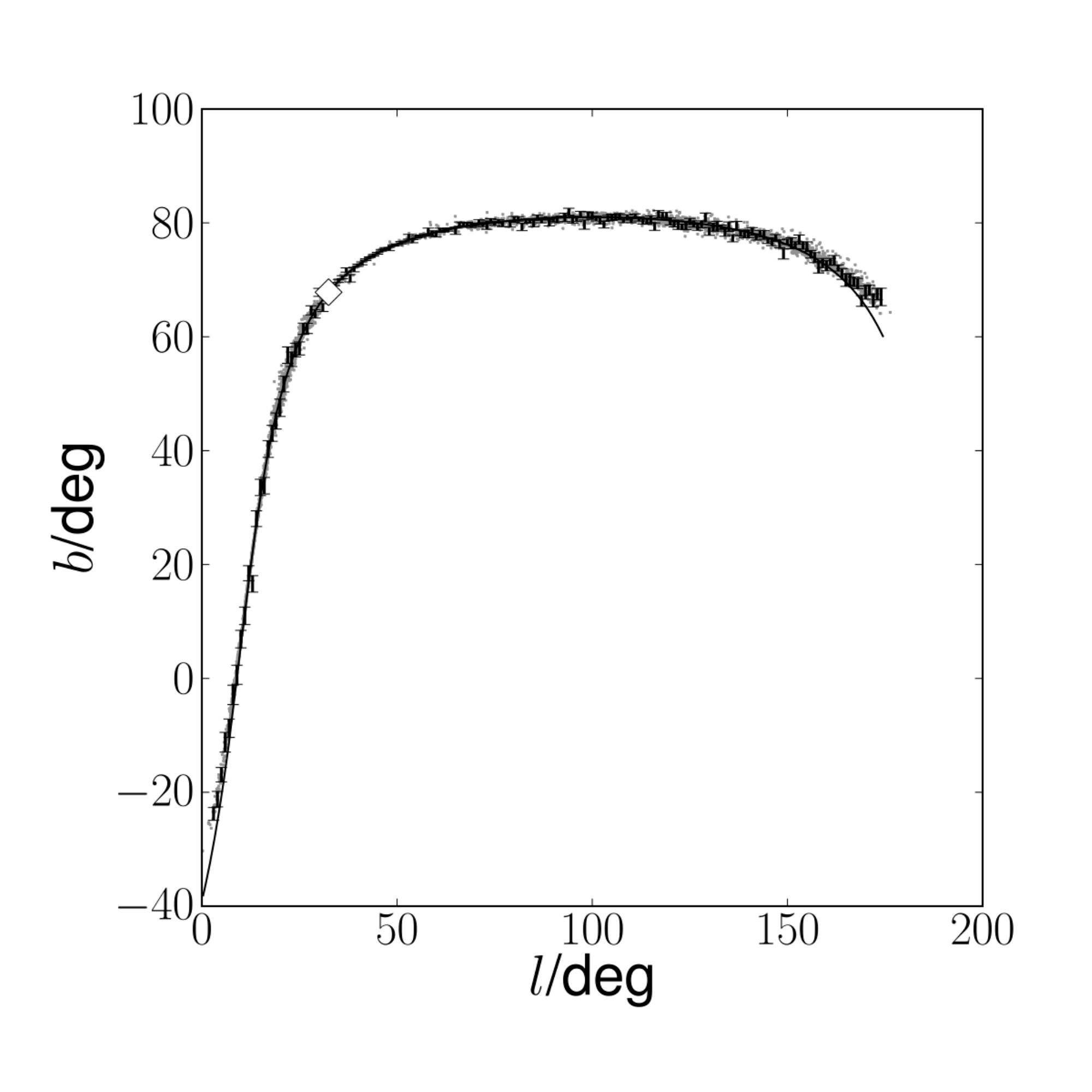}
\includegraphics[width=0.44\textwidth]{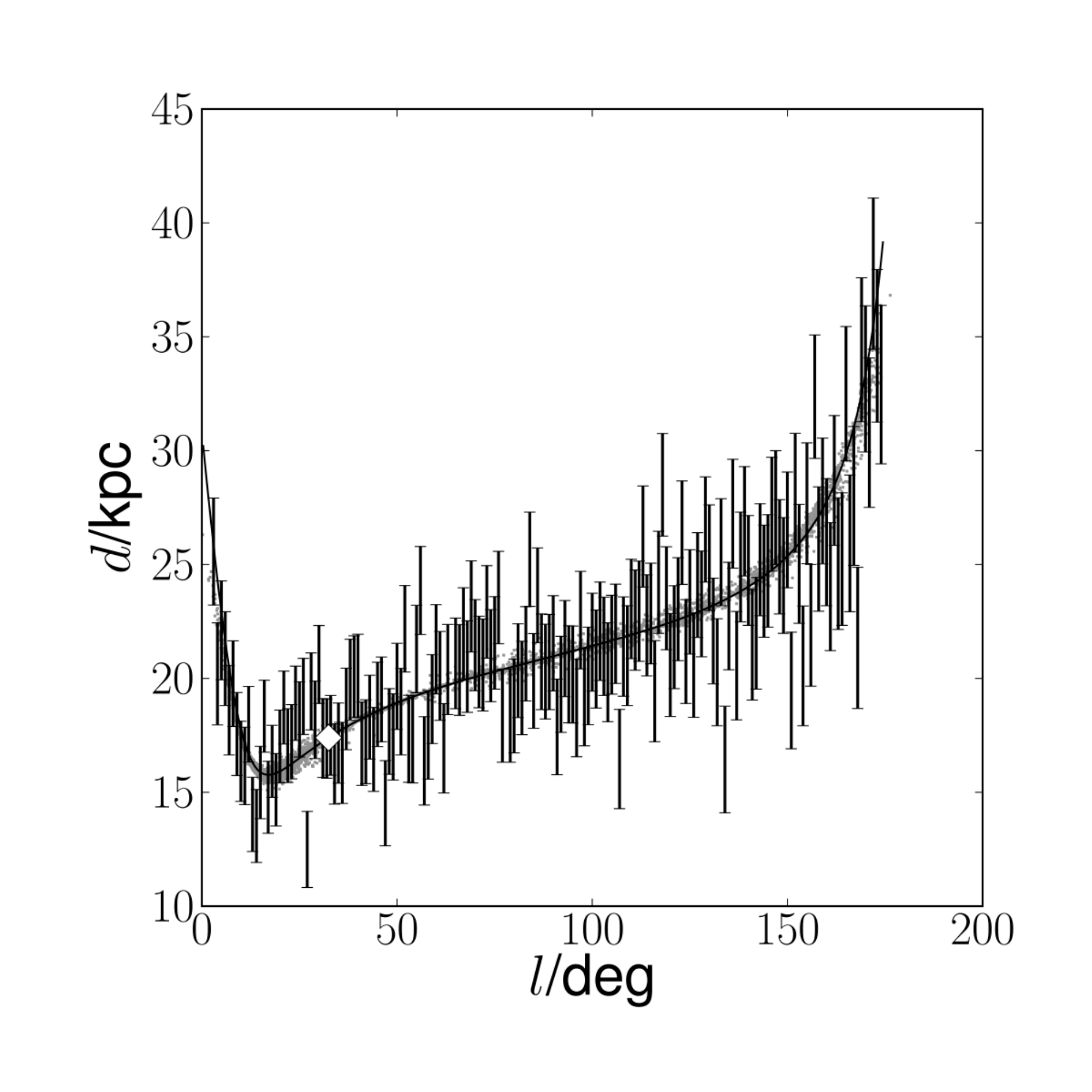}
\includegraphics[width=0.44\textwidth]{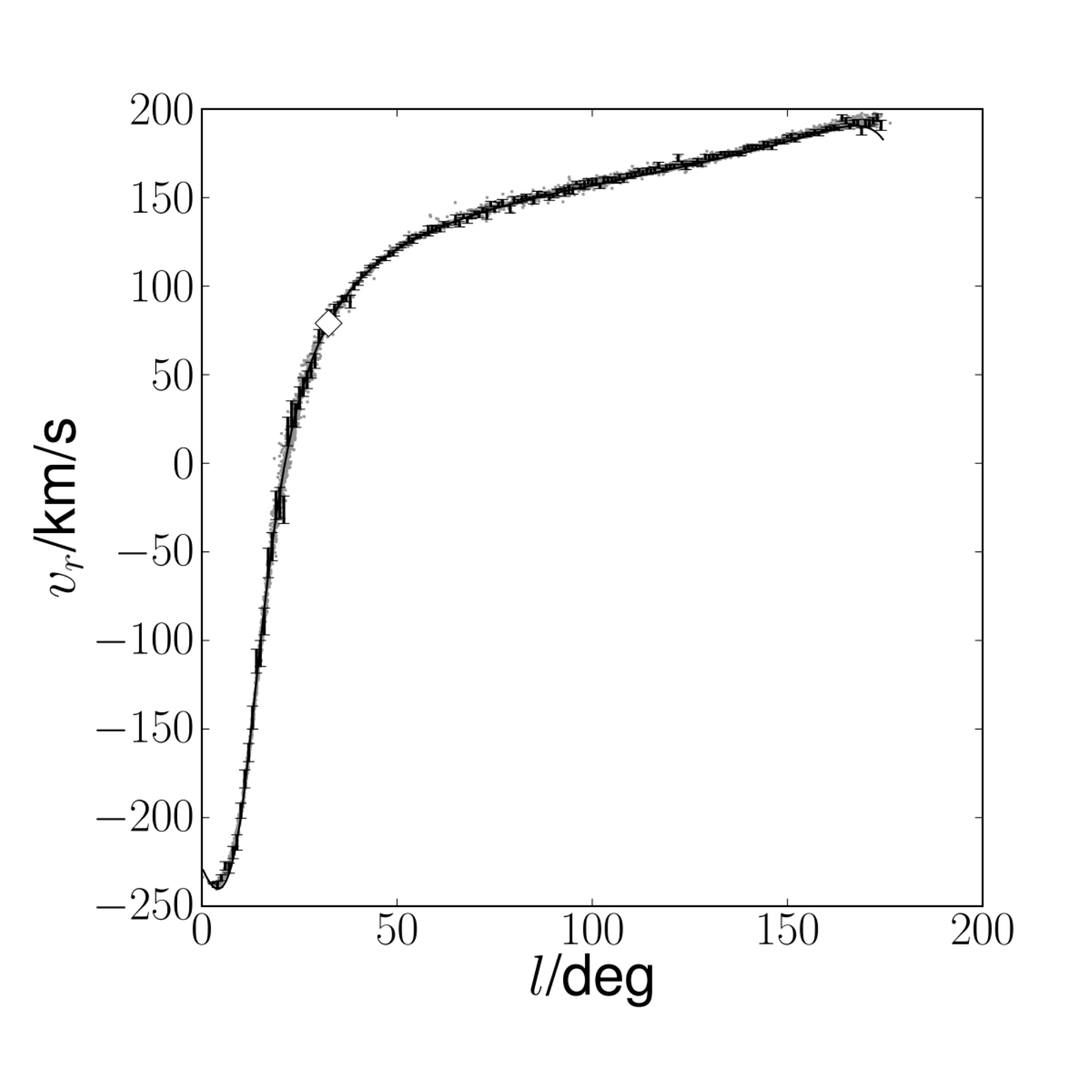}
\includegraphics[width=0.44\textwidth]{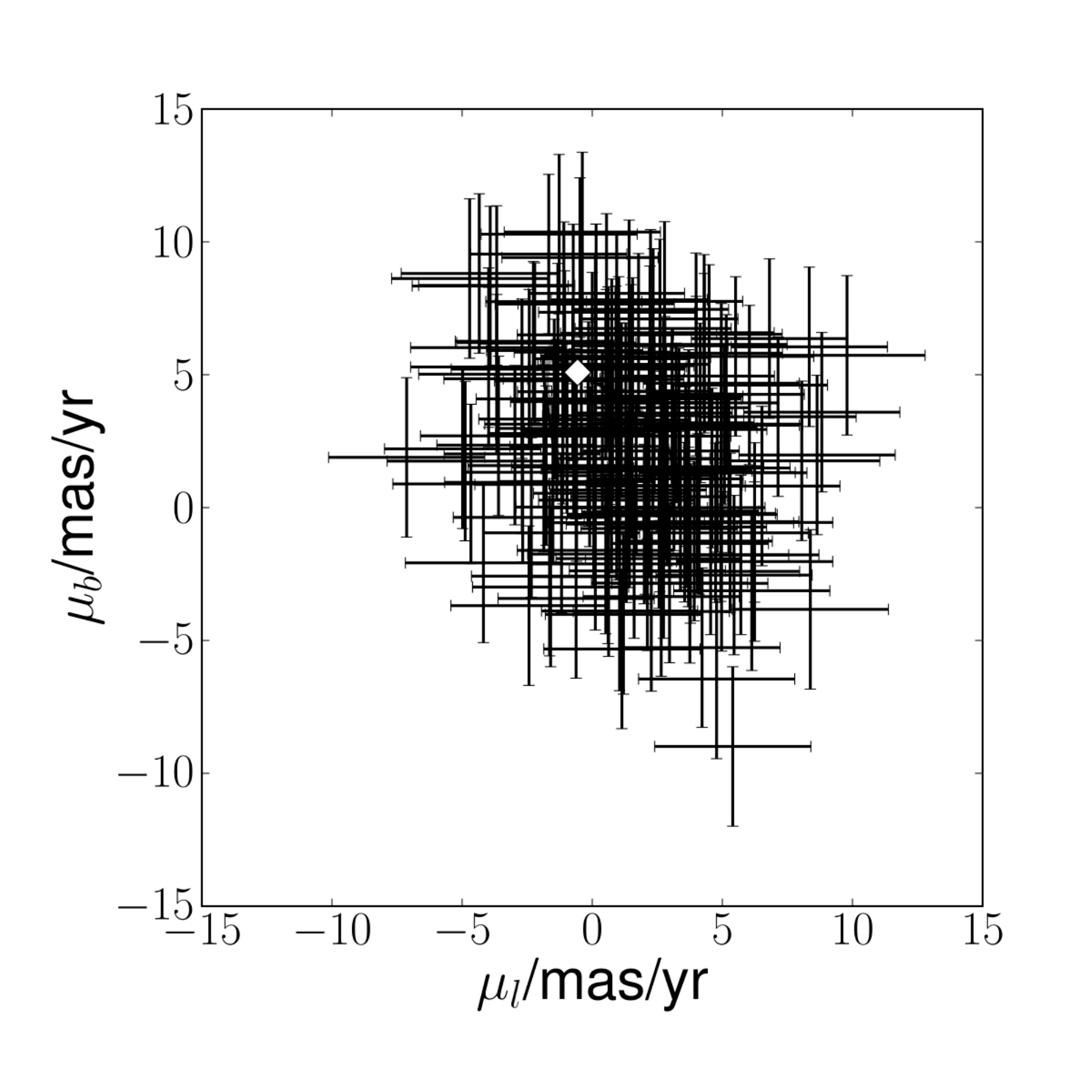}
\caption{The black error bars show the mock data for an NGC 5466 like stream with hi-res data (see \S\ref{sec:mockdata}). The data was sampled in $\Delta l=1^\circ$ bins. The plots show angular position in Galactocentric coordinates $(l,b)$ (top left); distance (top right); radial velocity (bottom left); and proper motion data (bottom right). The black line denotes the progenitor's orbit; the white diamond the position of the progenitor; and the grey dots the N-body particles representing the stream.}
\label{fig:NGCmock}
\end{figure*}

\begin{figure*}
\begin{center}
\includegraphics[width=0.3\textwidth]{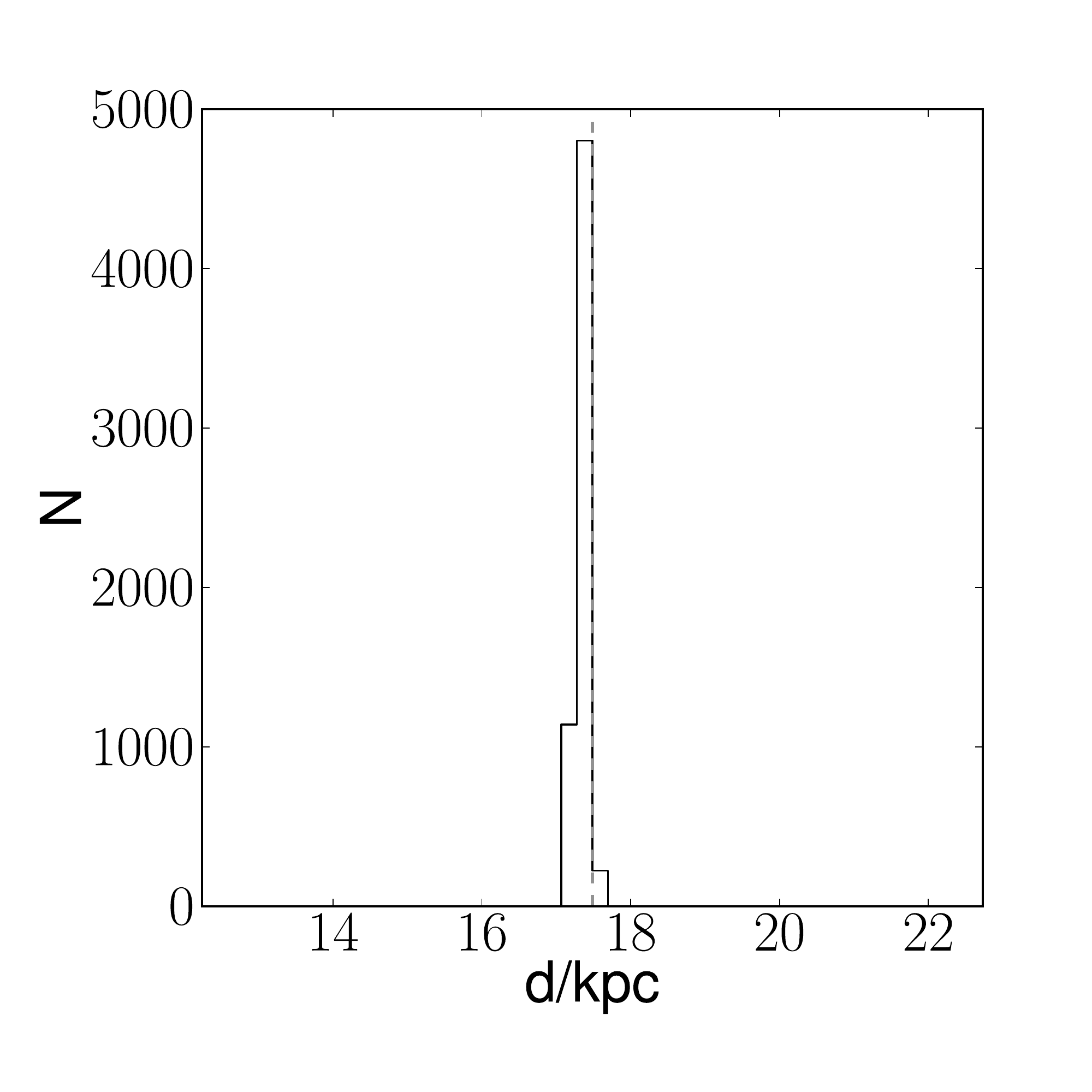}
\includegraphics[width=0.3\textwidth]{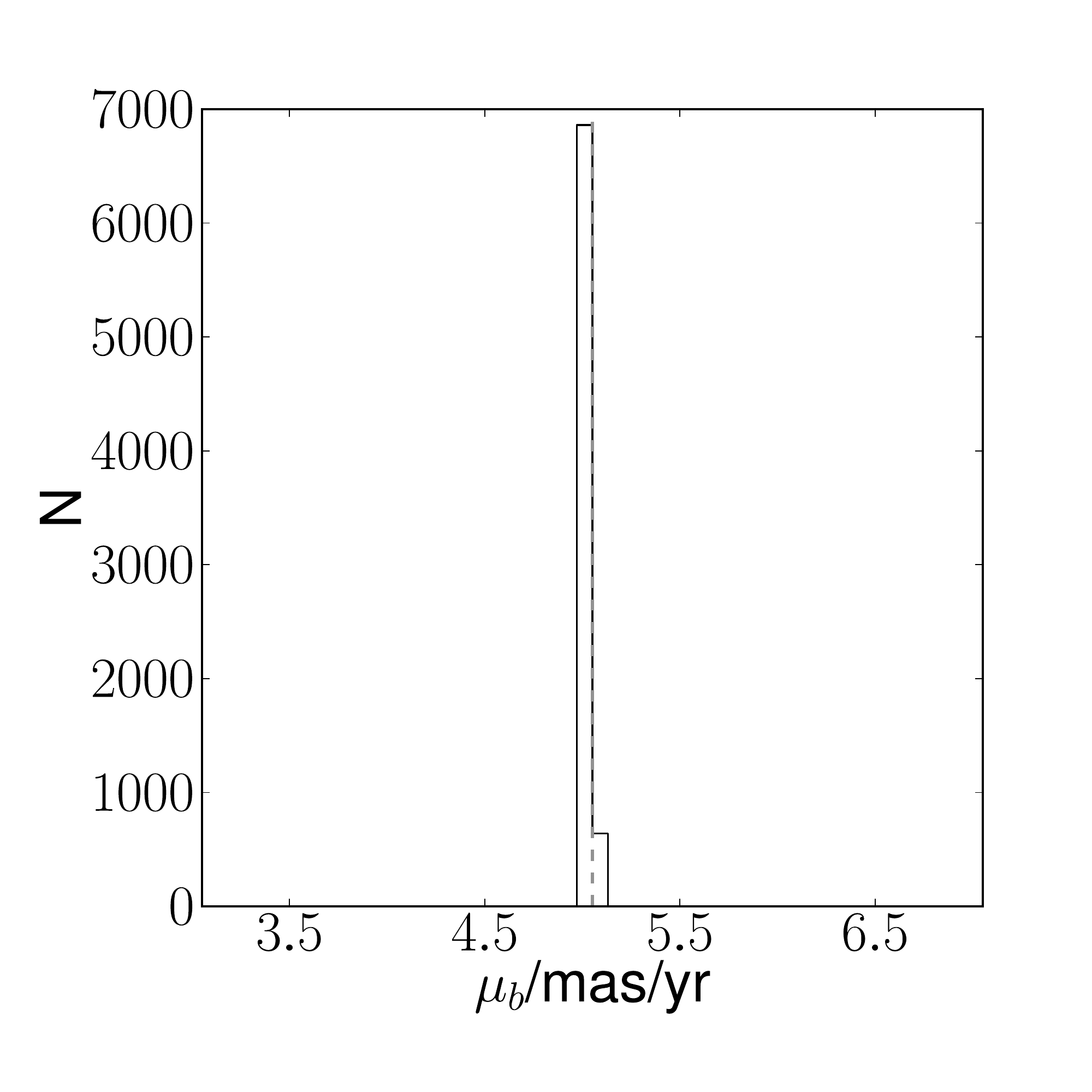}
\includegraphics[width=0.3\textwidth]{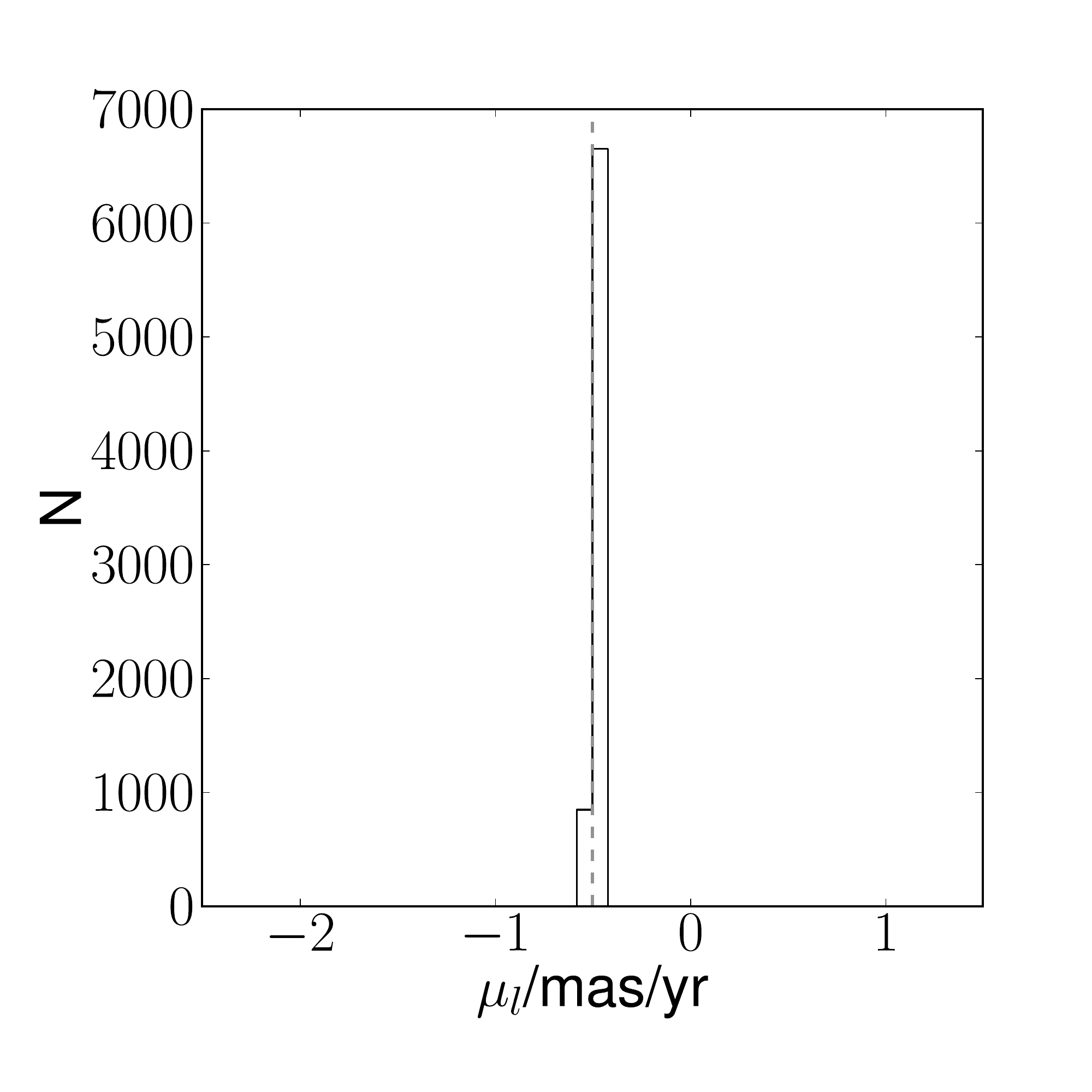}
\includegraphics[width=0.3\textwidth]{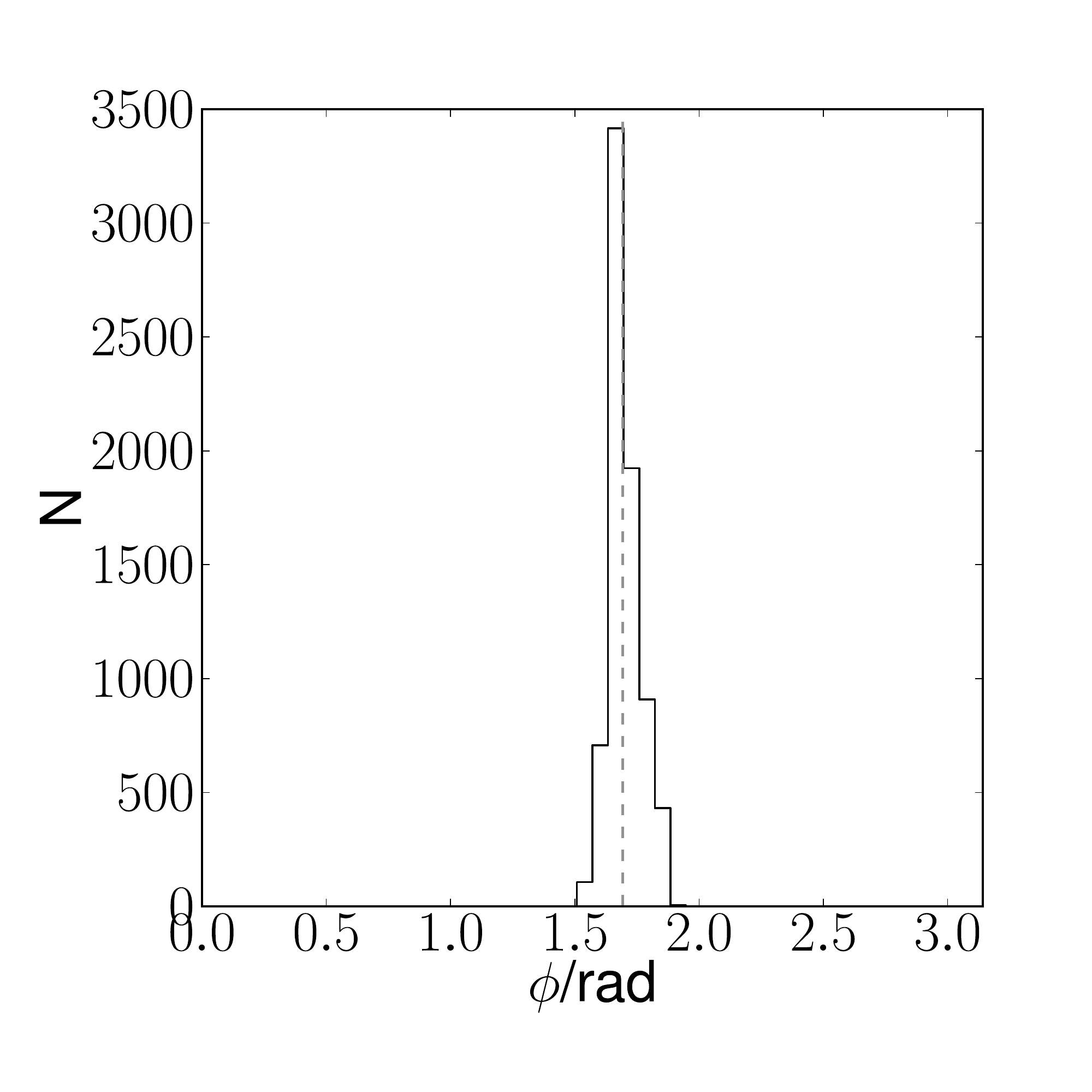}
\includegraphics[width=0.3\textwidth]{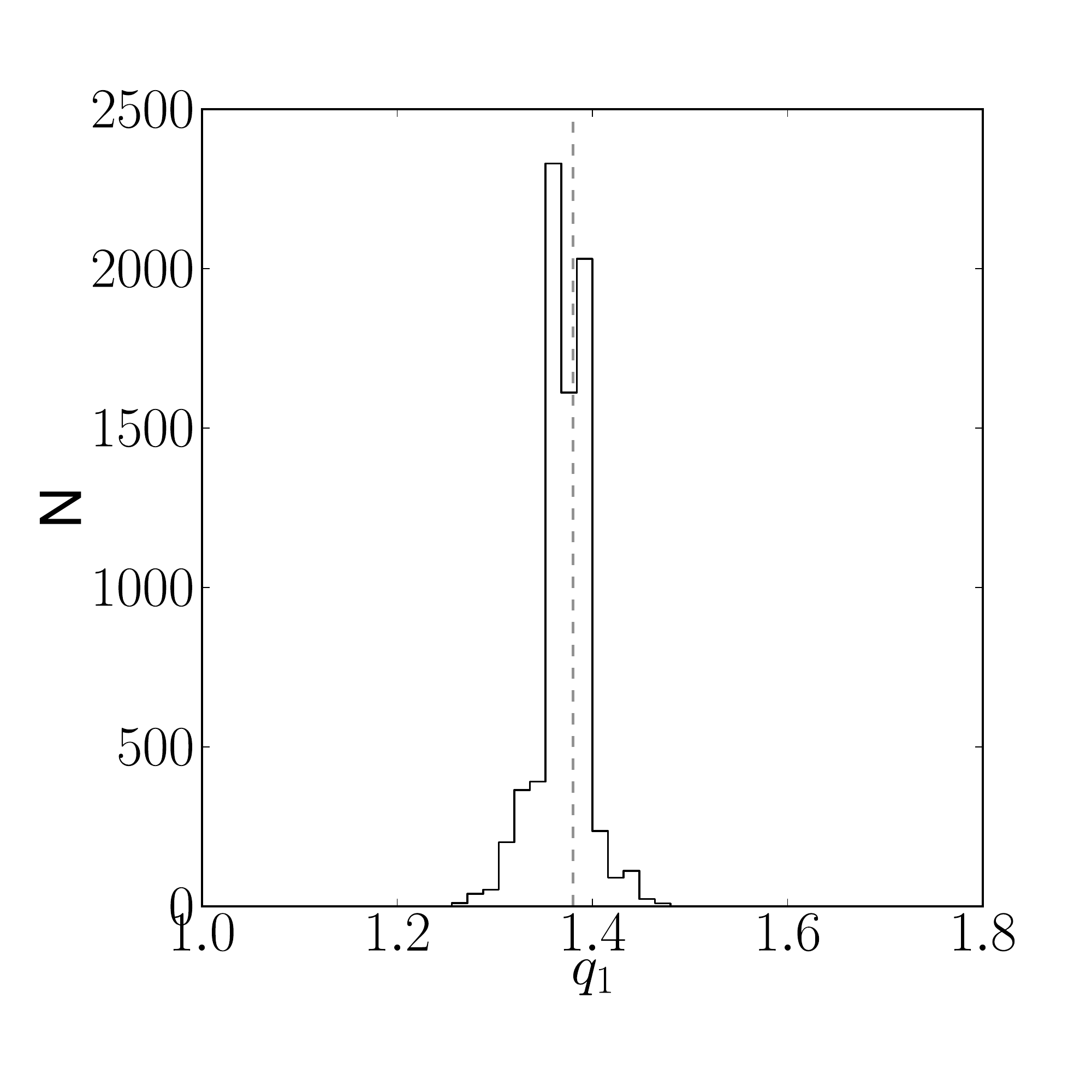}
\includegraphics[width=0.3\textwidth]{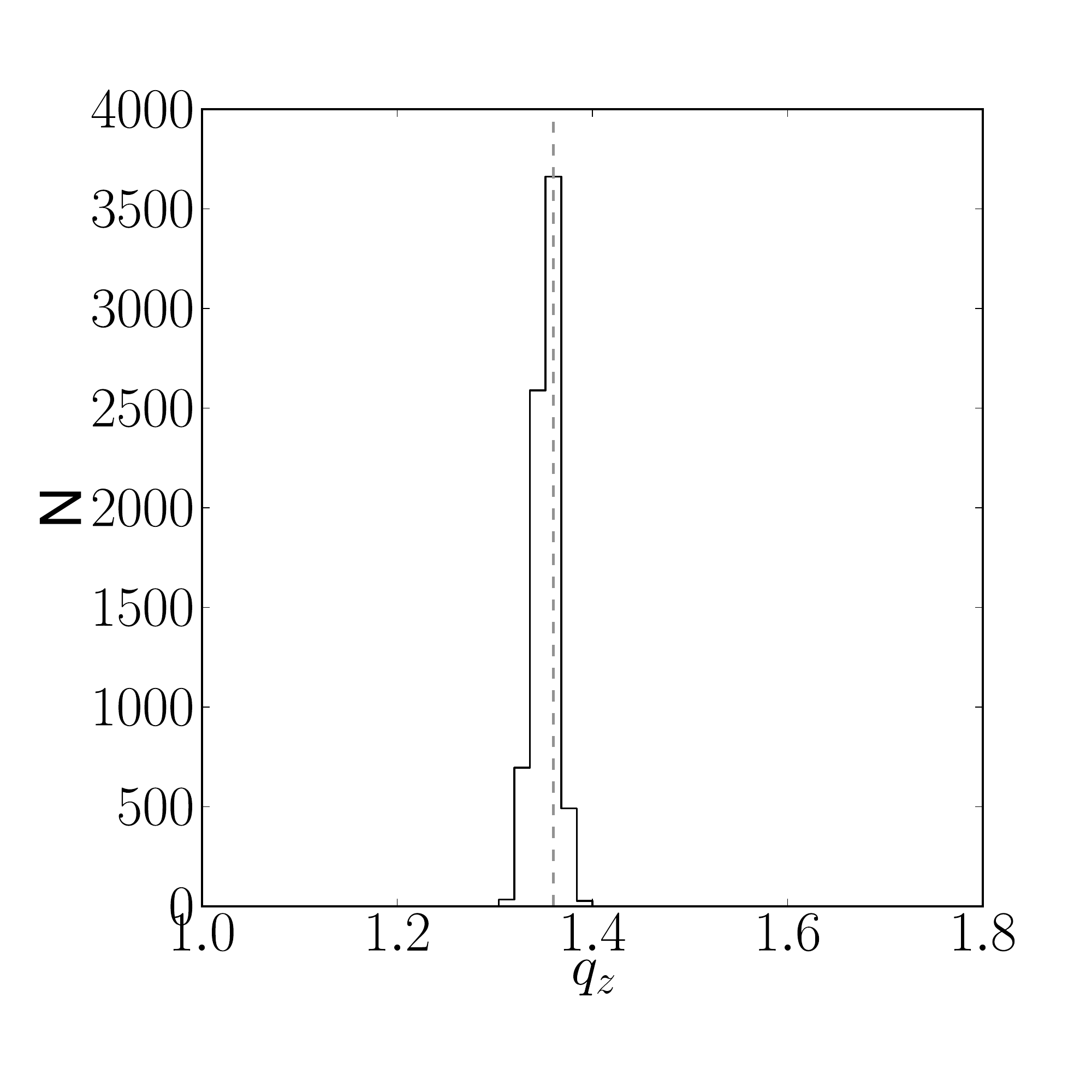}
\caption{MCMC results for fitting the {\it orbit} of an NGC 5466 like object with relatively good data along the full length of the equivalent mock stream. The parameter histograms (black) are shown for the distance $d$ and the proper motions $\mu_l$, $\mu_b$ of the initial conditions as well as the shape parameters of the halo potential $\phi$, $q_1$ and $q_z$. In each plot the correct value of the mock data model is marked by a grey dashed vertical line. Clearly, all parameters are recovered well within the uncertainties. This shows that the method works for a complete high resolution data set sampled around the original orbit.}
\label{fig:fitOrbit}
\end{center}
\end{figure*}


We typically run for $10^4$ iterations and discard the burn-in phase (initial phase where the $\chi^2$ values are higher than the 3$\sigma$ variation around the settled $\chi^2$ at the end or at least the first quarter of iterations) in our further analysis. For each analysis, we use at least 4 (but typically more) chains with different versions of starting parameters (upper/lower end and in the middle of the allowed range as well as completely randomly selected) and ensure that the final results agree between them -- i.e. that our chains are converged. 

\begin{figure*}
\begin{center} 
\includegraphics[width=0.3\textwidth]{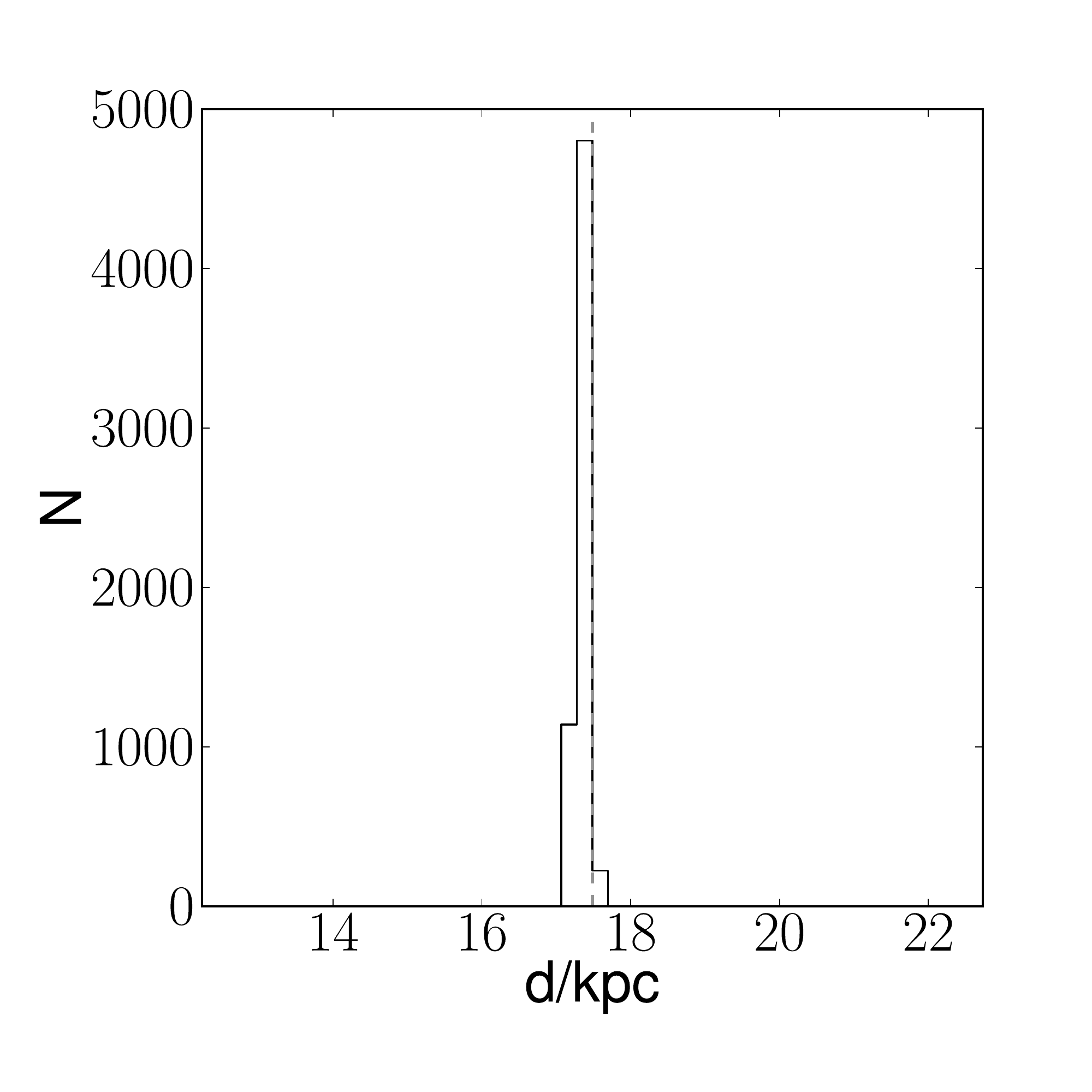}
\includegraphics[width=0.3\textwidth]{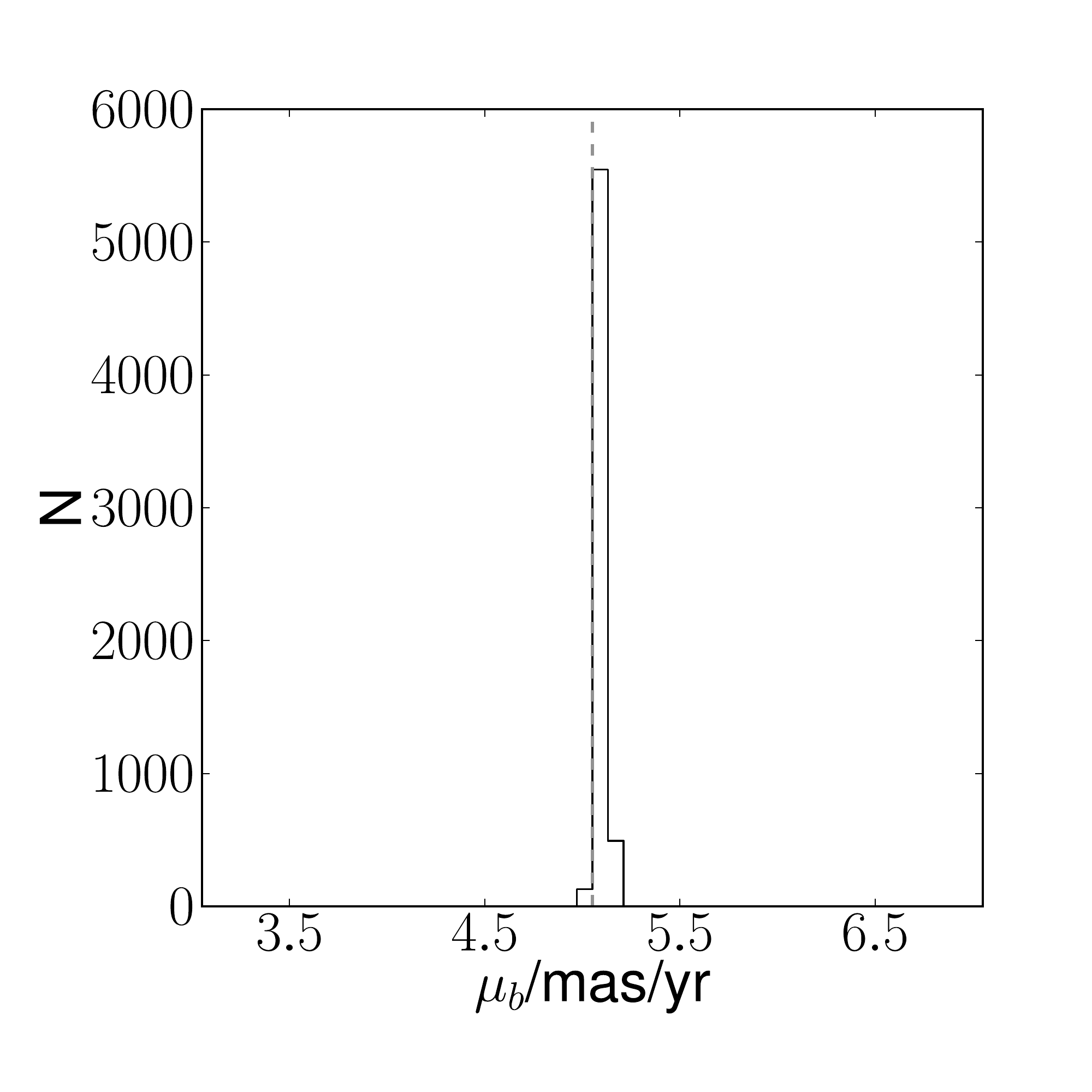}
\includegraphics[width=0.3\textwidth]{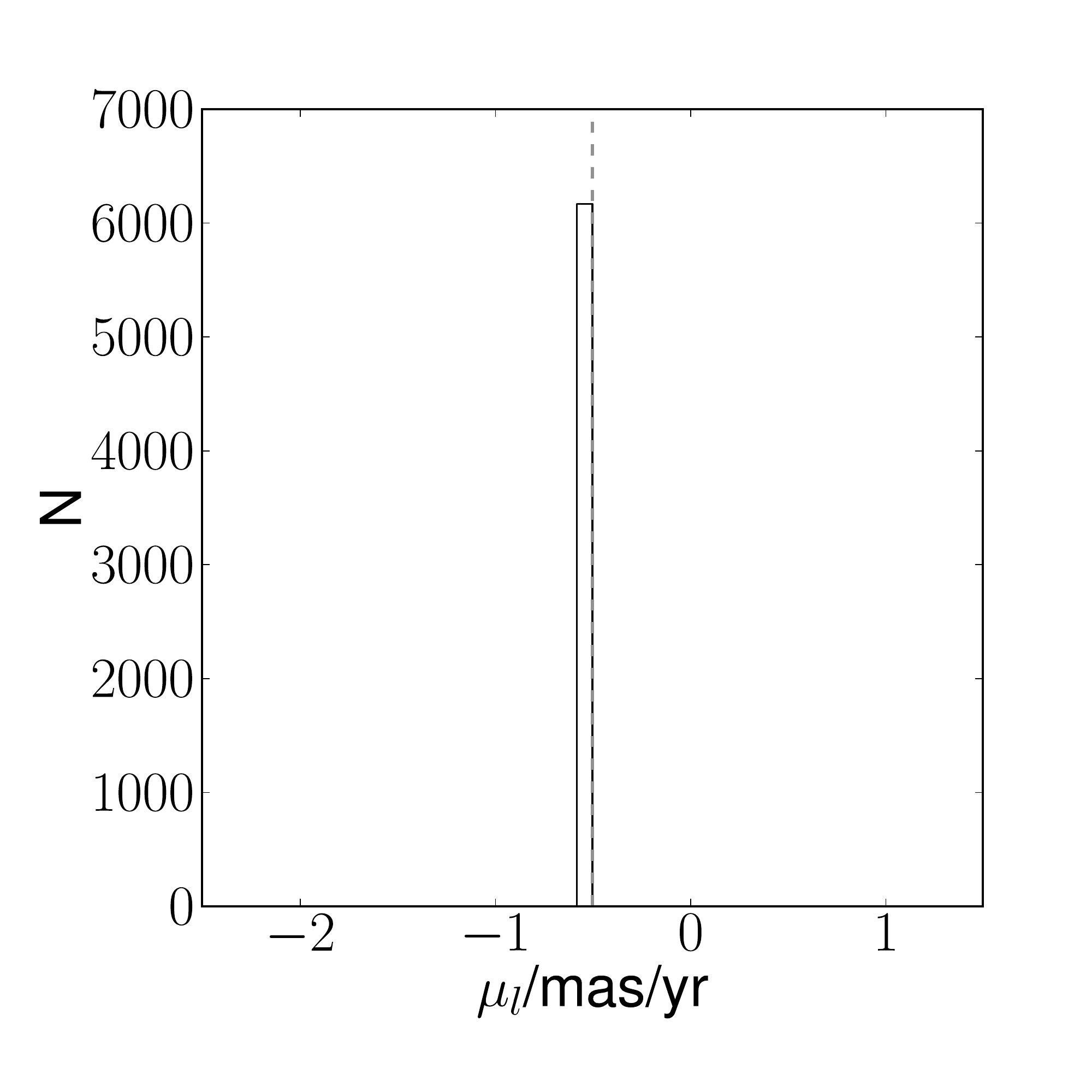}
\includegraphics[width=0.3\textwidth]{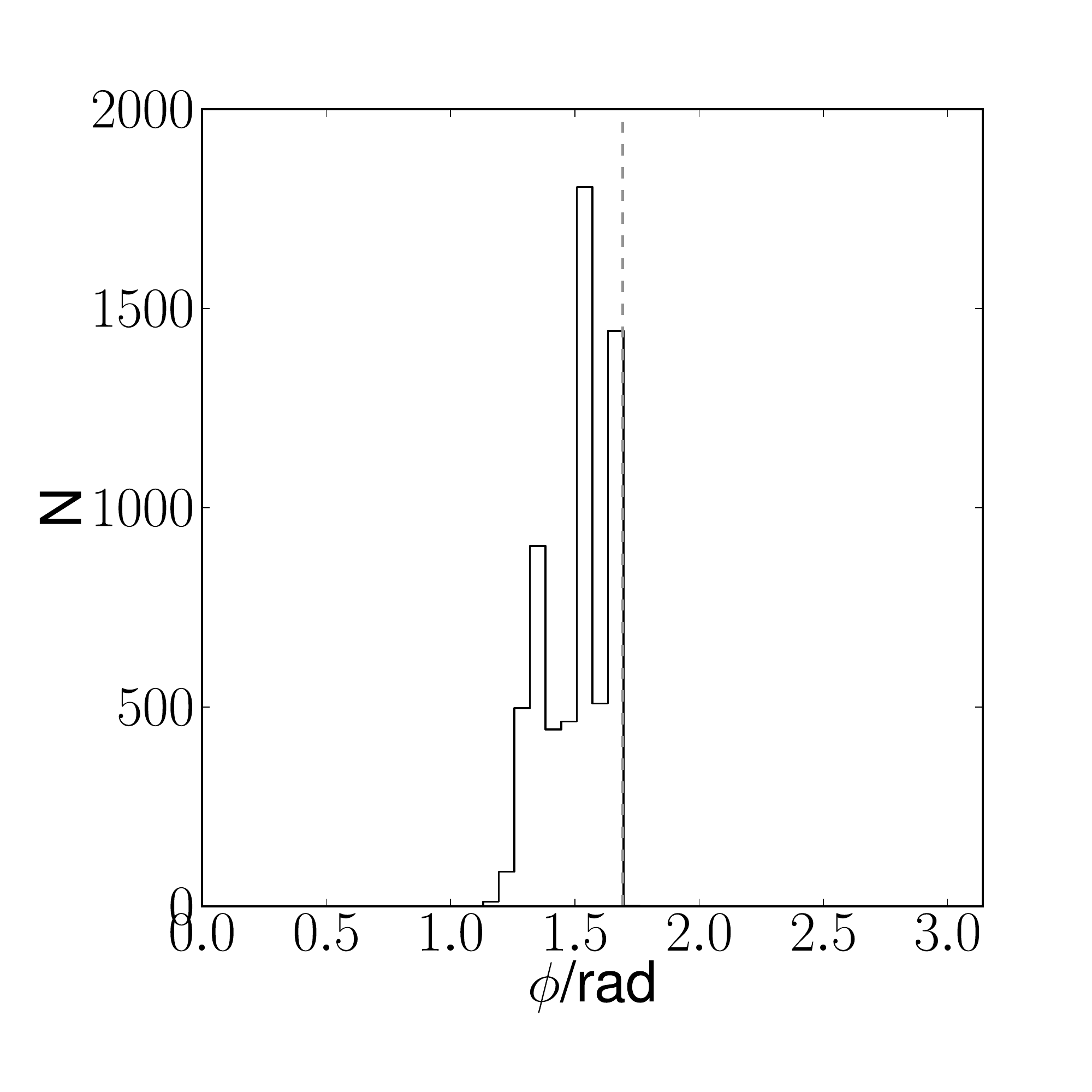}
\includegraphics[width=0.3\textwidth]{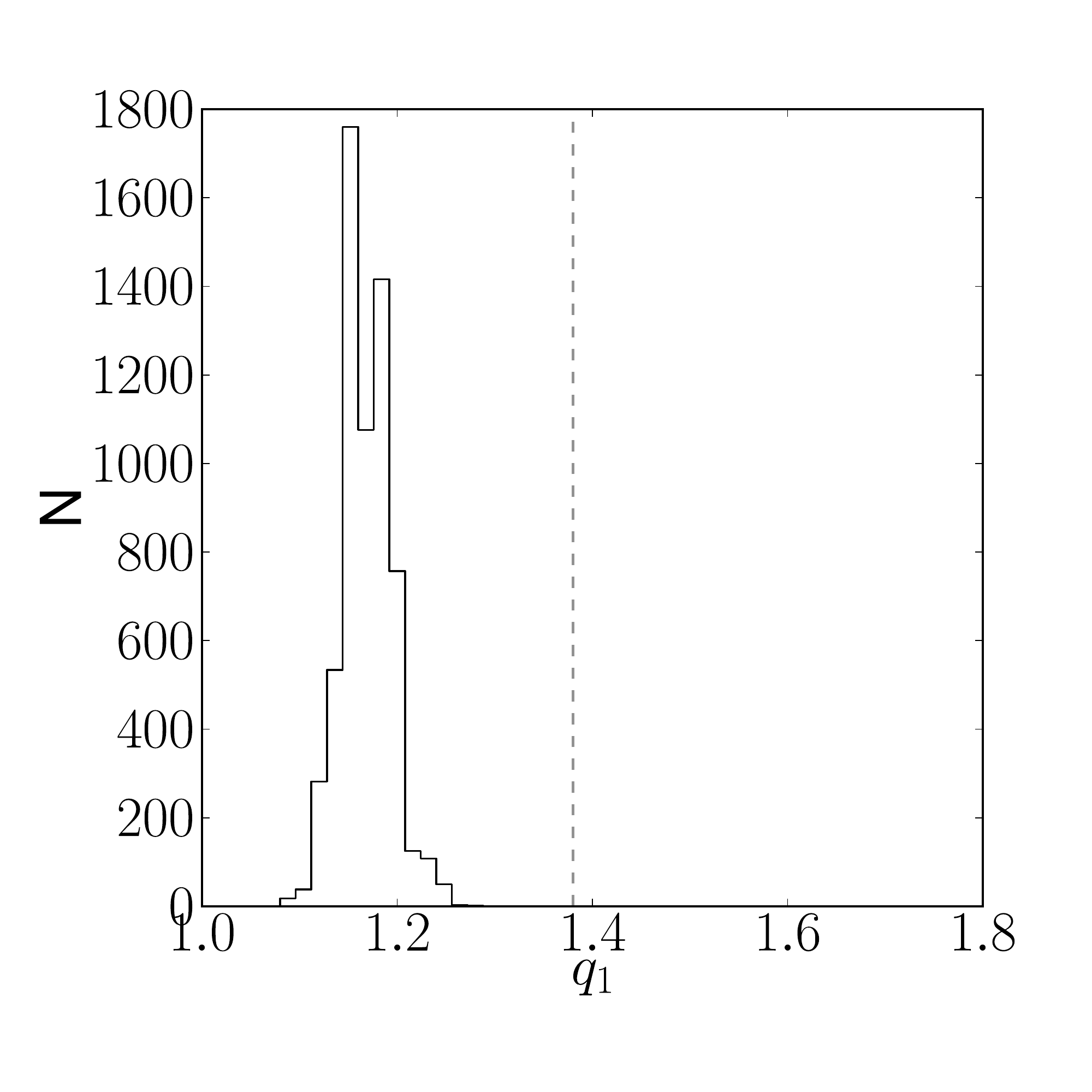}
\includegraphics[width=0.3\textwidth]{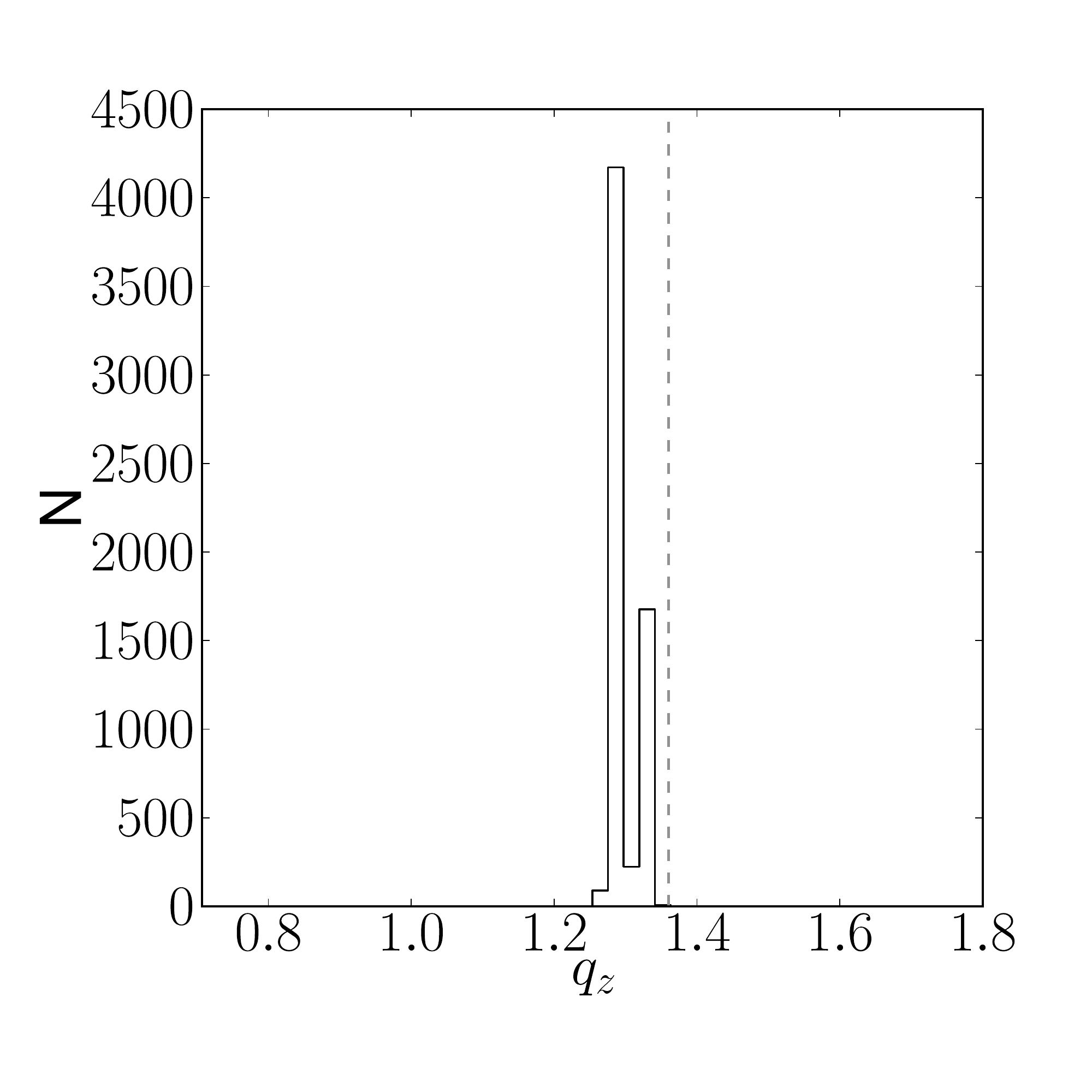}
\caption{MCMC parameter fit to a mock {\it stream} of an NGC 5466 like object. Shown are the results for the distance $d$ and the proper motions $\mu_l$, $\mu_b$ of the initial conditions as well as the shape parameters of the halo potential $\phi$, $q_1$ and $q_z$. The black histogram shows the MCMC results for each parameter, while the grey dashed line represent the correct value in the mock data model. The parameter of the initial conditions are recovered well, whereas the recovery of the halo shape parameters is biased for $q_1$ and broadened for $\phi$.}
\label{fig:fitStream}
\end{center}
\end{figure*}

\begin{figure*}
\begin{center} 
\includegraphics[width=0.3\textwidth]{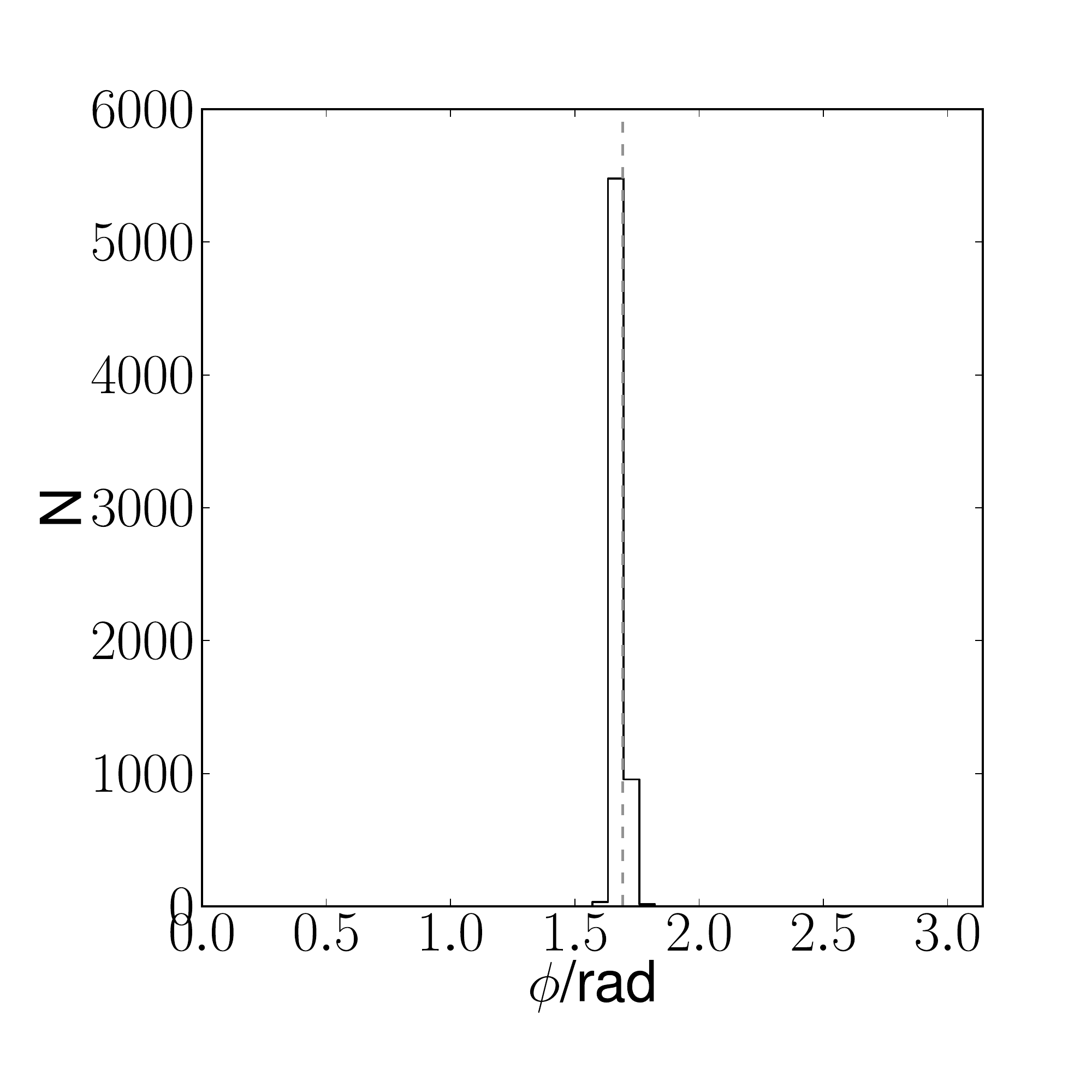}
\includegraphics[width=0.3\textwidth]{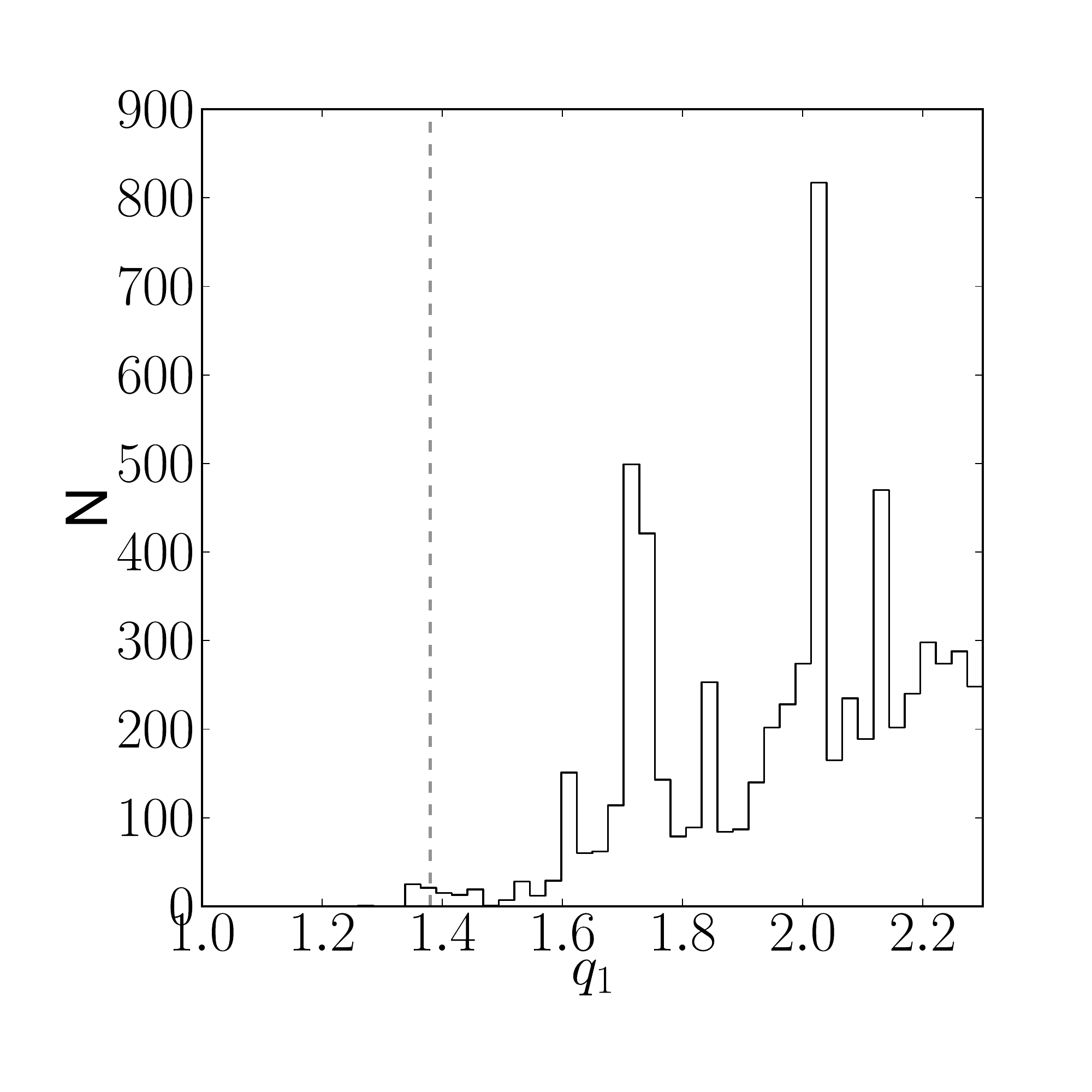}
\includegraphics[width=0.3\textwidth]{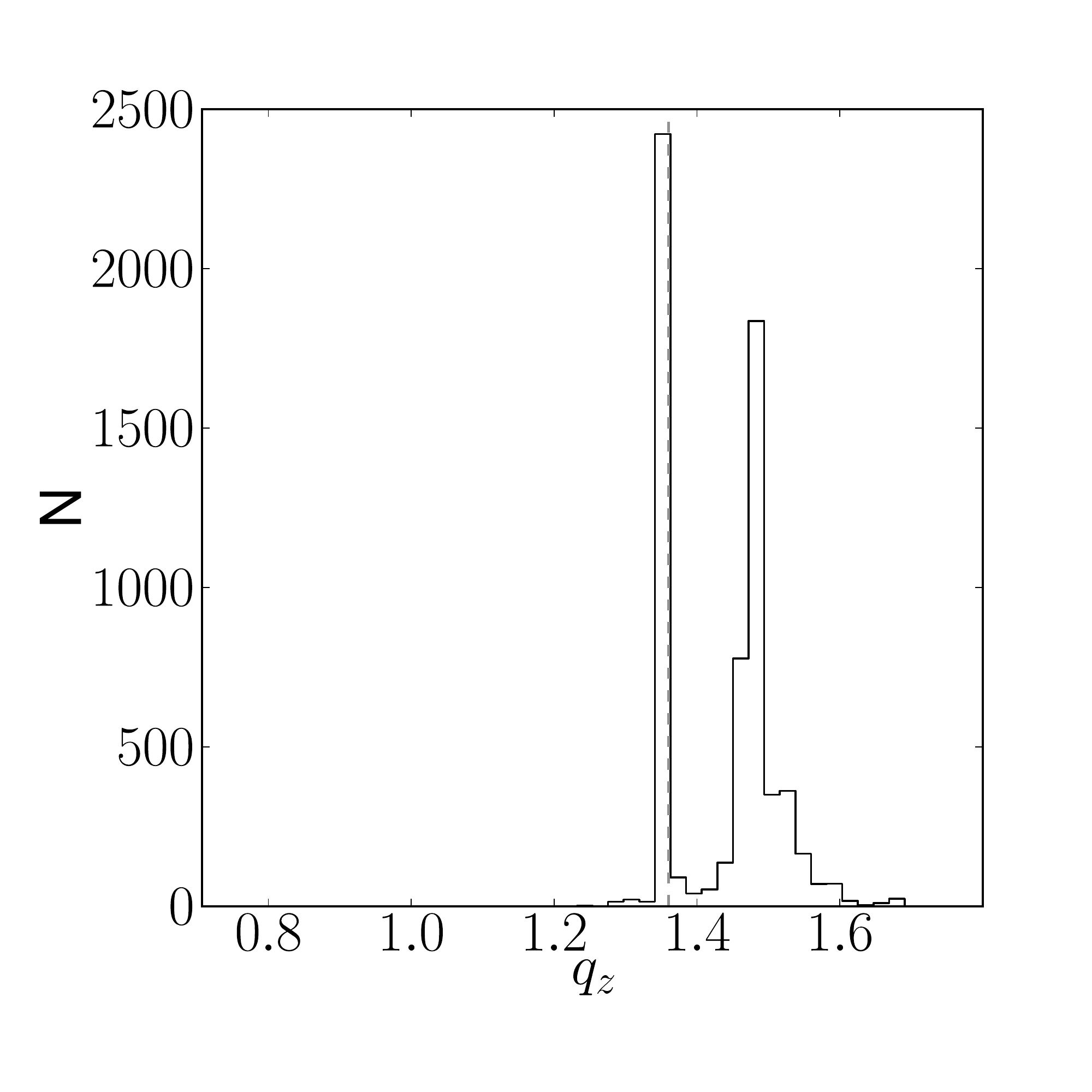}
\caption{As Figure \ref{fig:fitStream}, but for Pal 5. For brevity, we omit results for the well-constrained orbit parameters. Notice that $q_1$ is poorly constrained, while $q_z$ is double-peaked. As for NGC 5466, not modelling the stream-orbit offsets introduces a systematic error on $q_z$ of order $\sim 20$\%.}
\label{fig:Pal5fitstream}
\end{center}
\end{figure*}


\section{Results}\label{sec:results}

\subsection{Mock data}\label{sec:mockdata}
To determine what can be learned from fitting test particle orbits to thin stream data in the Milky Way halo, we created mock data sets to mimic the Pal 5 and NGC 5466 streams. We assume knowledge of all 6 data dimensions along the full length of the stream. The stream was simulated using the same technique as described in \S\ref{sec:tracing} using 10,000 particles initially in a Hernquist profile with mass $10^4$\,M$_\odot$ and scale radius $0.0075$\,kpc. The integration was run for 6.8\,Gyrs using our fiducial triaxial potential described in \S\ref{sec:method}. (We run for this length of time as over longer times than this, orbits are likely to be significantly affected by group infall or the time varying Galactic potential \citep{2010MNRAS.406.2312L}.) We then generated the mock data from these N-body data including the effect of measurement errors by sampling stream data from a Gaussian centred on the stream. We assume uncertainties equivalent to the stream width in Galactocentric coordinates $l$ and $b$; of order the radial velocity dispersion for $v_r$; distance errors of 10\% \citep[e.g.][]{2009ApJ...697..207W}; and proper motion errors of 3\,mas/yr \citep{2004AJ....127.3034M}. We bin the stream data in $\Delta l =1^\circ$ angular bins. An example plot showing our mock data for NGC 5466 is given in Figure \ref{fig:NGCmock}.

\subsection{Fitting orbits}
To test the correct functionality of our method we start by fitting data sampled around the {\it orbit} of an NGC 5466 like object to constrain the underlying potential. This is instructive as these data generated from the true orbit do not have any systematic offset (unlike the stream). Here, we know that the errors are sampled with a Gaussian distribution around the true orbit created in our fiducial potential (c.f. \S\ref{sec:mockdata}).

The results of our MCMC chain are shown in Figure \ref{fig:fitOrbit}. The histograms represent the recovered probability distribution of the respective parameter given the constraints from the data. The correct values are depicted by grey dashed vertical lines. As expected, all parameters are recovered within our quoted uncertainties.


\begin{figure*}
\begin{center}
$q_1 \in [1.0,1.8]$\\
\includegraphics[width=0.32\textwidth]{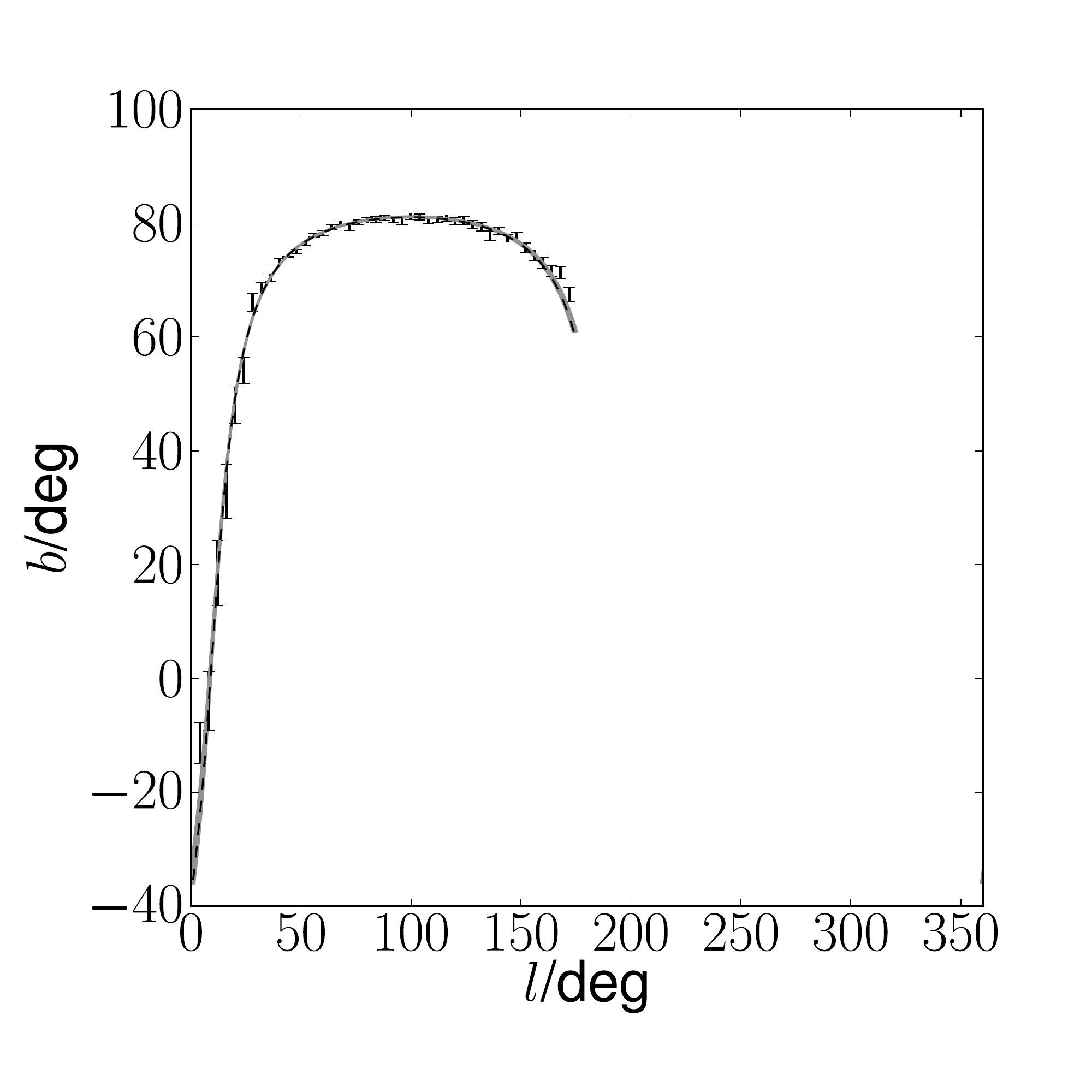}
\includegraphics[width=0.32\textwidth]{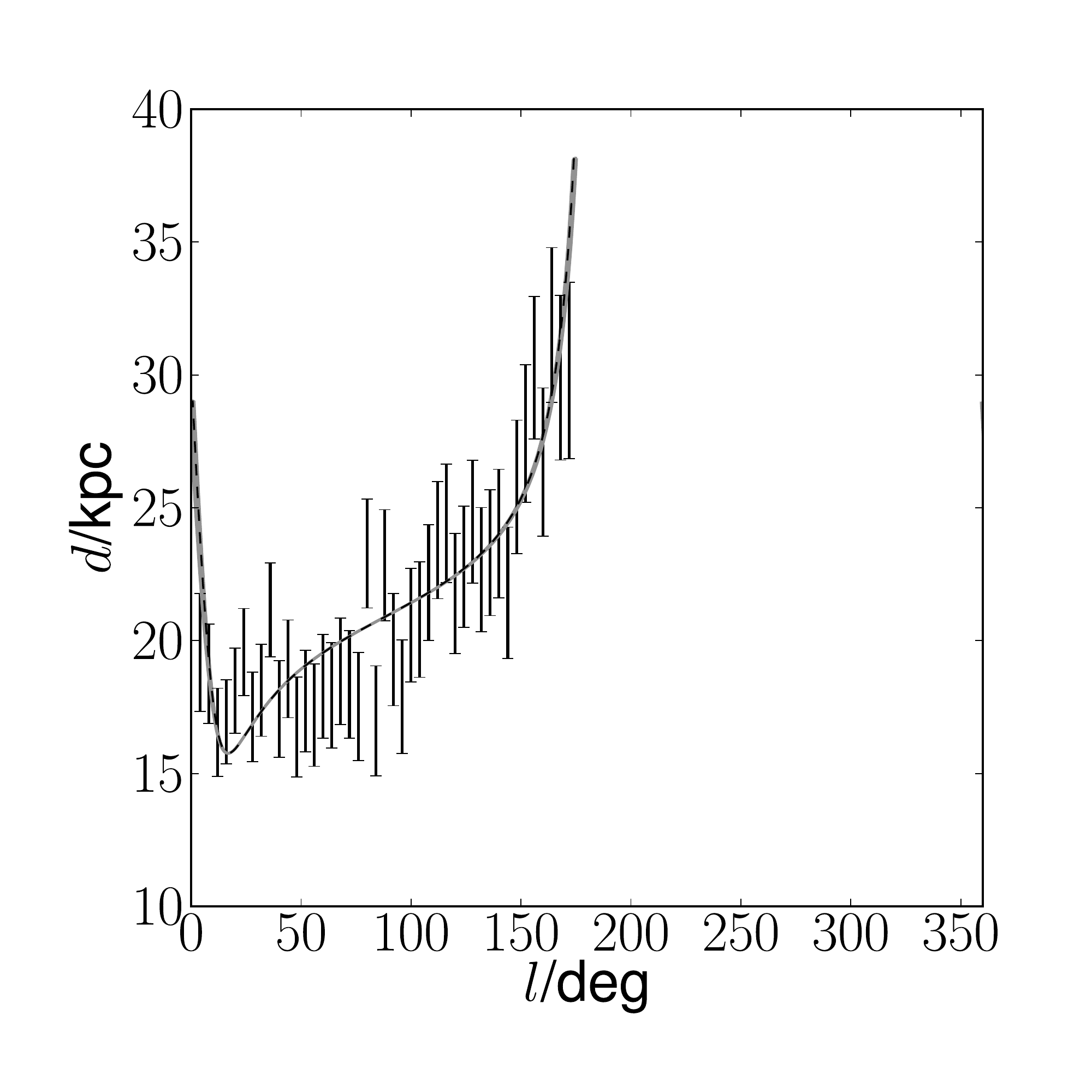}
\includegraphics[width=0.32\textwidth]{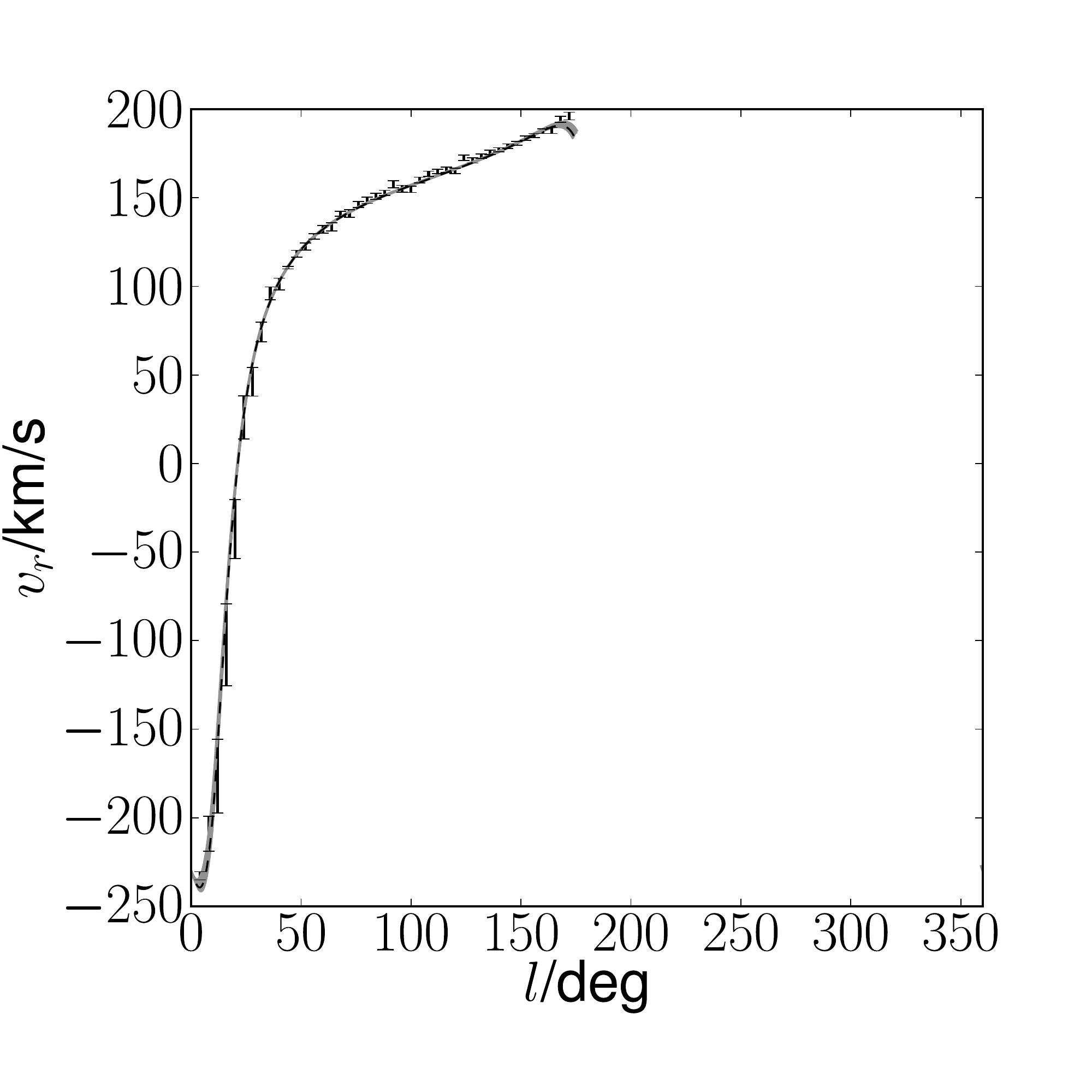}
$q_z \in [1.0,1.8]$\\
\includegraphics[width=0.32\textwidth]{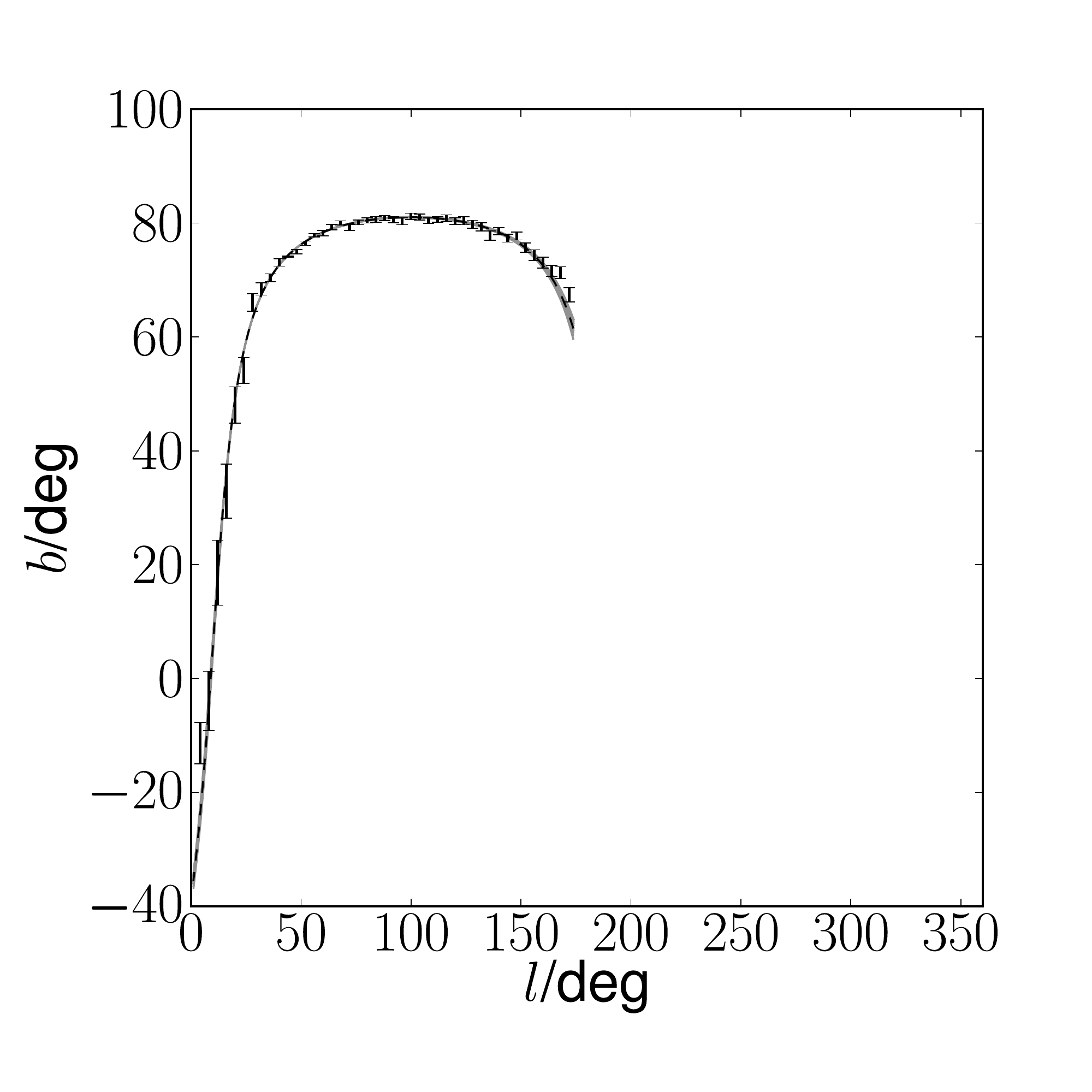}
\includegraphics[width=0.32\textwidth]{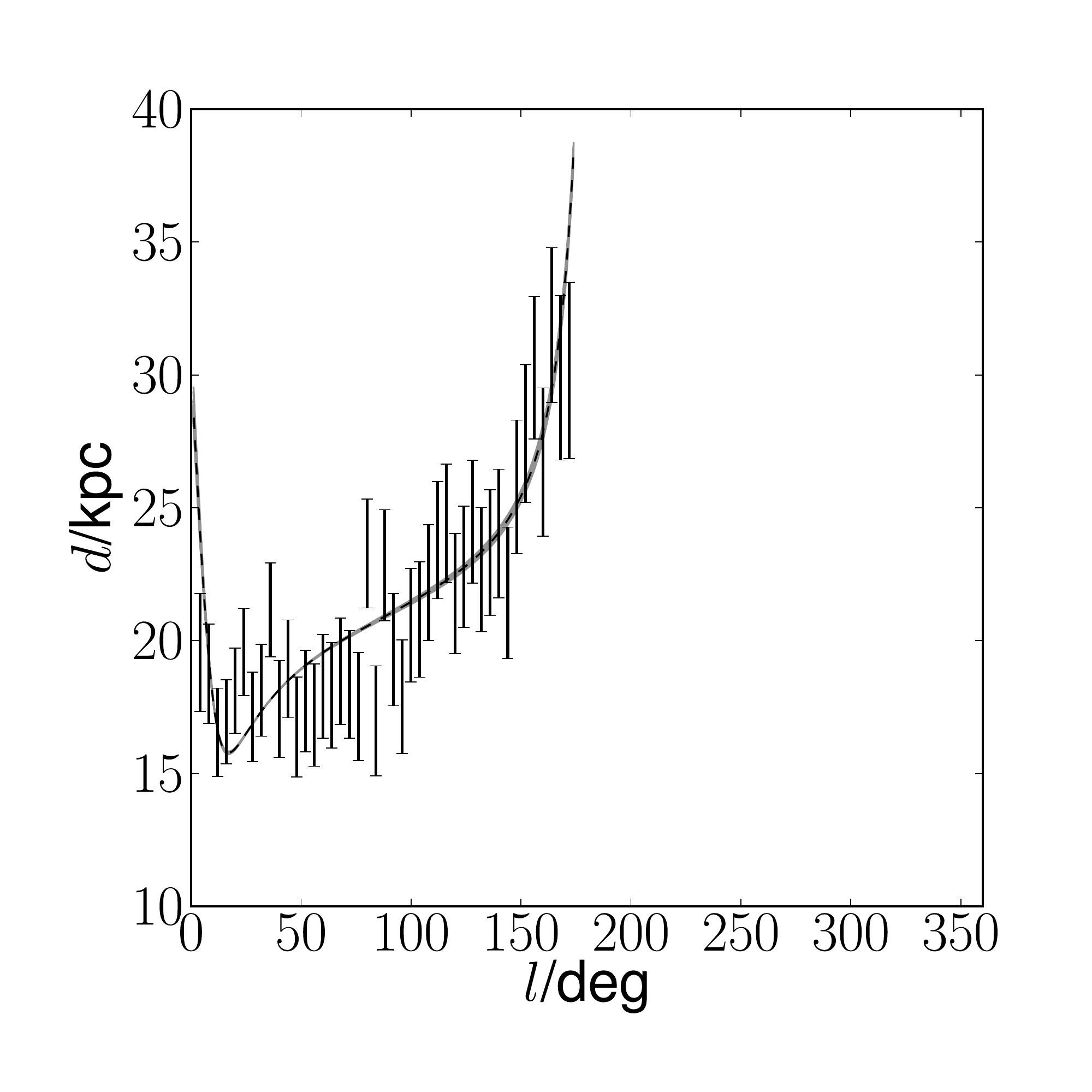}
\includegraphics[width=0.32\textwidth]{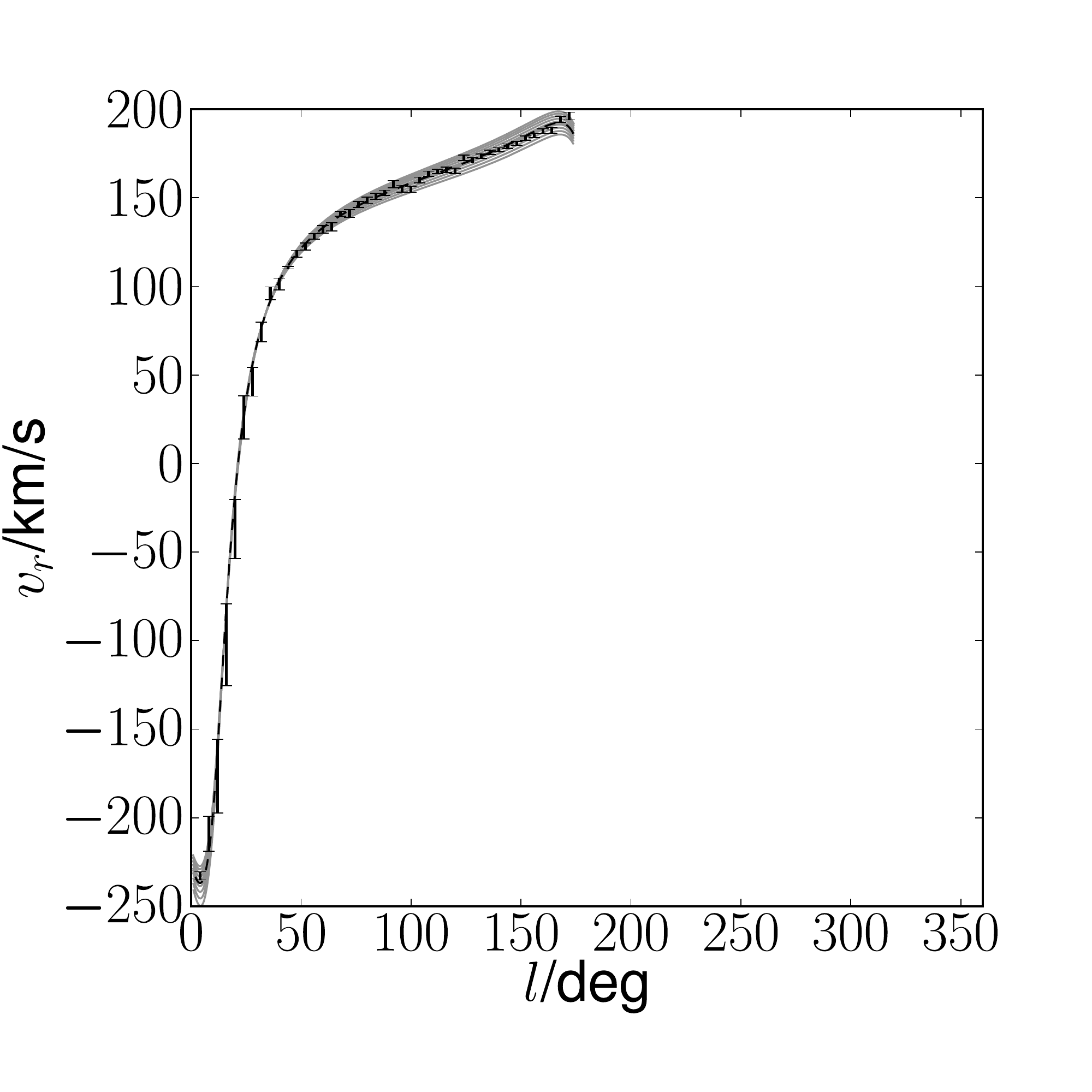}
\caption{NGC mock stream data (black error bars) and orbits for $q_1 \in [1.0,1.8]$ (upper panels) and $q_z \in [1.0,1.8]$ (lower panels). The black dashed orbit marks the true orbit of the mock model. The proper motion plot has been omitted for brevity. The stream is rather insensitive to changes in $q_1$ with respect to the assumed uncertainties, while being sensitive to changes in $q_z$ in the radial velocities. Note, that for these plots the orientation angle of the disc axis $\phi$ is fixed to our fiducial value, i.e. $q_1$ measures the length of the halo axis in the disk plane and perpendicular to the satellite's orbital plane.}
\label{fig:NGC_qs}
\end{center}
\end{figure*}

\begin{figure*}
\begin{center}
$q_1 \in [1.0,1.8]$ \\
\includegraphics[width=0.32\textwidth]{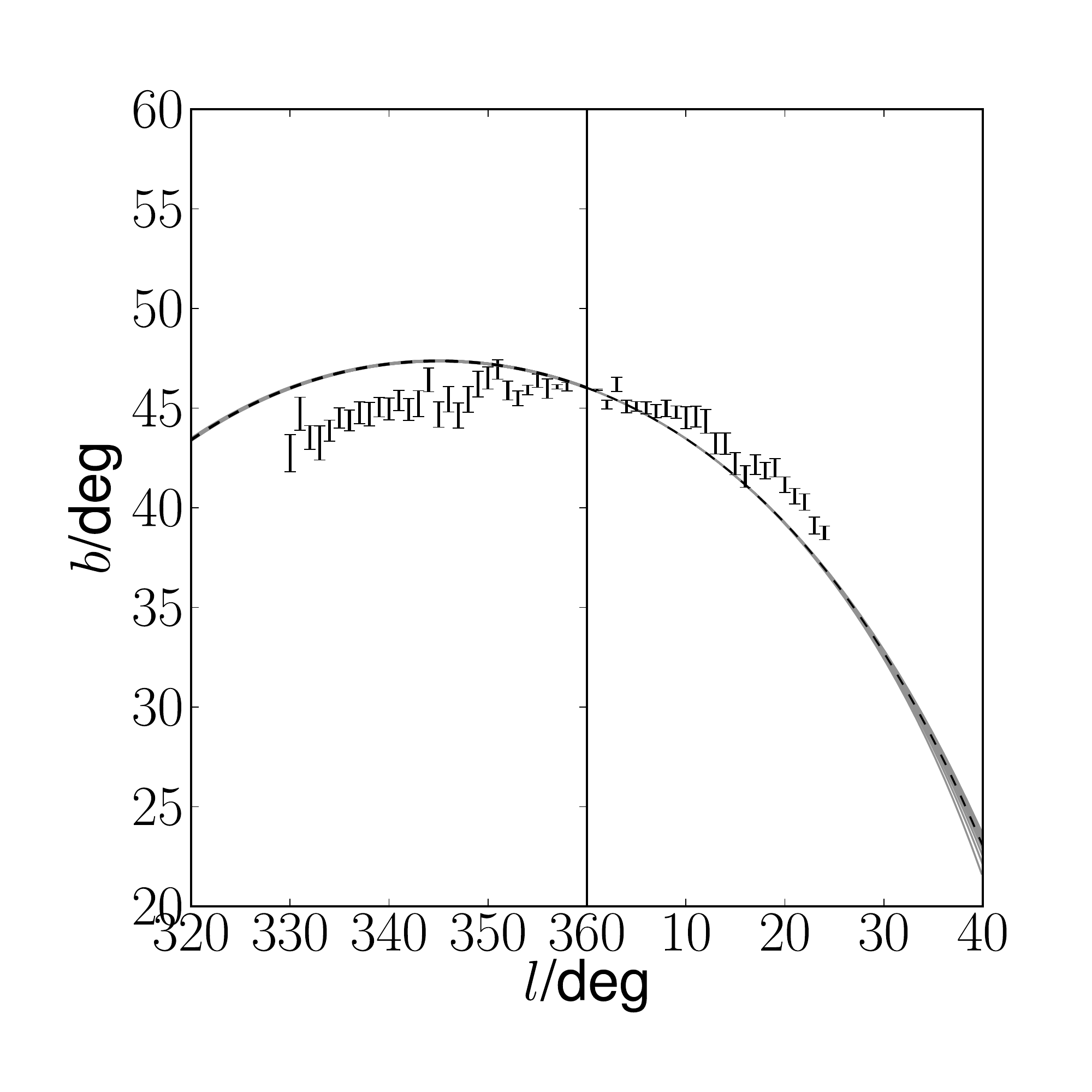}
\includegraphics[width=0.32\textwidth]{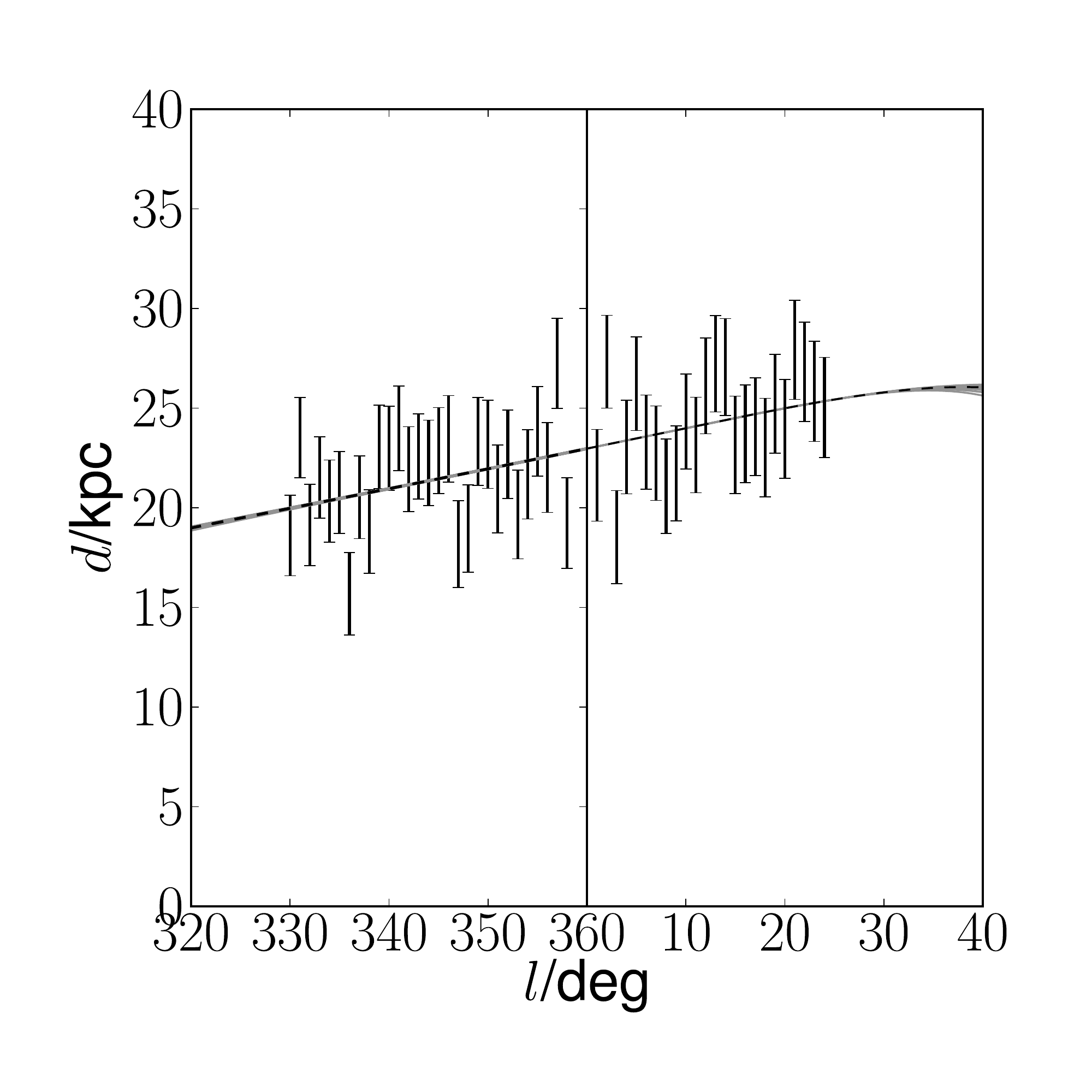}
\includegraphics[width=0.32\textwidth]{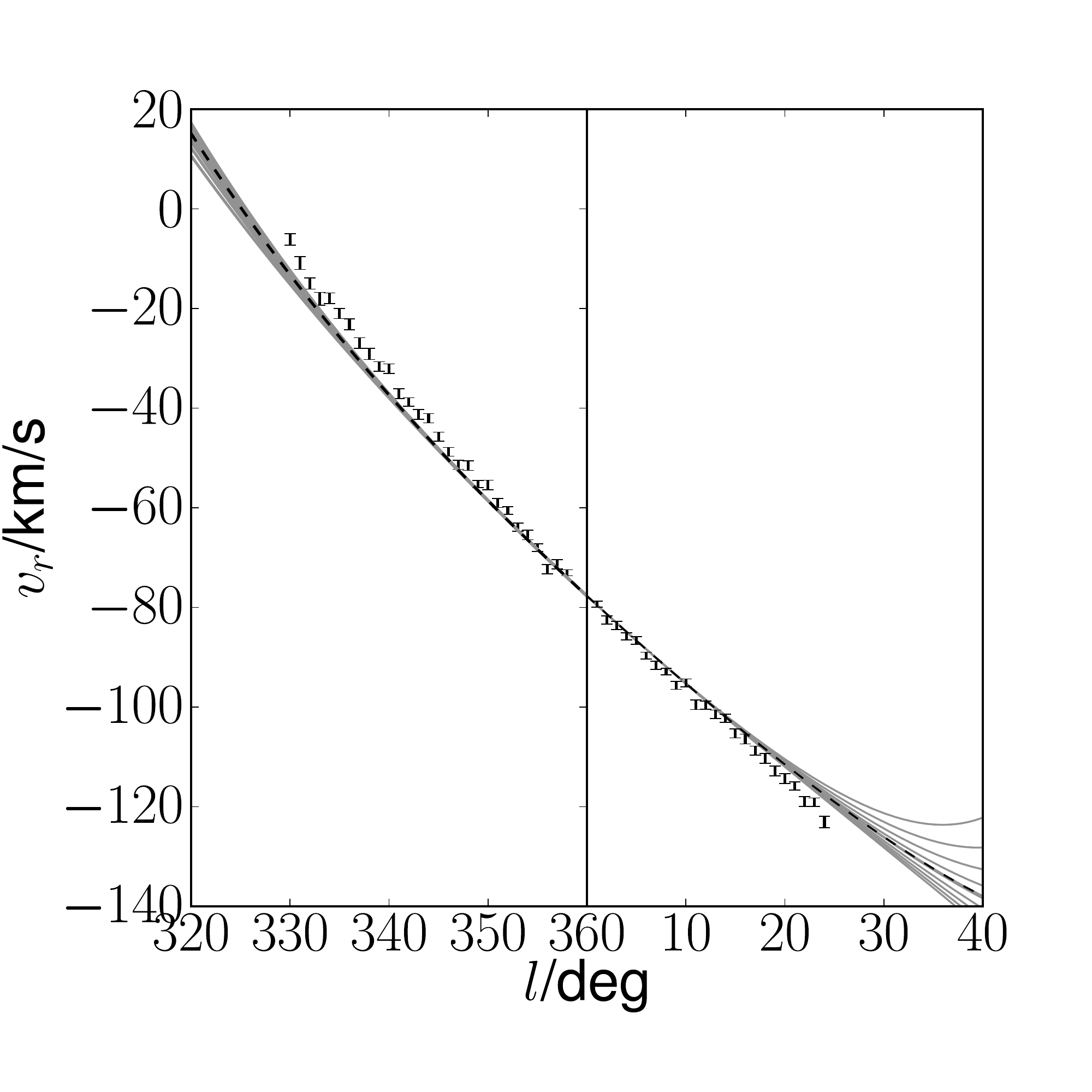}
$q_z \in [1.0,1.8]$\\
\includegraphics[width=0.32\textwidth]{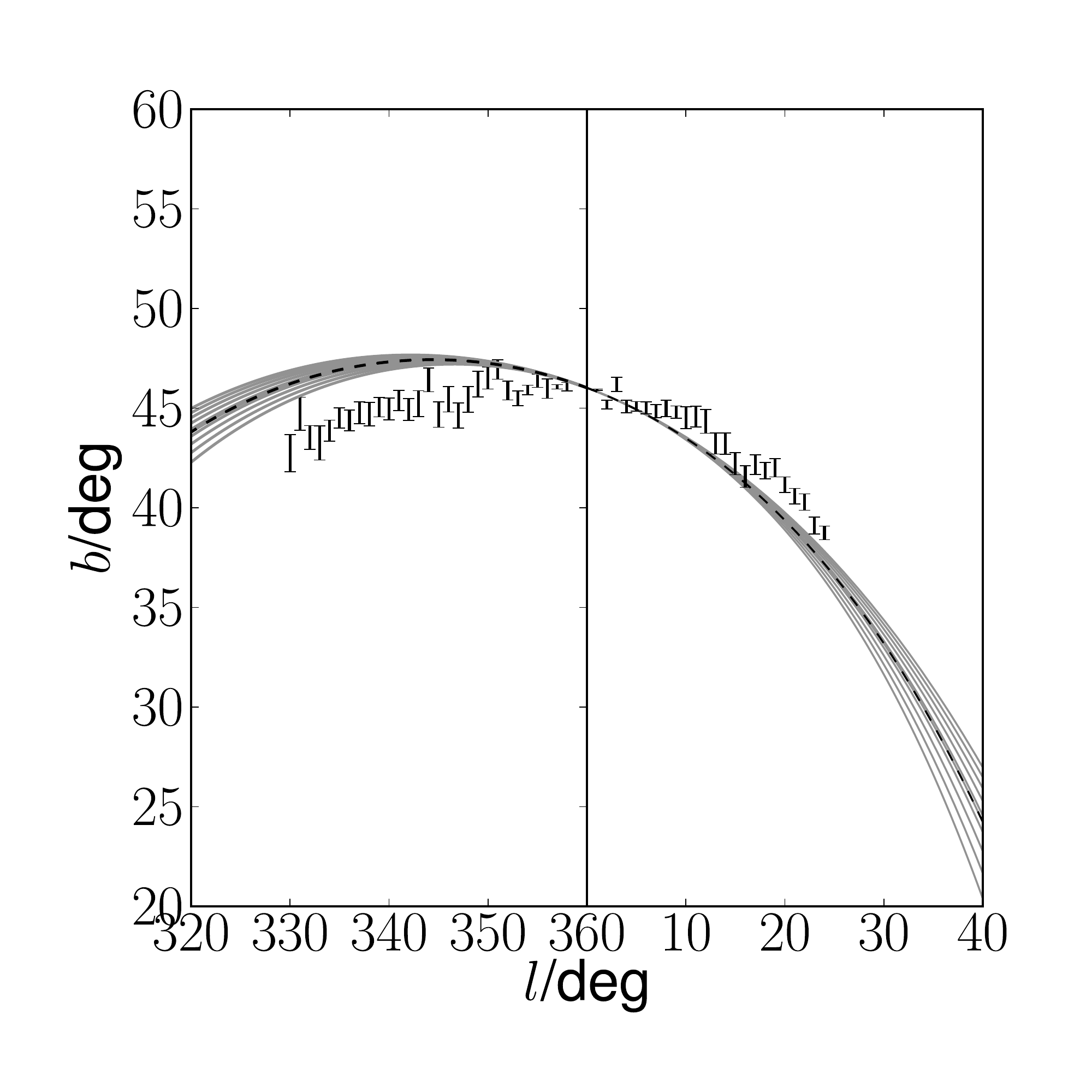}
\includegraphics[width=0.32\textwidth]{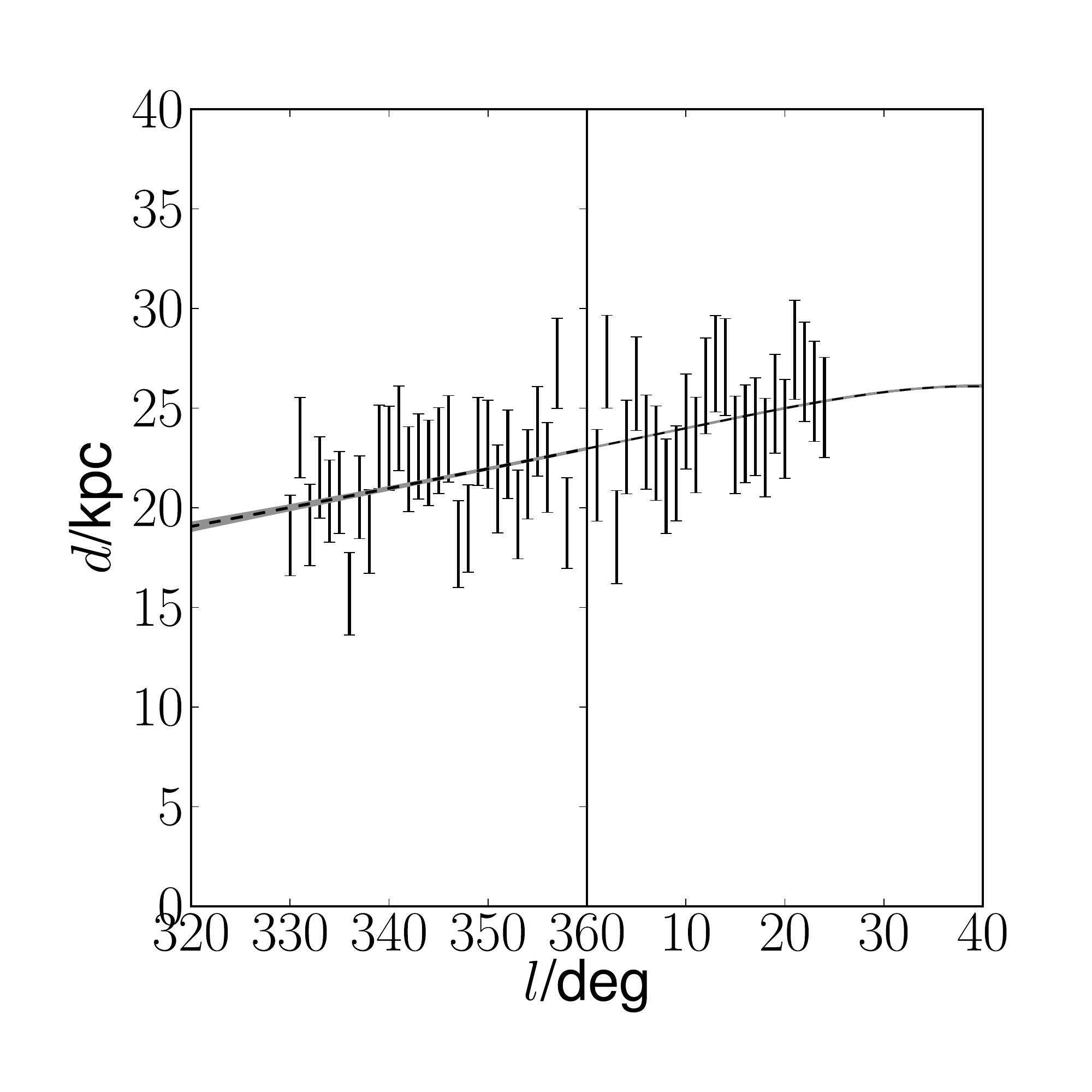}
\includegraphics[width=0.32\textwidth]{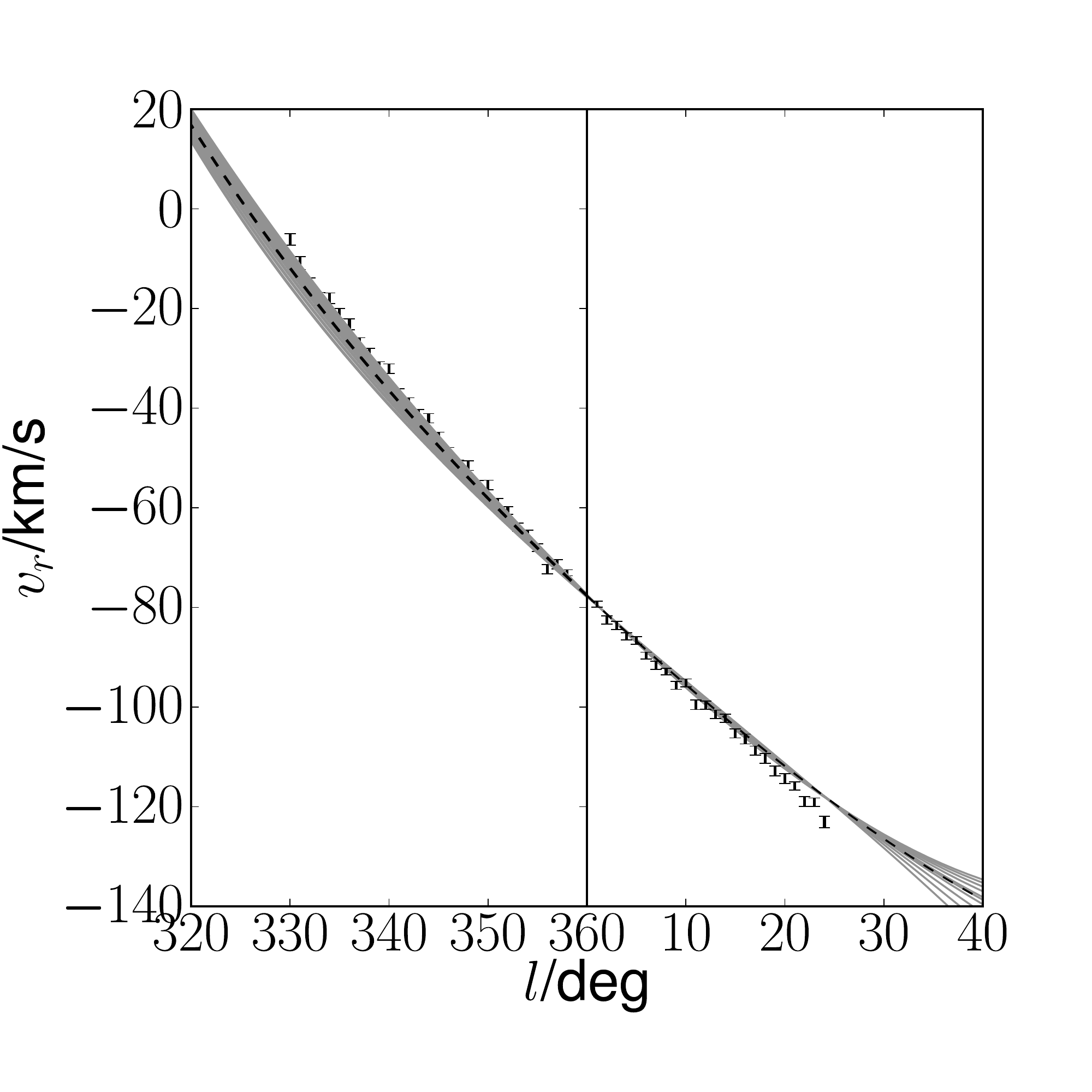}
\caption{Same as Figure \ref{fig:NGC_qs} for a Pal 5 mock stream. The stream is rather insensitive to changes in $q_1$ with respect to the assumed uncertainties, but quite sensitive to changes in $q_z$ -- particularly in the well measured on-sky positions and radial velocities. Note, that for these plots the orientation angle of the disc axis $\phi$ is fixed to our fiducial value, i.e. $q_1$ measures the length of the halo axis in the disk plane and in the satellite's orbital plane.}
\label{fig:P5_qs}
\end{center}
\end{figure*}


\subsection{Fitting streams}
We have shown in the previous section that our method works well for data sampling orbits; we now address the recovery of the halo shape from {\it stream} data that are offset from the true orbit (\S\ref{sec:tracing}). 

Figure \ref{fig:fitStream} shows the equivalent to Figure \ref{fig:fitOrbit} for an NGC 5466 like mock {\it stream}. While the parameters for the initial conditions of the stream and the halo shape parameter perpendicular to the disc are recovered well, the recovered halo shape parameters in the disc plane are biased by $\sim 20$\%; the biases perpendicular to the stream are smaller. The good recovery of the stream initial conditions is expected. Several previous papers have shown that missing data components along the stream can be derived from others without knowledge of the underlying gravitational potential \citep{2008MNRAS.383.1686J,2008MNRAS.386L..47B,2009MNRAS.399L.160E,2009MNRAS.400L..43J}. Furthermore, the initial conditions only represent one point along the stream, while the halo shape parameters are only constrained by the total data set along the whole stream. However, fitting a stream instead of an orbit -- even though its offset is small in comparison to our assumed uncertainties -- reduces our ability to recover the halo shape parameters. This indicates that when fitting simple test particle orbits, only some shape parameters can be robustly recovered. Similar conclusions were reached recently by \citep{2013MNRAS.433.1813S}. They find that errors can be as large as order unity. However, for the two streams we consider here, and our particular parameterisation of the Milky Way potential with fixed mass, we are likely underestimating the errors as discussed in \S\ref{sec:method}. Still there are also reasons why we might overestimate the bias of the stream: In near spherical potentials the stream-orbit-offset are mainly constrained to the orbital plane and hence the viewing angle of the stream can potentially influence the shape recovery. For example, one could argue that for the Sagitarius stream which is observed from nearly within its orbital plane and hence the offset is expected mainly in the badly constrained distances the bias is less significant than for Pal 5 which is observed face on and, hence, the offset is mainly in the well constrained angular position of the stream. This could explain why \cite{2010ApJ...714..229L} find similar results using an N-body model to \cite{2009ApJ...703L..67L} using a simple test particle method to model the Sagittarius stream. However, as there are still other inconsistencies with the results of \cite{2010ApJ...714..229L} as shown by \cite{2013arXiv1301.2670D}, unknown issues with the supposedly superior N-body approach cannot be excluded. 

Figure \ref{fig:Pal5fitstream} shows similar results for our Pal 5 mock data. Even with full 6D phase space data, the stream is not sensitive to $q_1$, but it is sensitive to $q_z$. As with NGC 5466, the un-modelled stream-orbit offsets introduce systematic errors of the order $\sim 20$\%.

\subsection{Pal 5 and NGC 5466: constraints on halo shape parameters}\label{sec:constpower}

We have demonstrated that it is insufficient to assume that thin streams exactly trace an orbit in order to accurately recover the potential, although the initial conditions for the orbit can be recovered. Here, we examine which data are likely to be most constraining, assuming that the stream-orbit offsets -- not modelled in our current method -- can be corrected for\footnote{Possible options for such methods are full N-body models or simpler correction methods as suggested by \citet{2011MNRAS.417..198V}.}. To assess this, we consider how small changes to the potential shape parameters $q_z$ and $q_1$ affect the underlying orbit. Figure \ref{fig:NGC_qs} shows results for our NGC 5466 mock data stream. We overlay a sequence of orbits (grey lines) for varying $q_1$ and $q_z$ at fixed $\phi$, as marked in the Figure caption. The stream is most sensitive to $q_z$ in the {\it radial velocities}, however the combination of data for this stream appears to be sensitive to both $q_1$ and $q_z$ as evidenced by our results in Figures \ref{fig:fitOrbit} and \ref{fig:fitStream}. Radial velocity data for this stream already gives constraints on $q_z$, while full 6D phase space information would give powerful constrains on the Milky Way potential. 

\begin{figure}
\begin{center}
\includegraphics[width=0.39\textwidth]{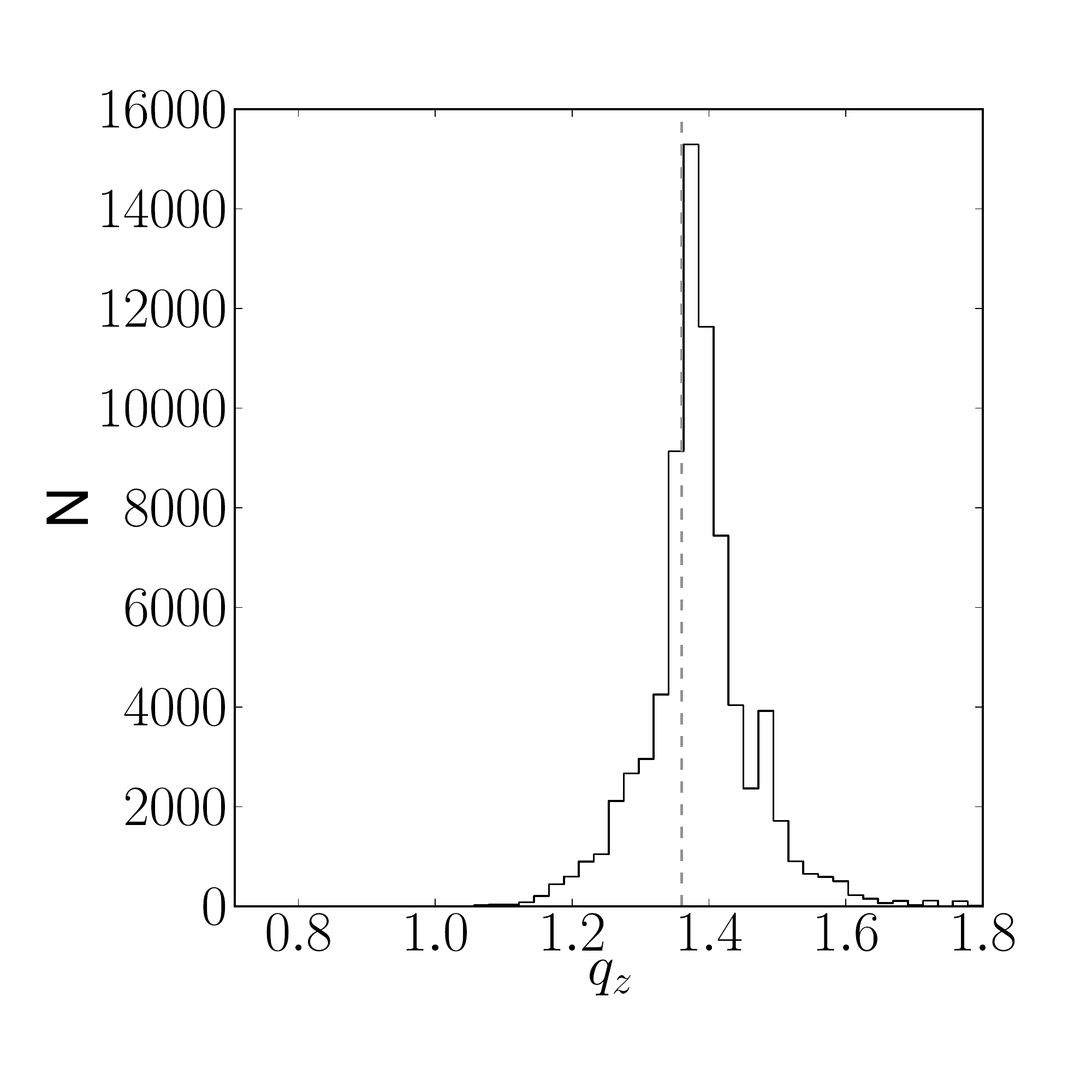}
\caption{The recovery of $q_z$ from mock Pal 5 orbit data, assuming only photometric and radial velocity data (with errors of order the stream velocity dispersion) along the stream. As in previous plots, the grey dashed line represents the correct model value of $q_z$. Such data are obtainable with current instrumentation and promise interesting new constraints on $q_z$.}
\label{fig:P5_qz_radv_recover}
\end{center}
\end{figure}

Figure \ref{fig:P5_qs} shows the similar results for our Pal 5 mock data stream. Note that in this Figure, $q_1$ measures the length of the axis in the disk plane and in Pal 5's orbital plane. We find that, as for NGC 5466, Pal 5 is mainly sensitive to changes in $q_z$ -- in particular in the well-measured angular positions and the radial velocities. Thus, Pal 5 promises good constraints on the MW halo shape perpendicular to the disc if stream-orbit offsets are correctly modelled. By contrast, Pal 5 is significantly less sensitive to $q_1$. There is hope of constraining $q_1$, however, if radial velocity data can be obtained at high $l \simgt 30^\circ$.

\subsection{Current and near-future constraints}

Finally, we consider whether current data, or data that could be easily obtained, for Pal 5 and NGC 5466 can constrain $q_z$ or $q_1$. To asses this, we degrade our mock {\it orbit} data (i.e. without considering the stream-orbit offsets) to be consistent with current observational uncertainties for each stream, as outlined in Table \ref{tab:streams}. To maximise constraints, we model both streams simultaneously. Unfortunately, we find that neither stream is at present constraining, even if modelled in combination. As noted in \S\ref{sec:method}, our variable step size MCMC performs poorly when the solution space becomes highly degenerate, settling on local minima that depend entirely on our starting position in parameter space for each chain. When running our chains on our current data quality mock orbit data, no global minimum in $q_z$ or $q_1$ is found. Thus, even if correctly modelling stream-orbit offsets, the currently available data for NGC 5466 and Pal 5 are not constraining. However, relatively small improvements in data quality promise new and interesting constraints on $q_1$ and $q_z$. As suggested by Figure \ref{fig:P5_qs}, good radial velocity data along the Pal 5 stream should provide constraints on $q_z$. To test this, we consider our mock orbit data for Pal 5 for `current' data quality, but adding in radial velocities with an error of order the radial velocity dispersion. The results for recovering $q_z$ are shown in Figure \ref{fig:P5_qz_radv_recover}. Notice that we obtain an excellent recovery of $q_z$ given only on-sky positions and radial velocity data for the Pal 5 stream. 

From Figure \ref{fig:NGC_qs}, it seems less likely that small improvements to the data available for NGC 5466 would yield significantly improved constraints on $q_1$ or $q_z$. However, for the real NGC 5466 data there is tentative evidence for a departure from a smooth orbit at the western edge of the stream \citep{2006ApJ...639L..17G}. We discuss this separately in a companion paper where we show that, if such a deviation in the orbit is present, NGC 5466 provides rather strong constraints on the MW halo shape parameters. Even if only photometric data along the stream are available, it is possible to rule out spherical or prolate halos \citep{2012MNRAS.tmpL.464L}. (That such ``turning points" in streams provide powerful constraints was already noted by \citet{2011MNRAS.417..198V}.)

\section{Conclusions}\label{sec:conclusion}
We have considered which Milky Way thin tidal streams promise useful constraints on the potential shape, assuming a simple triaxial potential model for the Milky Way. We introduced a test particle method to fit stream data, using a Markov Chain Monte Carlo technique to marginalise over uncertainties in the progenitor's orbit and the Milky Way halo shape parameters. We showed that, even for very cold streams, stream-orbit offsets -- not modelled in our simple method -- introduce systematic biases in the recovered shape parameters. For the streams that we considered, and our particular choice of potential parameterisation, these errors are of order $\sim 20$\% on the halo flattening parameters. However, larger systematic errors can arise for more general streams and potentials; such offsets need to be correctly modelled in order to obtain an unbiased recovery of the underlying potential. 

We used our new method to assess what type and quality of stream data are most constraining, under the assumption that stream-orbit offsets can be corrected for, e.g. by using N-body models or other methods as shown in \cite{2011MNRAS.417..198V}. Our key findings are as follows: 

\begin{enumerate} 
\item Of all known Milky Way thin streams, the globular cluster streams NGC 5466 and Pal 5 are the most promising. These form an interesting pair as their orbital planes are both approximately perpendicular to each other and to the disc, giving optimal constraints on the MW halo shape. 

\item Using orbit-models for the streams, we show that with current data quality and modelling both streams together, current constraints on the potential shape parameters are poor. 

\item Good radial velocity data along the Pal 5 stream will provide constraints on $q_z$ (see Figure \ref{fig:P5_qz_radv_recover}). Furthermore, for the real NGC 5466 data, there is tentative evidence for a departure from a smooth orbit for NGC 5466 at its western edge \citep{2006ApJ...639L..17G}. We discuss this separately in a companion paper where we show that, if such a deviation in the orbit is present, NGC 5466 provides rather strong constraints on the MW halo shape parameters. Even if only photometric data along the stream are available, it is possible to rule out spherical or prolate halos \citep{2012MNRAS.tmpL.464L}.

\item With full 6D phase space information, NGC 5466 is sensitive to both flattening in $q_z$ and $q_1$. Pal 5 is sensitive to $q_z$, but only sensitive to $q_1$ if radial velocity data at $l \simgt 30^\circ$ can be obtained.

\end{enumerate}

\section*{Acknowledgments}
The authors would like to thank the anonymous referee who improved this work with his/her advice. Furthermore, they are grateful for David Law's continuous support and would like to thank him for helpful discussions. HL also gratefully acknowledges helpful discussions with Frazer Pearce, Steven Bamford and Mike Merrifield. HL acknowledges a fellowship from the European CommissionÕs Framework Programme 7, through the Marie Curie Initial Training Network CosmoComp (PITN-GA-2009-238356). JIR would like to acknowledge support from SNF grant PP00P2\_128540 / 1. 

\bibliographystyle{mn2e}
\bibliography{StreamsMethodN_v1}

\begin{thebibliography}{}

\bibitem[\protect\citeauthoryear{{Aumer} \& {Binney}}{{Aumer} \&
  {Binney}}{2009}]{2009MNRAS.397.1286A}
{Aumer} M.,  {Binney} J.~J.,  2009, \mnras, 397, 1286

\bibitem[\protect\citeauthoryear{{Belokurov}, {Evans}, {Bell}, {Irwin},
  {Hewett}, {Koposov}, {Rockosi}, {Gilmore} \& {et al.}}{{Belokurov}
  et~al.}{2007}]{2007ApJ...657L..89B}
{Belokurov} V.,  {Evans} N.~W.,  {Bell} E.~F.,  {Irwin} M.~J.,  {Hewett} P.~C.,
   {Koposov} S.,  {Rockosi} C.~M.,  {Gilmore} G.,    {et al.} 2007, \apjl, 657,
  L89

\bibitem[\protect\citeauthoryear{{Belokurov}, {Evans}, {Irwin}, {Hewett} \&
  {Wilkinson}}{{Belokurov} et~al.}{2006}]{2006ApJ...637L..29B}
{Belokurov} V.,  {Evans} N.~W.,  {Irwin} M.~J.,  {Hewett} P.~C.,    {Wilkinson}
  M.~I.,  2006, \apjl, 637, L29

\bibitem[\protect\citeauthoryear{{Belokurov}, {Evans}, {Irwin}, {Lynden-Bell},
  {Yanny}, {Vidrih}, {Gilmore}, {Seabroke} \& {et al.}}{{Belokurov}
  et~al.}{2007}]{2007ApJ...658..337B}
{Belokurov} V.,  {Evans} N.~W.,  {Irwin} M.~J.,  {Lynden-Bell} D.,  {Yanny} B.,
   {Vidrih} S.,  {Gilmore} G.,  {Seabroke} G.,    {et al.} 2007, \apj, 658, 337

\bibitem[\protect\citeauthoryear{{Belokurov}, {Koposov}, {Evans},
  {Pe{\~n}arrubia}, {Irwin}, {Smith}, {Lewis}, {Gieles}, {Wilkinson},
  {Gilmore}, {Olszewski} \& {Niederste-Ostholt}}{{Belokurov}
  et~al.}{2013}]{2013arXiv1301.7069B}
{Belokurov} V.,  {Koposov} S.~E.,  {Evans} N.~W.,  {Pe{\~n}arrubia} J.,
  {Irwin} M.~J.,  {Smith} M.~C.,  {Lewis} G.~F.,  {Gieles} M.,  {Wilkinson} M.,
   {Gilmore} G.,  {Olszewski} E.~W.,    {Niederste-Ostholt} M.~N.,  2013, ArXiv
  e-prints

\bibitem[\protect\citeauthoryear{{Belokurov}, {Zucker}, {Evans}, {Gilmore},
  {Vidrih}, {Bramich}, {Newberg}, {Wyse} \& {et al.}}{{Belokurov}
  et~al.}{2006}]{2006ApJ...642L.137B}
{Belokurov} V.,  {Zucker} D.~B.,  {Evans} N.~W.,  {Gilmore} G.,  {Vidrih} S.,
  {Bramich} D.~M.,  {Newberg} H.~J.,  {Wyse} R.~F.~G.,    {et al.} 2006, \apjl,
  642, L137

\bibitem[\protect\citeauthoryear{{Binney}}{{Binney}}{2008}]{2008MNRAS.386L..47B}
{Binney} J.,  2008, \mnras, 386, L47

\bibitem[\protect\citeauthoryear{{Bonaca}, {Geha} \& {Kallivayalil}}{{Bonaca}
  et~al.}{2012}]{2012ApJ...760L...6B}
{Bonaca} A.,  {Geha} M.,    {Kallivayalil} N.,  2012, \apjl, 760, L6

\bibitem[\protect\citeauthoryear{{Casetti-Dinescu}, {Girard}, {Majewski},
  {Vivas}, {Wilhelm}, {Carlin}, {Beers} \& {van Altena}}{{Casetti-Dinescu}
  et~al.}{2009}]{2009ApJ...701L..29C}
{Casetti-Dinescu} D.~I.,  {Girard} T.~M.,  {Majewski} S.~R.,  {Vivas} A.~K.,
  {Wilhelm} R.,  {Carlin} J.~L.,  {Beers} T.~C.,    {van Altena} W.~F.,  2009,
  \apjl, 701, L29

\bibitem[\protect\citeauthoryear{{Casetti-Dinescu}, {Girard}, {Platais} \& {van
  Altena}}{{Casetti-Dinescu} et~al.}{2010}]{2010AJ....139.1889C}
{Casetti-Dinescu} D.~I.,  {Girard} T.~M.,  {Platais} I.,    {van Altena} W.~F.,
   2010, \aj, 139, 1889

\bibitem[\protect\citeauthoryear{{Deason}, {Belokurov}, {Evans} \&
  {An}}{{Deason} et~al.}{2012}]{2012MNRAS.424L..44D}
{Deason} A.~J.,  {Belokurov} V.,  {Evans} N.~W.,    {An} J.,  2012, \mnras,
  424, L44

\bibitem[\protect\citeauthoryear{{Debattista}, {Moore}, {Quinn}, {Kazantzidis},
  {Maas}, {Mayer}, {Read} \& {Stadel}}{{Debattista}
  et~al.}{2008}]{2008ApJ...681.1076D}
{Debattista} V.~P.,  {Moore} B.,  {Quinn} T.,  {Kazantzidis} S.,  {Maas} R.,
  {Mayer} L.,  {Read} J.,    {Stadel} J.,  2008, \apj, 681, 1076

\bibitem[\protect\citeauthoryear{{Debattista}, {Roskar}, {Valluri}, {Quinn},
  {Moore} \& {Wadsley}}{{Debattista} et~al.}{2013}]{2013arXiv1301.2670D}
{Debattista} V.~P.,  {Roskar} R.,  {Valluri} M.,  {Quinn} T.,  {Moore} B.,
  {Wadsley} J.,  2013, ArXiv e-prints

\bibitem[\protect\citeauthoryear{{D'Onghia} \& {Lake}}{{D'Onghia} \&
  {Lake}}{2008}]{2008ApJ...686L..61D}
{D'Onghia} E.,  {Lake} G.,  2008, \apjl, 686, L61

\bibitem[\protect\citeauthoryear{{Dubinski}}{{Dubinski}}{1994}]{1994ApJ...431..617D}
{Dubinski} J.,  1994, \apj, 431, 617

\bibitem[\protect\citeauthoryear{{Dubinski} \& {Carlberg}}{{Dubinski} \&
  {Carlberg}}{1991}]{1991ApJ...378..496D}
{Dubinski} J.,  {Carlberg} R.~G.,  1991, \apj, 378, 496

\bibitem[\protect\citeauthoryear{{Eyre} \& {Binney}}{{Eyre} \&
  {Binney}}{2009}]{2009MNRAS.399L.160E}
{Eyre} A.,  {Binney} J.,  2009, \mnras, 399, L160

\bibitem[\protect\citeauthoryear{{Eyre} \& {Binney}}{{Eyre} \&
  {Binney}}{2011}]{2011MNRAS.413.1852E}
{Eyre} A.,  {Binney} J.,  2011, \mnras, 413, 1852

\bibitem[\protect\citeauthoryear{{Fellhauer}, {Belokurov}, {Evans},
  {Wilkinson}, {Zucker}, {Gilmore}, {Irwin}, {Bramich}, {Vidrih}, {Wyse},
  {Beers} \& {Brinkmann}}{{Fellhauer} et~al.}{2006}]{2006ApJ...651..167F}
{Fellhauer} M.,  {Belokurov} V.,  {Evans} N.~W.,  {Wilkinson} M.~I.,  {Zucker}
  D.~B.,  {Gilmore} G.,  {Irwin} M.~J.,  {Bramich} D.~M.,  {Vidrih} S.,  {Wyse}
  R.~F.~G.,  {Beers} T.~C.,    {Brinkmann} J.,  2006, \apj, 651, 167

\bibitem[\protect\citeauthoryear{{Fellhauer}, {Evans}, {Belokurov}, {Wilkinson}
  \& {Gilmore}}{{Fellhauer} et~al.}{2007}]{2007MNRAS.380..749F}
{Fellhauer} M.,  {Evans} N.~W.,  {Belokurov} V.,  {Wilkinson} M.~I.,
  {Gilmore} G.,  2007, \mnras, 380, 749

\bibitem[\protect\citeauthoryear{G{\'o}mez}{G{\'o}mez}{2006}]{Gomez:2006p1638}
G{\'o}mez G.~C.,  2006, The Astronomical Journal, 132, 2376

\bibitem[\protect\citeauthoryear{{Grillmair}}{{Grillmair}}{2006}]{2006ApJ...651L..29G}
{Grillmair} C.~J.,  2006, \apjl, 651, L29

\bibitem[\protect\citeauthoryear{{Grillmair}}{{Grillmair}}{2009}]{2009ApJ...693.1118G}
{Grillmair} C.~J.,  2009, \apj, 693, 1118

\bibitem[\protect\citeauthoryear{{Grillmair}, {Carlin} \&
  {Majewski}}{{Grillmair} et~al.}{2008}]{2008ApJ...689L.117G}
{Grillmair} C.~J.,  {Carlin} J.~L.,    {Majewski} S.~R.,  2008, \apjl, 689,
  L117

\bibitem[\protect\citeauthoryear{{Grillmair} \& {Dionatos}}{{Grillmair} \&
  {Dionatos}}{2006a}]{2006ApJ...641L..37G}
{Grillmair} C.~J.,  {Dionatos} O.,  2006a, \apjl, 641, L37

\bibitem[\protect\citeauthoryear{{Grillmair} \& {Dionatos}}{{Grillmair} \&
  {Dionatos}}{2006b}]{2006ApJ...643L..17G}
{Grillmair} C.~J.,  {Dionatos} O.,  2006b, \apjl, 643, L17

\bibitem[\protect\citeauthoryear{{Grillmair} \& {Johnson}}{{Grillmair} \&
  {Johnson}}{2006}]{2006ApJ...639L..17G}
{Grillmair} C.~J.,  {Johnson} R.,  2006, \apjl, 639, L17

\bibitem[\protect\citeauthoryear{{Harrigan}, {Newberg}, {Newberg}, {Yanny},
  {Beers}, {Lee} \& {Re Fiorentin}}{{Harrigan}
  et~al.}{2010}]{2010MNRAS.405.1796H}
{Harrigan} M.~J.,  {Newberg} H.~J.,  {Newberg} L.~A.,  {Yanny} B.,  {Beers}
  T.~C.,  {Lee} Y.~S.,    {Re Fiorentin} P.,  2010, \mnras, 405, 1796

\bibitem[\protect\citeauthoryear{{Helmi}}{{Helmi}}{2004}]{2004ApJ...610L..97H}
{Helmi} A.,  2004, \apjl, 610, L97

\bibitem[\protect\citeauthoryear{{Helmi} \& {White}}{{Helmi} \&
  {White}}{1999}]{1999MNRAS.307..495H}
{Helmi} A.,  {White} S.~D.~M.,  1999, \mnras, 307, 495

\bibitem[\protect\citeauthoryear{{Hernquist}}{{Hernquist}}{1990}]{1990ApJ...356..359H}
{Hernquist} L.,  1990, \apj, 356, 359

\bibitem[\protect\citeauthoryear{{Hernquist} \& {Ostriker}}{{Hernquist} \&
  {Ostriker}}{1992}]{1992ApJ...386..375H}
{Hernquist} L.,  {Ostriker} J.~P.,  1992, \apj, 386, 375

\bibitem[\protect\citeauthoryear{{Ibata}, {Irwin}, {Lewis} \& {Stolte}}{{Ibata}
  et~al.}{2001}]{2001ApJ...547L.133I}
{Ibata} R.,  {Irwin} M.,  {Lewis} G.~F.,    {Stolte} A.,  2001, \apjl, 547,
  L133

\bibitem[\protect\citeauthoryear{{Ibata}, {Lewis}, {Irwin}, {Totten} \&
  {Quinn}}{{Ibata} et~al.}{2001}]{2001ApJ...551..294I}
{Ibata} R.,  {Lewis} G.~F.,  {Irwin} M.,  {Totten} E.,    {Quinn} T.,  2001,
  \apj, 551, 294

\bibitem[\protect\citeauthoryear{{Ibata}, {Lewis}, {Martin}, {Bellazzini} \&
  {Correnti}}{{Ibata} et~al.}{2013}]{2013ApJ...765L..15I}
{Ibata} R.,  {Lewis} G.~F.,  {Martin} N.~F.,  {Bellazzini} M.,    {Correnti}
  M.,  2013, \apjl, 765, L15

\bibitem[\protect\citeauthoryear{{Ibata}, {Irwin}, {Lewis}, {Ferguson} \&
  {Tanvir}}{{Ibata} et~al.}{2003}]{2003MNRAS.340L..21I}
{Ibata} R.~A.,  {Irwin} M.~J.,  {Lewis} G.~F.,  {Ferguson} A.~M.~N.,
  {Tanvir} N.,  2003, \mnras, 340, L21

\bibitem[\protect\citeauthoryear{{Jin} \& {Lynden-Bell}}{{Jin} \&
  {Lynden-Bell}}{2008}]{2008MNRAS.383.1686J}
{Jin} S.,  {Lynden-Bell} D.,  2008, \mnras, 383, 1686

\bibitem[\protect\citeauthoryear{{Jin} \& {Martin}}{{Jin} \&
  {Martin}}{2009}]{2009MNRAS.400L..43J}
{Jin} S.,  {Martin} N.~F.,  2009, \mnras, 400, L43

\bibitem[\protect\citeauthoryear{{Jing} \& {Suto}}{{Jing} \&
  {Suto}}{2002}]{2002ApJ...574..538J}
{Jing} Y.~P.,  {Suto} Y.,  2002, \apj, 574, 538

\bibitem[\protect\citeauthoryear{{Johnston}}{{Johnston}}{1998}]{1998ApJ...495..297J}
{Johnston} K.~V.,  1998, \apj, 495, 297

\bibitem[\protect\citeauthoryear{{Johnston}, {Bullock}, {Sharma}, {Font},
  {Robertson} \& {Leitner}}{{Johnston} et~al.}{2008}]{2008ApJ...689..936J}
{Johnston} K.~V.,  {Bullock} J.~S.,  {Sharma} S.,  {Font} A.,  {Robertson}
  B.~E.,    {Leitner} S.~N.,  2008, \apj, 689, 936

\bibitem[\protect\citeauthoryear{{Johnston}, {Law} \& {Majewski}}{{Johnston}
  et~al.}{2005}]{2005ApJ...619..800J}
{Johnston} K.~V.,  {Law} D.~R.,    {Majewski} S.~R.,  2005, \apj, 619, 800

\bibitem[\protect\citeauthoryear{{Johnston}, {Sackett} \& {Bullock}}{{Johnston}
  et~al.}{2001}]{2001ApJ...557..137J}
{Johnston} K.~V.,  {Sackett} P.~D.,    {Bullock} J.~S.,  2001, \apj, 557, 137

\bibitem[\protect\citeauthoryear{{Johnston}, {Spergel} \&
  {Hernquist}}{{Johnston} et~al.}{1995}]{1995ApJ...451..598J}
{Johnston} K.~V.,  {Spergel} D.~N.,    {Hernquist} L.,  1995, \apj, 451, 598

\bibitem[\protect\citeauthoryear{{Johnston}, {Zhao}, {Spergel} \&
  {Hernquist}}{{Johnston} et~al.}{1999}]{1999ApJ...512L.109J}
{Johnston} K.~V.,  {Zhao} H.,  {Spergel} D.~N.,    {Hernquist} L.,  1999,
  \apjl, 512, L109

\bibitem[\protect\citeauthoryear{{Juri{\'c}}, {Ivezi{\'c}}, {Brooks}, {Lupton},
  {Schlegel}, {Finkbeiner}, {Padmanabhan}, {Bond} \& {et al.}}{{Juri{\'c}}
  et~al.}{2008}]{2008ApJ...673..864J}
{Juri{\'c}} M.,  {Ivezi{\'c}} {\v Z}.,  {Brooks} A.,  {Lupton} R.~H.,
  {Schlegel} D.,  {Finkbeiner} D.,  {Padmanabhan} N.,  {Bond} N.,    {et al.}
  2008, \apj, 673, 864

\bibitem[\protect\citeauthoryear{{Kazantzidis}, {Abadi} \&
  {Navarro}}{{Kazantzidis} et~al.}{2010}]{2010ApJ...720L..62K}
{Kazantzidis} S.,  {Abadi} M.~G.,    {Navarro} J.~F.,  2010, \apjl, 720, L62

\bibitem[\protect\citeauthoryear{{Kollmeier}, {Gould}, {Shectman}, {Thompson},
  {Preston}, {Simon}, {Crane}, {Ivezi{\'c}} \& {Sesar}}{{Kollmeier}
  et~al.}{2009}]{2009ApJ...705L.158K}
{Kollmeier} J.~A.,  {Gould} A.,  {Shectman} S.,  {Thompson} I.~B.,  {Preston}
  G.~W.,  {Simon} J.~D.,  {Crane} J.~D.,  {Ivezi{\'c}} {\v Z}.,    {Sesar} B.,
  2009, \apjl, 705, L158

\bibitem[\protect\citeauthoryear{{Koposov}, {Rix} \& {Hogg}}{{Koposov}
  et~al.}{2010}]{2010ApJ...712..260K}
{Koposov} S.~E.,  {Rix} H.,    {Hogg} D.~W.,  2010, \apj, 712, 260

\bibitem[\protect\citeauthoryear{{Law} \& {Majewski}}{{Law} \&
  {Majewski}}{2010}]{2010ApJ...714..229L}
{Law} D.~R.,  {Majewski} S.~R.,  2010, \apj, 714, 229

\bibitem[\protect\citeauthoryear{{Law}, {Majewski} \& {Johnston}}{{Law}
  et~al.}{2009}]{2009ApJ...703L..67L}
{Law} D.~R.,  {Majewski} S.~R.,    {Johnston} K.~V.,  2009, \apjl, 703, L67

\bibitem[\protect\citeauthoryear{{Li} \& {Helmi}}{{Li} \&
  {Helmi}}{2008}]{2008MNRAS.385.1365L}
{Li} Y.-S.,  {Helmi} A.,  2008, \mnras, 385, 1365

\bibitem[\protect\citeauthoryear{{Lux}, {Read} \& {Lake}}{{Lux}
  et~al.}{2010}]{2010MNRAS.406.2312L}
{Lux} H.,  {Read} J.~I.,    {Lake} G.,  2010, \mnras, 406, 2312

\bibitem[\protect\citeauthoryear{{Lux}, {Read}, {Lake} \& {Johnston}}{{Lux}
  et~al.}{2012}]{2012MNRAS.tmpL.464L}
{Lux} H.,  {Read} J.~I.,  {Lake} G.,    {Johnston} K.~V.,  2012, \mnras,
  p.~L464

\bibitem[\protect\citeauthoryear{{Majewski}, {Kunkel}, {Law}, {Patterson},
  {Polak}, {Rocha-Pinto}, {Crane}, {Frinchaboy}, {Hummels}, {Johnston}, {Rhee},
  {Skrutskie} \& {Weinberg}}{{Majewski} et~al.}{2004}]{2004AJ....128..245M}
{Majewski} S.~R.,  {Kunkel} W.~E.,  {Law} D.~R.,  {Patterson} R.~J.,  {Polak}
  A.~A.,  {Rocha-Pinto} H.~J.,  {Crane} J.~D.,  {Frinchaboy} P.~M.,  {Hummels}
  C.~B.,  {Johnston} K.~V.,  {Rhee} J.,  {Skrutskie} M.~F.,    {Weinberg} M.,
  2004, \aj, 128, 245

\bibitem[\protect\citeauthoryear{{Majewski}, {Ostheimer}, {Rocha-Pinto},
  {Patterson}, {Guhathakurta} \& {Reitzel}}{{Majewski}
  et~al.}{2004}]{2004ApJ...615..738M}
{Majewski} S.~R.,  {Ostheimer} J.~C.,  {Rocha-Pinto} H.~J.,  {Patterson} R.~J.,
   {Guhathakurta} P.,    {Reitzel} D.,  2004, \apj, 615, 738

\bibitem[\protect\citeauthoryear{{Majewski}, {Skrutskie}, {Weinberg} \&
  {Ostheimer}}{{Majewski} et~al.}{2003}]{2003ApJ...599.1082M}
{Majewski} S.~R.,  {Skrutskie} M.~F.,  {Weinberg} M.~D.,    {Ostheimer} J.~C.,
  2003, \apj, 599, 1082

\bibitem[\protect\citeauthoryear{{Martin}, {Ibata} \& {Irwin}}{{Martin}
  et~al.}{2007}]{2007ApJ...668L.123M}
{Martin} N.~F.,  {Ibata} R.~A.,    {Irwin} M.,  2007, \apjl, 668, L123

\bibitem[\protect\citeauthoryear{{Mart{\'{\i}}nez-Delgado}, {Pe{\~n}arrubia},
  {Juri{\'c}}, {Alfaro} \& {Ivezi{\'c}}}{{Mart{\'{\i}}nez-Delgado}
  et~al.}{2007}]{2007ApJ...660.1264M}
{Mart{\'{\i}}nez-Delgado} D.,  {Pe{\~n}arrubia} J.,  {Juri{\'c}} M.,  {Alfaro}
  E.~J.,    {Ivezi{\'c}} Z.,  2007, \apj, 660, 1264

\bibitem[\protect\citeauthoryear{{Metz}, {Kroupa} \& {Jerjen}}{{Metz}
  et~al.}{2007}]{2007MNRAS.374.1125M}
{Metz} M.,  {Kroupa} P.,    {Jerjen} H.,  2007, \mnras, 374, 1125

\bibitem[\protect\citeauthoryear{{Miyamoto} \& {Nagai}}{{Miyamoto} \&
  {Nagai}}{1975}]{1975PASJ...27..533M}
{Miyamoto} M.,  {Nagai} R.,  1975, \pasj, 27, 533

\bibitem[\protect\citeauthoryear{{Munn}, {Monet}, {Levine}, {Canzian}, {Pier},
  {Harris}, {Lupton}, {Ivezi{\'c}}, {Hindsley}, {Hennessy}, {Schneider} \&
  {Brinkmann}}{{Munn} et~al.}{2004}]{2004AJ....127.3034M}
{Munn} J.~A.,  {Monet} D.~G.,  {Levine} S.~E.,  {Canzian} B.,  {Pier} J.~R.,
  {Harris} H.~C.,  {Lupton} R.~H.,  {Ivezi{\'c}} {\v Z}.,  {Hindsley} R.~B.,
  {Hennessy} G.~S.,  {Schneider} D.~P.,    {Brinkmann} J.,  2004, \aj, 127,
  3034

\bibitem[\protect\citeauthoryear{{Navarro}, {Frenk} \& {White}}{{Navarro}
  et~al.}{1996}]{1996ApJ...462..563N}
{Navarro} J.~F.,  {Frenk} C.~S.,    {White} S.~D.~M.,  1996, \apj, 462, 563

\bibitem[\protect\citeauthoryear{{Newberg}, {Willett}, {Yanny} \&
  {Xu}}{{Newberg} et~al.}{2010}]{2010ApJ...711...32N}
{Newberg} H.~J.,  {Willett} B.~A.,  {Yanny} B.,    {Xu} Y.,  2010, \apj, 711,
  32

\bibitem[\protect\citeauthoryear{{Newberg}, {Yanny}, {Rockosi}, {Grebel},
  {Rix}, {Brinkmann}, {Csabai}, {Hennessy}, {Hindsley}, {Ibata}, {Ivezi{\'c}},
  {Lamb}, {Nash}, {Odenkirchen}, {Rave}, {Schneider}, {Smith}, {Stolte} \&
  {York}}{{Newberg} et~al.}{2002}]{2002ApJ...569..245N}
{Newberg} H.~J.,  {Yanny} B.,  {Rockosi} C.,  {Grebel} E.~K.,  {Rix} H.,
  {Brinkmann} J.,  {Csabai} I.,  {Hennessy} G.,  {Hindsley} R.~B.,  {Ibata} R.,
   {Ivezi{\'c}} Z.,  {Lamb} D.,  {Nash} E.~T.,  {Odenkirchen} M.,  {Rave}
  H.~A.,  {Schneider} D.~P.,  {Smith} J.~A.,  {Stolte} A.,    {York} D.~G.,
  2002, \apj, 569, 245

\bibitem[\protect\citeauthoryear{{Newberg}, {Yanny} \& {Willett}}{{Newberg}
  et~al.}{2009}]{2009ApJ...700L..61N}
{Newberg} H.~J.,  {Yanny} B.,    {Willett} B.~A.,  2009, \apjl, 700, L61

\bibitem[\protect\citeauthoryear{{Odenkirchen}, {Grebel}, {Dehnen}, {Rix} \&
  {Cudworth}}{{Odenkirchen} et~al.}{2002}]{2002AJ....124.1497O}
{Odenkirchen} M.,  {Grebel} E.~K.,  {Dehnen} W.,  {Rix} H.,    {Cudworth}
  K.~M.,  2002, \aj, 124, 1497

\bibitem[\protect\citeauthoryear{{Odenkirchen}, {Grebel}, {Dehnen}, {Rix},
  {Yanny}, {Newberg}, {Rockosi}, {Mart{\'{\i}}nez-Delgado}, {Brinkmann} \&
  {Pier}}{{Odenkirchen} et~al.}{2003}]{2003AJ....126.2385O}
{Odenkirchen} M.,  {Grebel} E.~K.,  {Dehnen} W.,  {Rix} H.,  {Yanny} B.,
  {Newberg} H.~J.,  {Rockosi} C.~M.,  {Mart{\'{\i}}nez-Delgado} D.,
  {Brinkmann} J.,    {Pier} J.~R.,  2003, \aj, 126, 2385

\bibitem[\protect\citeauthoryear{{Odenkirchen}, {Grebel}, {Kayser}, {Rix} \&
  {Dehnen}}{{Odenkirchen} et~al.}{2009}]{2009AJ....137.3378O}
{Odenkirchen} M.,  {Grebel} E.~K.,  {Kayser} A.,  {Rix} H.,    {Dehnen} W.,
  2009, \aj, 137, 3378

\bibitem[\protect\citeauthoryear{{Odenkirchen}, {Grebel}, {Rockosi}, {Dehnen},
  {Ibata}, {Rix}, {Stolte}, {Wolf} \& {et. al.}}{{Odenkirchen}
  et~al.}{2001}]{2001ApJ...548L.165O}
{Odenkirchen} M.,  {Grebel} E.~K.,  {Rockosi} C.~M.,  {Dehnen} W.,  {Ibata} R.,
   {Rix} H.,  {Stolte} A.,  {Wolf} C.,    {et. al.} 2001, \apjl, 548, L165

\bibitem[\protect\citeauthoryear{{Pe{\~n}arrubia}, {Belokurov}, {Evans},
  {Mart{\'{\i}}nez-Delgado}, {Gilmore}, {Irwin}, {Niederste-Ostholt} \&
  {Zucker}}{{Pe{\~n}arrubia} et~al.}{2010}]{2010MNRAS.408L..26P}
{Pe{\~n}arrubia} J.,  {Belokurov} V.,  {Evans} N.~W.,
  {Mart{\'{\i}}nez-Delgado} D.,  {Gilmore} G.,  {Irwin} M.,
  {Niederste-Ostholt} M.,    {Zucker} D.~B.,  2010, \mnras, 408, L26

\bibitem[\protect\citeauthoryear{{Pe{\~n}arrubia}, {Zucker}, {Irwin}, {Hyde},
  {Lane}, {Lewis}, {Gilmore}, {Wyn Evans} \& {Belokurov}}{{Pe{\~n}arrubia}
  et~al.}{2011}]{2011ApJ...727L...2P}
{Pe{\~n}arrubia} J.,  {Zucker} D.~B.,  {Irwin} M.~J.,  {Hyde} E.~A.,  {Lane}
  R.~R.,  {Lewis} G.~F.,  {Gilmore} G.,  {Wyn Evans} N.,    {Belokurov} V.,
  2011, \apjl, 727, L2+

\bibitem[\protect\citeauthoryear{{Prior}, {Da Costa}, {Keller} \&
  {Murphy}}{{Prior} et~al.}{2009}]{2009ApJ...691..306P}
{Prior} S.~L.,  {Da Costa} G.~S.,  {Keller} S.~C.,    {Murphy} S.~J.,  2009,
  \apj, 691, 306

\bibitem[\protect\citeauthoryear{{Read}, {Lake}, {Agertz} \&
  {Debattista}}{{Read} et~al.}{2008}]{2008MNRAS.389.1041R}
{Read} J.~I.,  {Lake} G.,  {Agertz} O.,    {Debattista} V.~P.,  2008, \mnras,
  389, 1041

\bibitem[\protect\citeauthoryear{{Read} \& {Moore}}{{Read} \&
  {Moore}}{2005}]{2005MNRAS.361..971R}
{Read} J.~I.,  {Moore} B.,  2005, \mnras, 361, 971

\bibitem[\protect\citeauthoryear{{Rocha-Pinto}, {Majewski}, {Skrutskie},
  {Crane} \& {Patterson}}{{Rocha-Pinto} et~al.}{2004}]{2004ApJ...615..732R}
{Rocha-Pinto} H.~J.,  {Majewski} S.~R.,  {Skrutskie} M.~F.,  {Crane} J.~D.,
  {Patterson} R.~J.,  2004, \apj, 615, 732

\bibitem[\protect\citeauthoryear{{Sales}, {Helmi}, {Starkenburg}, {Morrison},
  {Engle}, {Harding}, {Mateo}, {Olszewski} \& {Sivarani}}{{Sales}
  et~al.}{2008}]{2008MNRAS.389.1391S}
{Sales} L.~V.,  {Helmi} A.,  {Starkenburg} E.,  {Morrison} H.~L.,  {Engle} E.,
  {Harding} P.,  {Mateo} M.,  {Olszewski} E.~W.,    {Sivarani} T.,  2008,
  \mnras, 389, 1391

\bibitem[\protect\citeauthoryear{{Sandage} \& {Hartwick}}{{Sandage} \&
  {Hartwick}}{1977}]{1977AJ.....82..459S}
{Sandage} A.,  {Hartwick} F.~D.~A.,  1977, \aj, 82, 459

\bibitem[\protect\citeauthoryear{{Sanders} \& {Binney}}{{Sanders} \&
  {Binney}}{2013}]{2013MNRAS.433.1813S}
{Sanders} J.~L.,  {Binney} J.,  2013, \mnras, 433, 1813

\bibitem[\protect\citeauthoryear{{Sesar}, {Ivezi{\'c}}, {Lupton}, {Juri{\'c}},
  {Gunn}, {Knapp}, {DeLee}, {Smith} \& {et al.}}{{Sesar}
  et~al.}{2007}]{2007AJ....134.2236S}
{Sesar} B.,  {Ivezi{\'c}} {\v Z}.,  {Lupton} R.~H.,  {Juri{\'c}} M.,  {Gunn}
  J.~E.,  {Knapp} G.~R.,  {DeLee} N.,  {Smith} J.~A.,    {et al.} 2007, \aj,
  134, 2236

\bibitem[\protect\citeauthoryear{{Sesar}, {Vivas}, {Duffau} \&
  {Ivezi{\'c}}}{{Sesar} et~al.}{2010}]{2010ApJ...717..133S}
{Sesar} B.,  {Vivas} A.~K.,  {Duffau} S.,    {Ivezi{\'c}} {\v Z}.,  2010, \apj,
  717, 133

\bibitem[\protect\citeauthoryear{Simard, Willmer, Vogt, Sarajedini, Phillips,
  Weiner, Koo, Im, Illingworth \& Faber}{Simard
  et~al.}{2002}]{Simard:2002p2258}
Simard L.,  Willmer C. N.~A.,  Vogt N.~P.,  Sarajedini V.~L.,  Phillips A.~C.,
  Weiner B.~J.,  Koo D.~C.,  Im M.,  Illingworth G.~D.,    Faber S.~M.,  2002,
  The Astrophysical Journal Supplement Series, 142, 1

\bibitem[\protect\citeauthoryear{{Sofue}, {Honma} \& {Omodaka}}{{Sofue}
  et~al.}{2009}]{2009PASJ...61..227S}
{Sofue} Y.,  {Honma} M.,    {Omodaka} T.,  2009, \pasj, 61, 227

\bibitem[\protect\citeauthoryear{{Varghese}, {Ibata} \& {Lewis}}{{Varghese}
  et~al.}{2011}]{2011MNRAS.417..198V}
{Varghese} A.,  {Ibata} R.,    {Lewis} G.~F.,  2011, \mnras, 417, 198

\bibitem[\protect\citeauthoryear{{Vivas}, {Jaff{\'e}}, {Zinn}, {Winnick},
  {Duffau} \& {Mateu}}{{Vivas} et~al.}{2008}]{2008AJ....136.1645V}
{Vivas} A.~K.,  {Jaff{\'e}} Y.~L.,  {Zinn} R.,  {Winnick} R.,  {Duffau} S.,
  {Mateu} C.,  2008, \aj, 136, 1645

\bibitem[\protect\citeauthoryear{{Watkins}, {Evans}, {Belokurov}, {Smith},
  {Hewett}, {Bramich}, {Gilmore}, {Irwin}, {Vidrih}, {Wyrzykowski} \&
  {Zucker}}{{Watkins} et~al.}{2009}]{2009MNRAS.398.1757W}
{Watkins} L.~L.,  {Evans} N.~W.,  {Belokurov} V.,  {Smith} M.~C.,  {Hewett}
  P.~C.,  {Bramich} D.~M.,  {Gilmore} G.~F.,  {Irwin} M.~J.,  {Vidrih} S.,
  {Wyrzykowski} {\L}.,    {Zucker} D.~B.,  2009, \mnras, 398, 1757

\bibitem[\protect\citeauthoryear{{Willett}, {Newberg}, {Zhang}, {Yanny} \&
  {Beers}}{{Willett} et~al.}{2009}]{2009ApJ...697..207W}
{Willett} B.~A.,  {Newberg} H.~J.,  {Zhang} H.,  {Yanny} B.,    {Beers} T.~C.,
  2009, \apj, 697, 207

\bibitem[\protect\citeauthoryear{{Xue}, {Rix}, {Zhao}, {Re Fiorentin}, {Naab},
  {Steinmetz}, {van den Bosch}, {Beers}, {Lee}, {Bell}, {Rockosi}, {Yanny},
  {Newberg}, {Wilhelm}, {Kang}, {Smith} \& {Schneider}}{{Xue}
  et~al.}{2008}]{2008ApJ...684.1143X}
{Xue} X.~X.,  {Rix} H.~W.,  {Zhao} G.,  {Re Fiorentin} P.,  {Naab} T.,
  {Steinmetz} M.,  {van den Bosch} F.~C.,  {Beers} T.~C.,  {Lee} Y.~S.,  {Bell}
  E.~F.,  {Rockosi} C.,  {Yanny} B.,  {Newberg} H.,  {Wilhelm} R.,  {Kang} X.,
  {Smith} M.~C.,    {Schneider} D.~P.,  2008, \apj, 684, 1143

\bibitem[\protect\citeauthoryear{{Yanny}, {Newberg}, {Grebel}, {Kent},
  {Odenkirchen}, {Rockosi}, {Schlegel}, {Subbarao}, {Brinkmann}, {Fukugita},
  {Ivezic}, {Lamb}, {Schneider} \& {York}}{{Yanny}
  et~al.}{2003}]{2003ApJ...588..824Y}
{Yanny} B.,  {Newberg} H.~J.,  {Grebel} E.~K.,  {Kent} S.,  {Odenkirchen} M.,
  {Rockosi} C.~M.,  {Schlegel} D.,  {Subbarao} M.,  {Brinkmann} J.,  {Fukugita}
  M.,  {Ivezic} {\v Z}.,  {Lamb} D.~Q.,  {Schneider} D.~P.,    {York} D.~G.,
  2003, \apj, 588, 824

\bibitem[\protect\citeauthoryear{{Yanny}, {Newberg}, {Grebel}, {Kent},
  {Odenkirchen}, {Rockosi}, {Schlegel}, {Subbarao}, {Brinkmann}, {Fukugita},
  {Ivezic}, {Lamb}, {Schneider} \& {York}}{{Yanny}
  et~al.}{2004}]{2004ApJ...605..575Y}
{Yanny} B.,  {Newberg} H.~J.,  {Grebel} E.~K.,  {Kent} S.,  {Odenkirchen} M.,
  {Rockosi} C.~M.,  {Schlegel} D.,  {Subbarao} M.,  {Brinkmann} J.,  {Fukugita}
  M.,  {Ivezic} {\v Z}.,  {Lamb} D.~Q.,  {Schneider} D.~P.,    {York} D.~G.,
  2004, \apj, 605, 575

\bibitem[\protect\citeauthoryear{{Yanny}, {Newberg}, {Johnson}, {Lee}, {Beers},
  {Bizyaev}, {Brewington}, {Fiorentin}, {Harding}, {Malanushenko},
  {Malanushenko}, {Oravetz}, {Pan}, {Simmons} \& {Snedden}}{{Yanny}
  et~al.}{2009}]{2009ApJ...700.1282Y}
{Yanny} B.,  {Newberg} H.~J.,  {Johnson} J.~A.,  {Lee} Y.~S.,  {Beers} T.~C.,
  {Bizyaev} D.,  {Brewington} H.,  {Fiorentin} P.~R.,  {Harding} P.,
  {Malanushenko} E.,  {Malanushenko} V.,  {Oravetz} D.,  {Pan} K.,  {Simmons}
  A.,    {Snedden} S.,  2009, \apj, 700, 1282

\bibitem[\protect\citeauthoryear{{Zhao}}{{Zhao}}{2004}]{2004MNRAS.351..891Z}
{Zhao} H.,  2004, \mnras, 351, 891

\end{thebibliography}

\end{document}